\documentclass[a4paper,11pt]{article}
\pdfoutput=1 

\usepackage{jheppub} 

\usepackage[T1]{fontenc} 

\usepackage[utf8]{inputenc} 
\usepackage[english]{babel}
\usepackage{tikz-cd}
\usepackage[bbgreekl]{mathbbol}
\usepackage{amsmath,amsfonts,amssymb,amsthm}
\usepackage{csquotes}
\usepackage{cancel,verbatim,slashed,todonotes}
\usepackage{xcolor,mdframed,graphicx,color}
\usepackage{dsfont}
\usepackage{mathtools}
\usepackage{multirow,stmaryrd}
\usepackage{array}
\usepackage{makecell}
\usepackage{upgreek}
\usepackage{braket}
\DeclareSymbolFontAlphabet{\mathbbvar}{bbold}
\DeclareSymbolFontAlphabet{\mathbb}{AMSb}

\newcommand{\di}{\mathrm{d}}

\newcommand{\Coo}{\mathcal{C}^{\infty}}

\preprint{QMUL-PH-21-04}

\title{\boldmath Double Field Theory and Geometric Quantisation}

 \author{Luigi Alfonsi}
 \author{and David S. Berman}
 \affiliation{Centre for Research in String Theory, School of Physics and Astronomy,\\ Queen Mary University of London, 327 Mile End Road, London E1 4NS, UK}

\emailAdd{l.alfonsi@qmul.ac.uk}
\emailAdd{d.s.berman@qmul.ac.uk}

\abstract{We examine various properties of double field theory and the doubled string sigma model in the context of geometric quantisation. In particular we look at
T-duality as the symplectic transformation related to an alternative choice of polarisation
in the construction of the quantum bundle for the string. Following this perspective we
adopt a variety of techniques from geometric quantisation to study the doubled space. One
application is the construction of the "double coherent state" that provides the shortest
distance in any duality frame and a "stringy deformed" Fourier transform.}


\begin{document}
\maketitle
\section{Introduction}

Geometric quantisation provides an approach to quantisation that is underpinned by the symplectic geometry of phase space. Its emergence in the 1970s from the work of Kostant and Souriau has produced a geometric approach to quantisation that provides numerous insights into the quantisation procedure. In particular, it showed how the symplectic symmetry of phase space is broken in naive quantisation methods even though the physics is left invariant and how the underlying symplectic symmetry maybe restored (or even extended to the metaplectic group). In more mundane language, classical Hamiltonian physics is invariant under canonical transformations and yet the wavefunctions of quantum mechanics are functions of just half the coordinates of phase space and thus not symplectic representations. A key part of quantum mechanics is that physics cannot depend on the choice of basis of wavefunctions. We can transform between the coordinate and momentum basis and the physics is invariant. In fact, the coordinate and momentum representations are mutually non-local and to move between different bases requires a nonlocal transformation (this is the Fourier transform, in a free theory). This is highly analogous to duality symmetries in field theories. To take the example of Maxwell theory, electromagnetic duality is manifest in the Hamiltonian form of the theory, the transformation between electric and magnetic variables maybe generated by symplectic transformations and yet from the Lagrangian perspective (where the duality is not manifest) the electric and magnetic variables are mutually non-local and related via a Fourier transform \cite{Lozano:1995aq}.

The recovery of manifest symplectic symmetry in quantisation was an early motivator to introduce new quantisations methods. First, Weyl \cite{Weyl} in 1927 and then Groenewold \cite{Groenewold:1946kp} and Moyal \cite{Moyal:1949sk} in the 1940s produced reformulations of quantum mechanics to make the symplectic symmetry of the classical Hamiltonian form a manifest symmetry in quantisation. Out of this came the ideas of deformation quantisation where the  algebra of functions is deformed.

From a contemporary perspective the emergence of duality in string theory has thrown up a similar set of challenges. As is well known, the Hamiltonian of the string in toroidal backgrounds has a symmetry that is not manifest in the Lagrangian formulation. T-duality may be generated by canonical transformations of the string phase space, see \cite{AAL94, Lozano:1996sc} but from the traditional Lagrangian world sheet perspective the world sheet fields and their T-duals are mutually non-local.

Double Field theory, first introduced by Siegel \cite{Siegel:1993bj, Siegel:1993th} and then as a truncation of string field theory by Hull and Zwiebach \cite{Hull:2009mi}, reformulates the string background spacetime in a way to make T-duality a manifest symmetry. It does so, as the eponymous title of the theory suggests, by doubling the dimensions of space time to include both duality perspectives into one object. See \cite{Berman:2013eva, Hohm:2013bwa,Aldazabal:2013sca} for reviews of this approach. From the perspective of this paper, the doubling will be just returning to the original phase space of the string. This is related to the work of Tseytlin \cite{Tseytlin:1990nb,Tseytlin:1990va,Tseytlin:1991wr,Tseytlin:1996ne} and Duff \cite{Duff:1984hn,Duff:1989tf,Duff:1990hn} where the world sheet theory was reformulated in a doubled space and T-duality a manifest symmetry. 

Here we will examine string quantisation using the ideas and techniques of geometric quantisation. This will produce an infinite dimensional phase space associated to the loop space of the string. The zero modes of the loop space will give us the doubled space of DFT and the choice of polarisation in the quantisation will provide the T-duality frame. We will construct transformations between T-dual descriptions based on the transformations induced by different choices of polarisation. This will lead to the idea of a coherent state that saturates the uncertainty bound on distance in the doubled theory i.e. the shortest distance in any duality frame. 

This quantisation procedure will produce a noncommutative algebra associated to the doubled phase space. As usual the noncommutative nature of position and momentum is controlled by the dimensionful deformation parameter $\hbar$. Now though in addition we will also have an additional deformation parameter given by $\hbar \alpha'$ which controls the noncommutativity of coordinates $x^i$ and their "duals" $\tilde{x}_j$. $\alpha'$ is the square of the string length scale. Thus the space has two types of noncommutativity or quantisations, one has its origins in the usual phase space that that requires the traditional choice of polarisation to picking the Lagrangian submanifold of phase space and another which requires picking a Lagrangian submanifold in the doubled space or in simpler language picking the "duality frame".

To make the paper as self contained as possible we will first review the aspects of geometric quantisation that we will use for the string and hopefully foreshadow aspects double field theory.

We will then move on to the string loop space description and its phase space followed by the quantisation and the identification of the doubled space. We will then be equipped to use more of the machinery of geometric quantisation and produce the double coherent state and the transforms between T-dual frames.

We will end with a discussion on further directions and the implications for objects like T-folds.

Note that there has been other related work looking at double field theory from a world sheet perspective in \cite{Berman:2007xn, Lee:2013hma,Park:2016sbw,Basile:2019pic}  and also \cite{DeAngelis:2013wba}. Of particular overlap with the approach in this paper is the work \cite{Marotta:2018myj,Marotta:2019eqc}.

\subsection{Fundamentals of geometric quantisation}
Much of this can be found in any one of the books by \cite{WeiGQ, Kostant, Souriau, Woodhouse:1980pa} or the recent review by \cite{Nair:2016ufy}.

\paragraph{Hamiltonian mechanics.}
Let us recall that a symplectic manifold $(\mathcal{P},\omega)$ is defined as a smooth manifold $\mathcal{P}$ equipped with a closed non-degenerate $2$-form $\omega\in\Omega^2(M)$, called the symplectic form.
In Hamiltonian mechanics, a \textit{classical system} $(\mathcal{P},\omega,H)$ is defined by a symplectic manifold $(\mathcal{P},\omega)$, describing the  \textit{phase space} of the system, and a smooth function $H\in\mathcal{C}^\infty(\mathcal{P})$, called the \textit{Hamiltonian}. In Newtonian terms, the phase space and the Hamiltonian encode respectively the kinematics and the dynamics of a classical system. The equations of motion are described Hamilton's equation as follows:
\begin{equation}
     \iota_{X_H}\omega \,=\, \di H.
\end{equation}
A vector field $X_H\in\mathfrak{X}(\mathcal{P})$ which solves the Hamilton equation is called \textit{Hamiltonian vector} for the Hamiltonian $H$. The flow of a Hamiltonian vector fields describes the motion of the classical system on the phase space. This means if we choose a starting point $\gamma_0\in M$ in phase space, the motion of the classical system will be given by the path
\begin{equation}
\begin{aligned}
    \gamma: \mathbb{R} \;&\longrightarrow\; \mathcal{P} \\
     \tau \;&\longmapsto\; \gamma(\tau) = e^{\tau X_H}\gamma_0  \, 
\end{aligned}
\end{equation}
where $\tau\in\mathbb{R}$ is a $1$-dimensional parameter.

Locally, on an simply connected open subset $U\subset \mathcal{P}$, we can apply Poincar\'{e} lemma to the symplectic form and find $\omega = \di \theta$, where the local $1$-form $\theta\in\Omega^1(U)$ is called Liouville potential. The definition of the Liouville potential is gauge dependent, meaning that any other choice of potential $\theta' = \theta + \di \lambda$ with $\lambda\in\Coo(U)$ equally satisfies $\omega= \di \theta'$. 
Now, given a path $\gamma:\mathbb{R}\rightarrow \mathcal{P}$ on the phase space, we define the Lagrangian $L_H\in\Omega^1(\mathbb{R})$ by
\begin{equation}
    L_H \;=\; \gamma^\ast\theta - H\di\tau \, .
\end{equation}
Where we denote the pull-back of the Liouville one-form $\theta$ to the curve $\gamma$ by $\gamma^\ast\theta$.
The action $S_H[\gamma(\tau)]$ associated to such a Lagrangian will be given by
\begin{equation}
    S_H[\gamma(\tau)] \;=\; \int_\mathbb{R}(\gamma^\ast\theta - H\di\tau)  \, .
\end{equation}
For future use, let us notice that we can rewrite $\gamma^\ast\theta=\iota_{X_H}\theta\,\di\tau$, when restricted on the path $\gamma$. Thus note that the choice of Liouville potential effects the Lagrangian description.

\paragraph{Classical algebra of observables.}
An observable is defined as a smooth function $f\in\mathcal{C}^\infty(\mathcal{P})$ of the phase space.
Crucially, a symplectic manifold $(\mathcal{P},\omega)$ is canonically also a Poisson manifold $(\mathcal{P}, \{-,-\})$, where the Poisson bracket is given as it follows:
\begin{equation}\label{eq:gq1}
    \{f,g\} \;:=\; \omega(X_f,X_g)
\end{equation}
for any pair of observables $f,g\in\mathcal{C}^\infty(\mathcal{P})$. In other words, this means that the observables of a symplectic manifold $(M,\omega)$ constitute a Poisson algebra:
\begin{equation}\label{eq:gq2}
    [X_f,X_g] \;=\; X_{\{f,g\}}  \, .
\end{equation}
Thus, the Hamiltonian vector fields on a symplectic manifold $(\mathcal{P},\omega)$ constitute a Lie algebra, which we will denote as $\mathfrak{ham}(\mathcal{P},\omega)$.

\paragraph{Prequantum geometry.}
Let us consider the Lie group $U(1)_\hbar:=\mathbb{R}/2\pi\hbar\mathbb{Z}$. The \textit{prequantum bundle} $\mathcal{Q}\twoheadrightarrow \mathcal{P}$ is defined as the principal $U(1)_\hbar$-bundle, whose first Chern class $\mathrm{c}_1(\mathcal{Q})\in H^2(M,\mathbb{Z})$ is the image of the element $[\omega]\in H^2(M,\mathbb{R})$ of the de Rham cohomology group. We can now define the associated bundle $\mathcal{E}\twoheadrightarrow \mathcal{P}$ to the prequantum bundle with fibre $\mathbb{C}$, i.e.
\begin{equation}
    \mathcal{E} \,:=\, \mathcal{Q}\times_{U(1)_\hbar}\!\mathbb{C},
\end{equation}
where the natural action $U(1)_\hbar\times \mathbb{C}\rightarrow\mathbb{C}$ is given by the map $(\phi,z)\mapsto e^{\frac{i}{\hbar}\phi}z$. Now, the \textit{prequantum Hilbert space} of the system is defined by
\begin{equation}
    \mathbf{H}_{\mathrm{pre}}\;:=\; \mathrm{L}^{\!2}(\mathcal{P},\mathcal{E}),
\end{equation}
i.e. the Hilbert space of $\mathrm{L}^2$-integrable sections of the bundle $\mathcal{E}$ on the base manifold $\mathcal{P}$. Whenever the first Chern class of $\mathcal{Q}$ is trivial, then the bundle $\mathcal{E}=\mathcal{P}\times\mathbb{C}$ is trivial and the prequantum Hilbert space reduces to $\mathbf{H}_{\mathrm{pre}} = \mathrm{L}^{\!2}(\mathcal{P};\mathbb{C})$, i.e. the Hilbert space of $\mathrm{L}^2$-integrable complex functions.

\paragraph{Quantum algebra of observables.}
In geometric quantisation, given a classical observable $f\in\Coo(\mathcal{P})$, we define a \textit{quantum observable} $\hat{f}\in\mathrm{Aut}(\mathbf{H}_{\mathrm{pre}})$ by the expression
\begin{equation}
    \hat{f} \;:=\; i\hbar \nabla_{V_f} + f,
\end{equation}
where $V_f\in\mathfrak{ham}(\mathcal{P},\omega)$ is the Hamiltonian vector of the Hamiltonian function $f\in\Coo(\mathcal{P})$ and $\nabla$ is the connection on $T\mathcal{P}$ given by the Liouville potential $\theta$.
By using equations \eqref{eq:gq1} and \eqref{eq:gq2}, we find that the commutator of two quantum observables closes. In particular, given $\hat{f}$ and $\hat{g}$, we obtain the commutator
\begin{equation}
    [\hat{f},\hat{g}] \;=\; i\hbar\, \widehat{\{f,g\}}
\end{equation}
where the classical observable of $\widehat{\{f,g\}}$ is the Poisson bracket $\{f,g\}$ of the classical observables $f$ and $g$. Thus, we can use this fact to define the Heisenberg Lie algebra $\mathfrak{heis}(\mathcal{P},\omega)$ of quantum observables on our phase space $(\mathcal{P},\omega)$.

\paragraph{Quantum geometry.}
Denote the tangent bundle of phase space by  $T\mathcal{P}$.
A \textit{polarisation} of the phase space $(\mathcal{P},\omega)$ is an involutive Lagrangian subbundle $L\subset T\mathcal{P}$, i.e. an $n$-dimensional subbundle of $T\mathcal{P}$ such that $\omega|_{L}=0$ and $[V,W]\subset L $ for any pair of vectors $V,W\in L$.

The \textit{square root bundle} of a line bundle $\mathcal{B}\twoheadrightarrow\mathcal{M}$ is defined as a complex line bundle, which we will denote as $\sqrt{\mathcal{B}}\twoheadrightarrow\mathcal{M}$, equipped with a bundle isomorphism $\sqrt{\mathcal{B}}\otimes \sqrt{\mathcal{B}} \;\xrightarrow{\;\;\simeq\;\;}\; \mathcal{B}$ which sends sections $\sqrt{s}\in\Gamma(\mathcal{M},\sqrt{\mathcal{B}})$ to $\sqrt{s}\otimes\sqrt{s}\mapsto s\in\Gamma(\mathcal{M},\mathcal{B})$.

Let us consider the determinant bundle $\mathrm{det}(L):=\wedge^nL^\ast_{\mathbb{C}}$ of a Lagrangian subbundle $L\subset T\mathcal{P}$ of our phase space, where $n=\mathrm{rank}(L)$. We need now to consider the square root bundle $\sqrt{\mathrm{det}(L)}$ of the determinant bundle, which comes equipped with the isomorphism
\begin{equation}
    \sqrt{\mathrm{det}(L)} \otimes \sqrt{\mathrm{det}(L)} \;\xrightarrow{\;\;\simeq\;\;}\;   \mathrm{det}(L)
\end{equation}
The choice of square root bundle $\sqrt{\mathrm{det}(L)}$ is also related to the \textit{metaplectic correction} which leads to the quantum theory forming a representation of the metaplectic group rather than the symplectic group..
The \textit{quantum Hilbert space} is defined by the following space of sections:
\begin{equation}\label{eq:quantumhilbert}
    \mathbf{H} \;:=\; \Big\{\; \psi\in\mathrm{L}^{\!2}\big(\mathcal{P},\,\mathcal{E}\otimes\sqrt{\mathrm{det}(L)}\big) \;\Big|\; \nabla_V\psi=0\;\;\,\forall\,V\in L \; \Big\}.
\end{equation}
If the Lagrangian subbundle $L$ is integrable, we can write $L=T\mathcal{M}$ for some $n$-dimensional submanifold $\mathcal{M}\subset \mathcal{P}$ of the phase space. Then, quantum states $\ket{\psi}\in\mathbf{H}$ can be uniquely chosen of the form
\begin{equation}
    \ket{\psi} = \psi\otimes \sqrt{\mathrm{vol}_\mathcal{M}},
\end{equation}
where $\psi\in\mathrm{L}^{\!2}(\mathcal{M},\mathcal{E})$ is a polarised section and $\sqrt{\mathrm{vol}_\mathcal{M}}\in\Gamma(\mathcal{P},\sqrt{\mathrm{det}(L)})$ is the half-form whose square is a fixed volume form $\mathrm{vol}_\mathcal{M}\in\Omega^n(\mathcal{M})$.
The inner product of the Hilbert space is given by the integral
\begin{equation}
    \Braket{\psi_1|\psi_2} \;=\; \int_\mathcal{M}\psi_1^\ast\psi_2\,\mathrm{vol}_\mathcal{M}
\end{equation}
for any couple of quantum states $\ket{\psi_1} = \psi_1\otimes \sqrt{\mathrm{vol}_\mathcal{M}}$ and $\ket{\psi_2} = \psi_2\otimes \sqrt{\mathrm{vol}_\mathcal{M}}\in\mathbf{H}$.

Crucially, the Hilbert space defined in \eqref{eq:quantumhilbert} does not depend on the choice of Lagrangian subbundle $L$. If we call $\mathbf{H}_L$ the quantum Hilbert space polarised along the Lagrangian subbundle $L$ and $\mathbf{H}_{L'}$ the one along another Lagrangian subbundle $L'$, we have a canonical isomorphism $\mathbf{H}_L\cong\mathbf{H}_{L'}$. At the end of this section we will explain why this is the case.

\paragraph{Example: wave-functions of QM.}
To illustrate the ideas in this section lets look at a simple example with $(M,\omega)=(\mathbb{R}^{2n},\,\di p_\mu \wedge \di x^\mu)$, where $\{p_\mu,x^\mu\}$ are Darboux coordinates on $\mathbb{R}^{2n}$. We can now choose the gauge $\theta = p_\mu\di x^\mu$ for the Liouville potential. We have two perpendicular polarisations defined by the Lagrangian fibrations $L_p:=\mathrm{Span}\!\left(\frac{\partial}{\partial p_\mu}\right)$ and $L_x:=\mathrm{Span}\!\left(\frac{\partial}{\partial x^\mu}\right)$. Recall that the covariant derivative is related to the Liouville potential by $\nabla_V = V - \frac{i}{\hbar}\iota_V\theta$. In our case, this implies
\begin{equation}
    \begin{aligned}
    \nabla_{\!\frac{\partial}{\partial x^\mu}} \;&=\; \frac{\partial}{\partial x^\mu} - \frac{i}{\hbar} p_\mu \\[0.4ex]
    \nabla_{\!\frac{\partial}{\partial p_\mu}} \;&=\; \frac{\partial}{\partial p_\mu}  .
    \end{aligned}
\end{equation}
Therefore, for the polarisation $L_p$ and $L_x$, we obtain respectively the sections 
\begin{equation}
    \begin{aligned}
    \Ket{\psi} \;&=\; \psi(p)e^{-ip_\mu x^\mu} \otimes \sqrt{\di^n p} \\
    \Ket{\psi} \;&=\; \psi(x) \otimes \sqrt{\di^n x},
    \end{aligned}
\end{equation}
where $\sqrt{\di^n p}$ is the half form such that $\sqrt{\di^n p}\otimes \sqrt{\di^n p}=\di^n p$ and analogously for $\sqrt{\di^n x}$.

\subsection{Canonical transformations and polarisations}

\paragraph{Canonical transformations as symplectomorphisms.}
Let us recall that a \textit{symplectomorphism} between two manifolds $(\mathcal{P},\omega)\xrightarrow{\;f\;}(\mathcal{P}',\omega')$ is a diffeomorphism $f:\mathcal{P}\rightarrow \mathcal{P}'$ which maps the symplectic form of the first manifold into the symplectic form of the second one, i.e. such that it satisfies $\omega = f^\ast\omega'$.
According to \cite{Arn89}, what in Hamiltonian physics is known under the name of \textit{canonical transformation} with generating function $F$ is equivalently a symplectomorphism $f:(\mathcal{P},\omega)\rightarrow(\mathcal{P}',\omega')$ such that the Liouville potential is gauge-transformed by $\theta-f^\ast\theta' = \di F$. However, this first formalisation can be significantly refined.

\paragraph{Lagrangian correspondence.}
Following \cite{WeinLag}, there exists a powerful way to formalise a canonical transformation by using the notion of \textit{Lagrangian correspondence}. To define a Lagrangian correspondence we first need to introduce the \textit{graph} of a symplectomorphism $f:(\mathcal{P},\omega)\rightarrow(\mathcal{P}',\omega')$, which is the submanifold of the product space $\mathcal{P}\times \mathcal{P}'$ given by
\begin{equation}
    \Gamma_f \;:=\; \big\{ (a,b)\in \mathcal{P}\times\mathcal{P}' \;\big|\; b=f(a)\big\}.
\end{equation} 
Let us call $\iota:\Gamma_f\hookrightarrow\mathcal{P}\times \mathcal{P}'$ the inclusion in the product space. Now, a Lagrangian correspondence is defined a correspondence diagram of the form
\begin{equation}\label{diag:lagrangiancorr}
    \begin{tikzcd}[row sep={13ex,between origins}, column sep={13ex,between origins}]
    &  (\mathcal{P}\times \mathcal{P}',\, \pi^\ast\omega-\pi^{\prime\ast}\omega') \arrow[rd, "\pi'", two heads]\arrow[ld, "\pi"', two heads] & \\
    (\mathcal{P},\,\omega) \arrow[rr, "f"] && (\mathcal{P}',\,\omega')
    \end{tikzcd}
\end{equation}
where $f$ is a symplectomorphism and $\pi,\pi'$ are the canonical projections of $\mathcal{P}\times \mathcal{P}'$ onto $\mathcal{P}, \mathcal{P}'$ respectively. The submanifold $\Gamma_f\subset\mathcal{P}\times\mathcal{P}'$ can be immediately recognised as a Lagrangian submanifold of $(\mathcal{P}\times \mathcal{P}',\, \pi^\ast\omega-\pi^{\prime\ast}\omega')$, i.e. the total symplectic form vanishes when restricted on $\Gamma_f$. In other words, we have
\begin{equation}
    \iota^\ast\!\left(\pi^\ast\omega-\pi^{\prime\ast}\omega'\right) \;=\; 0.
\end{equation}
To formalise a canonical transformation, we need to add another condition: the correspondence space $(\mathcal{P}\times \mathcal{P}',\, \pi^\ast\omega-\pi^{\prime\ast}\omega')$ must be symplectomorphic to a symplectic manifold $(T^\ast\mathcal{M},\, \omega_\mathrm{can})$ for some manifold $\mathcal{M}$, where $\omega_\mathrm{can}\in\Omega^2(T^\ast\mathcal{M})$ is just the canonical symplectic form of the cotangent bundle.

This implies that we can write the combination of Liouville potentials $\pi^\ast\theta - \pi^{\prime\ast}\theta^{\prime}$ as the Liouville 1-form on $\mathcal{P}\times \mathcal{P}'\cong T^\ast\mathcal{M}$. Since $\Gamma_f$ is Lagrangian, the Liouville potential can be trivialised on $T\Gamma_f$. In other words, we have the equation
\begin{equation}
    \pi^\ast\theta - \pi^{\prime\ast}\theta^{\prime} \;=\; \di (F\circ \Pi) \qquad \text{on }\;T\Gamma_f,
\end{equation}
where $\Pi:T^\ast\mathcal{M}\twoheadrightarrow\mathcal{M}$ is the canonical projection and where the function $F\in\Coo(\mathcal{M})$ can be interpreted as the \textit{generating function} of the canonical transformation associated to the symplectomorphism $f$.

\paragraph{Example: canonical transformations.}
For clarity, let us consider a simple example. Let us start from symplectic manifolds which are cotangent bundles of configuration spaces, i.e. $\mathcal{P}=T^\ast M$ and $\mathcal{P}'=T^\ast M'$. Thus we can write the Liouville potential as

\begin{equation}\label{eq:extype1}
    p_\mu \di x^\mu -  p'_\mu\di x^{\prime\mu} \,=\, \di F
\end{equation}
in local coordinates on the correspondence space $\mathcal{P}\times\mathcal{P}' = T^\ast(M\times M')$. We immediately notice that, in the notation of the previous paragraph, we have $\mathcal{M}:=M\times M'$. Now the generating function $F=F(x,x')$ of the canonical transformation can be properly seen as the pullback of a function of the product manifold $M\times M'$. Equation \eqref{eq:extype1} can be equivalently written as
\begin{equation}
    p_\mu \,=\, \frac{\partial F}{\partial x^\mu}, \qquad p'_\mu \,=\, -\frac{\partial F}{\partial x^{\prime\mu}}.
\end{equation}
In particular, If we choose $M,M'=\mathbb{R}^d$ and $F(x,x')= \delta_{\mu\nu} x^\mu x^{\prime\nu}$, we recover the symplectic linear transformation $(x,p)\mapsto f(x,p) = (p,-x)$.

\paragraph{Canonical transformation on the Hilbert space.}
So far we formalised canonical transformations as symplectomorphisms. Now, we need to show how these symplectomorphisms give rise to isomorphisms of the corresponding quantum Hilbert spaces.

First of all, we must fix a symplectomorphism $f:\mathcal{P}\rightarrow\mathcal{P}'$, then we must choose two polarizations $L\subset T\mathcal{P}$ and $L'\subset T\mathcal{P}'$ which satisfy $L=f^\ast(L')$. Let us call $\mathbf{H}_{L}$ and $\mathbf{H}_{L'}$ the quantum Hilbert spaces corresponding respectively to the $L$ and $L'$ polarisations of the phase space.

Now, notice that $T\Gamma_f\subset T(\mathcal{P}\times\mathcal{P}')$ is a Lagrangian submanifold of the Lagrangian correspondence space $(\mathcal{P}\times\mathcal{P}',\,\pi^\ast\omega- \pi^{\prime\ast}\omega')$.
As observed by \cite{WeiGQ}, The $T\Gamma_f$-polarised Hilbert space $\mathbf{H}_{T\Gamma_{\!f}}$ of the correspondence space $(\mathcal{P}\times\mathcal{P}',\,\pi^\ast\omega- \pi^{\prime\ast}\omega')$ is isomorphic to the topological tensor product $\mathbf{H}_{T\Gamma_{\!f}}\cong\mathbf{H}_{L}\,\widehat{\otimes}\, \mathbf{H}_{L'}^\ast$, which is nothing but the space of Hilbert–Schmidt operators $\mathbf{H}_{L'}\longrightarrow \mathbf{H}_{L}$.
Then, we will obtain the following diagram:
\begin{equation}\label{diag:hilbertcorr}
    \begin{tikzcd}[row sep={10ex,between origins}, column sep={10ex,between origins}]
    &  \mathbf{H}_{T\Gamma_{\!f}}  \arrow[rd, "\pi^{\prime\ast}", hookleftarrow]\arrow[ld, "\pi^\ast"', hookleftarrow] & \\
    \mathbf{H}_{L}  && \mathbf{H}_{L'} \arrow[ll, "f^\ast"]
    \end{tikzcd}
\end{equation}
Now we can lift sections $\psi\in\mathbf{H}_{L}$ and $\psi'\in\mathbf{H}_{L'}$ to the Hilbert space $\mathbf{H}_{T\Gamma_f}$ and consider their products $\Braket{\pi^\ast\psi|\pi^{\prime\ast}\psi'}$ in this space. This is then naturally defines a pairing $(\!(\;\cdot\;,\;\cdot\;)\!):\mathbf{H}_{L}\times \mathbf{H}_{L'}\rightarrow\mathbb{C}$ between the two polarised Hilbert spaces given by
\begin{equation}
    (\!(\;\cdot\;,\;\cdot\;)\!)\; :=\; \Braket{\pi^\ast\;\cdot\;|\pi^{\prime\ast}\;\cdot\;}
\end{equation}
But any such pairing is equivalently a linear isomorphism  $f^\ast:\mathbf{H}_{L'} \xrightarrow{\;\cong\;} \mathbf{H}_{L}$ such that
\begin{equation}
    (\!(\;\cdot\;,\;\cdot\;)\!)\; =\; \Braket{\;\cdot\;|f^\ast\;\cdot\;}
\end{equation}
where this time the product on the right hand side is the hermitian product of the first Hilbert space $\mathbf{H}_{L}$. 

Let us workout what this means in coordinates.
Recall that on $T\Gamma_f\subset T(\mathcal{P}\times\mathcal{P}')$ we have the gauge transformation $\pi^{\prime\ast}\theta^\prime = \pi^\ast\theta - \di F $, where $F(x,x')$ is the generating function of the symplectomorphism. Therefore, wave-functions $\psi(x)$ of $\mathbf{H}_L$ will be lifted by $\psi(x)\mapsto \psi(x,x')=\psi(x)$ and wave-functions $\psi'(x')$ of $\mathbf{H}_{L'}$ will be lifted by $\psi'(x')\mapsto \psi'(x,x')=\psi(x)e^{-\frac{i}{\hbar}F(x,x')}$ to wave-functions of $\mathbf{H}_{T\Gamma_f}$.
Thus, the pairing will be given by
\begin{equation}
    (\!(\psi,\psi')\!) \;=\; \int_\mathcal{M}\di^n x\,\di^n x^{\prime }\, \psi^\dagger(x)\psi'(x')e^{-\frac{i}{\hbar}F(x,x')}
\end{equation}
where we called $\mathcal{M}$ the manifold such that $T^\ast\mathcal{M}\cong \mathcal{P}\times \mathcal{P}$. Finally the isomorphism $f^\ast:\mathbf{H}_{L}\rightarrow \mathbf{H}_{L'}$ induced by the diffeomorphism $f$ will be given in coordinates by
\begin{equation}
    (f^\ast\psi')(x) \;=\; \int_{M'}\di^nx\, \psi'(x')e^{-\frac{i}{\hbar} F(x,x')}
\end{equation}
where we called $M'$ the manifold such that $T^\ast M' \cong \mathcal{P}'$. Therefore we have a natural isomoprhism $\mathbf{H}_L\cong\mathbf{H}_{L'}$ any time there is a canonical transformation mapping the Lagrangian subbundle $L$ into $L'$ and thus we are allowed to write just $\mathbf{H}$ for the Hilbert space of a quantum system, without specifying the polarization. We will write just
\begin{equation}
    \Ket{\psi} \,\in\,\mathbf{H}
\end{equation}
for an abstract element of the Hilbert space, independent from the polarization.

\paragraph{Example: quantum canonical transformations.}
To give some intuition for this idea, let us consider a simple example. Choose $M,M'=\mathbb{R}^n$ and let the symplectomorphism $f:(\mathbb{R}^{2n},\,\di p_\mu\wedge \di x^\mu)\rightarrow(\mathbb{R}^{2n},\,\di p'_\mu\wedge \di x^{\prime\mu})$ be the linear transformation $f(x,p)=(p,-x)$. This is generated by generating function $F(x,x')=\delta_{\mu\nu}x^\mu x^{\prime\nu}$. Thus, if we substitute the expression $(x',p')=f(x,p)= (p,-x)$, we recover that $(f^\ast)^{-1}$ is exactly the Fourier transformation of wave-functions:
\begin{equation}
    \begin{aligned}
    (f^\ast\psi')(x) \;&=\; \int_{M'}\di^np\, \psi'(p)e^{-\frac{i}{\hbar} p_\mu x^\mu} \\
    \left((f^\ast)^{-1}\psi\right)\!(p) \;&=\; \int_M\di^nx\, \psi(x)e^{\frac{i}{\hbar} p_\mu x^\mu}
    \end{aligned}
\end{equation}
Thus the same quantum state $\Ket{\psi} \,\in\,\mathbf{H}$ can be represented as a wave-function $\Braket{x|\psi}=\psi(x)$ or as its Fourier transform $\Braket{p|\psi}=\psi(p)$ in the two basis $\big\{\Bra{x}\big\}_{x\in M}$ and $\big\{\Bra{p}\big\}_{p\in M'}$ given by the Lagrangian correspondence.

\paragraph{Generating function and polarisations.}
Let the symplectomorphism $f:\mathcal{P}\rightarrow \mathcal{P}'$ be the change of polarisation from $L$ to $L'$. Now recall that the symplectic form satisfies the equation $\iota^\ast\!\left(\pi^\ast\omega-\pi^{\prime\ast}\omega'\right) = 0$ on the graph $\Gamma_f$. It is not easy that we can relate the symplectic form to the generating function by 
\begin{equation}\label{eq:genfunandpol}
    \omega \;=\; \di_{L}\di_{L'}F(x,x')|_{\Gamma_f},
\end{equation}
where $\di_{L}$ and $\di_{L'}$ are respectively the differentials of the polarisations $L$ and $L'$.
For example, the canonical symplectic form on the vector space $\mathbb{R}^{2n}$ can be expressed by the equation $\di p_\mu \wedge \di x^\mu = \di_{L}\di_{L'} (\delta_{\mu\nu} x^\mu x^{\prime \nu})|_{\{x'=p\}}$, where $F(x,x')=\delta_{\mu\nu} x^\mu x^{\prime \nu}$ is the generating function of the symplectomorphism $f(x,p)=(p,-x)$.

\section{Quantum geometry of the closed string}

\subsection{The phase space of the closed string}\label{sectionphasespace}

\paragraph{The configuration space of $\sigma$-models.}
The fields $X^\mu(\sigma,\tau)$ are embeddings from a surface $\Sigma$ into a target space $M$, i.e. smooth maps $\Coo(\Sigma, M)$, denoted by 
\begin{equation}
    \begin{aligned}
    X^\mu:\, \Sigma \;&\lhook\joinrel\longrightarrow\; M \\
    (\sigma,\tau) \;&\longmapsto\; X^\mu(\sigma,\tau)
    \end{aligned}
\end{equation}

\paragraph{The configuration space of the closed string.}
Consider a surface of the form $\Sigma\simeq\mathbb{R}\times S^1$ with coordinates $\sigma\in[0,2\pi)$ and $\tau\in\mathbb{R}$. The fields $X^\mu(\sigma,\tau)$ of the $\sigma$-model can now be seen as curves $\Coo(\mathbb{R},\mathcal{L}M)$ on the \textit{free loop space} $\mathcal{L}M := \Coo(S^1,M)$ of the original manifold $M$. This will be denoted as follows:
\begin{equation}
    \begin{aligned}
    X^\mu(\sigma):\, \mathbb{R} \;&\lhook\joinrel\longrightarrow\; \mathcal{L}M \\
    \tau \;&\longmapsto\; X^\mu(\sigma,\tau)
    \end{aligned}
\end{equation}
where $X^\mu(\sigma,\tau)$ is a loop for any fixed $\tau\in\mathbb{R}$. In other words we have
\begin{equation}
    \Coo(\mathbb{R},\mathcal{L}M) \;\cong\; \Coo(\Sigma,M)
\end{equation}
This is why the configuration space for the closed string can be identified with the free loop space $\mathcal{L}M$ of the spacetime manifold $M$.

Fortunately, for any given smooth manifold $M$, the free loop space $\mathcal{L}M$ is a Fr\'{e}chet manifold and this assures that there will be a well-defined notion of differential geometry on it. 


For any loop $X(\sigma):S^1\rightarrow M$ of $\mathcal{L}M$ we can consider the space of sections $\Gamma(S^1,X^\ast TM)$. This is homeomorphic to the loop space $\mathcal{L}\mathbb{R}^n$ where $n=\mathrm{dim}(M)$, which, thus, plays the role analogous to a local patch.

Since the points of the loop space are loops $X(\sigma)$ in $M$, a smooth function $F\in\Coo(\mathcal{L}M)$ can be identified with a functional $F[X(\sigma)]$. Similarly, a vector field $V\in T(\mathcal{L}M)$ will be given by a functional operator of the form
\begin{equation}
    V[X(\sigma)] \;=\; \oint\di\sigma \, V^\mu[X(\sigma)](\sigma)\, \frac{\delta}{\delta X^\mu(\sigma)}.
\end{equation}
For a wider and deeper exploration of use of loop spaces to formalise some kinds of path integrals in physics, see \cite{Sza96}.

\paragraph{The transgression functor.}
For any $n$-form $\xi=\frac{1}{n!}\xi_{\mu_1\dots\mu_n}\di x^{\mu_1}\wedge\cdots\wedge\di x^{\mu_n}$, given in local coordinates $\{x^\mu\}$ of $M$, there exist a map, named \textit{transgression functor}, from the cochain complex of differential forms on $M$ to the one of the differential forms on the loop space $\mathcal{L}M$:
\begin{equation}
    \begin{aligned}
    \mathfrak{T}:\; \Omega^n(M) \;&\longrightarrow\; \Omega^{n-1}(\mathcal{L}M)\\
    \xi \;&\longmapsto\; \oint\di\sigma \frac{1}{(n-1)!}\xi_{\mu_1\dots\mu_{n}}\!(X(\sigma))\,\frac{\partial X^{\mu_1}(\sigma)}{\partial \sigma}\,\delta X^{\mu_2}(\sigma)\wedge\cdots\wedge\delta X^{\mu_{n}}(\sigma)
    \end{aligned}
\end{equation}
Crucially, it satisfies the following functorial property:
\begin{equation}
    \delta\, \mathfrak{T} \,=\, \mathfrak{T}\,\di
\end{equation}

\paragraph{The phase space of the closed string.}
Thus the choice for phase space of a string on spacetime $M$ will be the free loop space of $T^\ast M$. By definition, this can be used as a definition of the cotangent bundle of $\mathcal{L}M$, i.e.
\begin{equation}
    T^\ast\mathcal{L}M \,:=\, \mathcal{L}(T^\ast M)
\end{equation}
i.e. the smooth space of loops $(X(\sigma),P(\sigma))$ in the cotangent bundle of $T^\ast M$. This space comes equipped with a canonical symplectic form:
\begin{equation}\label{eq:omega}
    \Omega \,:=\, \oint\di\sigma \, \delta P_\mu(\sigma) \wedge \delta X^\mu(\sigma) \;\;\in\,\Omega^2(T^\ast\mathcal{L}M)
\end{equation}
In the next paragraph we will illustrate that the phase space of the closed string is exactly the infinite-dimensional symplectic manifold $(T^\ast\mathcal{L}M,\,\Omega)$ we just described.

We can now define a Liouville potential $\Theta$ such that its derivative is the canonical symplectic form $\Omega\in\Omega^2(T^\ast\mathcal{L}M)$. Thus we have
\begin{equation}
    \Theta \;:=\, \oint\di\sigma\, P_\mu(\sigma)\, \delta X^\mu(\sigma)\;\;\in\,\Omega^1(T^\ast\mathcal{L}U)
\end{equation}
We can verify that $\Omega = \delta\Theta$ by calculating:
\begin{equation}
    \begin{aligned}
    \delta\Theta \,&=\, \oint\di\sigma\left(\delta X^\mu(\sigma)\wedge\frac{\delta \Theta}{\delta X^\mu(\sigma)} + \delta P_\mu(\sigma)\wedge\frac{\delta \Theta}{\delta P_\mu(\sigma)}\right) \\
    &= \oint\di\sigma\oint\di\sigma'\, \delta(\sigma-\sigma')\,\delta P_\mu(\sigma) \wedge \delta X^\mu(\sigma') \\
    &= \oint\di\sigma \, \delta P_\mu(\sigma) \wedge \delta X^\mu(\sigma) \;=\; \Omega
    \end{aligned}
\end{equation}

\paragraph{The closed string as classical system.}
Let us start from the action of the closed string
\begin{equation}\label{eq:closedaction}
\begin{aligned}
    S[X(\sigma,\tau),P(\sigma,\tau)] \;&=\; \frac{1}{2}\int\di\tau \oint\di\sigma \Big( P_\mu \dot{X}^\mu \,+\, \\
    & -\; g^{\mu\nu}(X)\big(P_\mu-B_{\mu\lambda}(X)X^{\prime\lambda}\big)\big(P_\nu-B_{\nu\lambda}(X)X^{\prime\lambda}\big) + g_{\mu\nu}(X)X^{\prime\mu}X^{\prime\nu} \Big).
\end{aligned}
\end{equation}
Recall that the $\sigma$-model of a closed string $\big(X(\sigma,\tau),P(\sigma,\tau)\big):\Sigma\simeq \mathbb{R}\times S^1 \rightarrow T^\ast M$ can be equivalently expressed as a path $\mathbb{R}\rightarrow T^\ast\mathcal{L}M$ on the cotangent bundle of the loop space. Recall also that the Lagrangian density $\mathfrak{L}\in\Omega^1(\mathbb{R})$ of a classical system, consisting of a phase space $(T^\ast\mathcal{L}M,\Omega)$ and a Hamiltonian function $H\in\Coo(T^\ast\mathcal{L}M)$, is given by $\mathfrak{L}=(\iota_{V_H}\Theta - H)\di\tau$, where $V_H\in\mathfrak{ham}(T^\ast\mathcal{L}M,\Omega)$ is the Hamiltonian vector of $H$. This means that the action will be
\begin{equation}\label{eq:aaction}
    S[X(\sigma,\tau),P(\sigma,\tau)] \;=\; \int_{\tau_0}^{\tau_1}\di\tau \big(\iota_{V_H}\Theta - H \big) 
\end{equation}
where $\Theta$ is the Liouville potential of the symplectic form $\Omega$.
Now we want to check that the symplectic structure $\Omega\in\Omega^2(T^\ast\mathcal{L}M)$ of the phase space of the closed string is exactly the canonical symplectic structure \eqref{eq:omega} on $T^\ast\mathcal{L}M$. To do that, we can assume that the Hamiltonian vector field is just the translation along proper time. In other words we impose
\begin{equation}\label{eq:ttrans}
\begin{aligned}
    V_H \;&:=\; \frac{\di}{\di\tau} \\
    \;&=\; \oint\di\sigma\left( \dot{X}^\mu(\sigma)\frac{\delta}{\delta X^\mu(\sigma)} + \dot{P}_\mu(\sigma)\frac{\delta}{\delta P_\mu(\sigma)} \right)
\end{aligned}
\end{equation}
By putting together definition \eqref{eq:ttrans} and equation \eqref{eq:aaction} we immediately get the equation
\begin{equation}
    \iota_{V_H}\Theta \;=\; \oint\di\sigma\,\dot{X}^\mu(\sigma) P_\mu(\sigma),
\end{equation}
which is solved by the Liouville potential
\begin{equation}
    \Theta \;=\; \oint\di\sigma\, P_\mu(\sigma)\, \delta X^\mu(\sigma)\;\;\in\,\Omega^1(T^\ast\mathcal{L}U).
\end{equation}
Its differential is, indeed, exactly the canonical symplectic form \eqref{eq:omega}, i.e.
\begin{equation}
    \begin{aligned}
    \Omega \;&=\; \delta\Theta \\
    &=\; \oint\di\sigma\,\delta P_\mu(\sigma)\wedge \delta X^\mu(\sigma)
    \end{aligned}
\end{equation}
Moreover, by combining the equation \eqref{eq:aaction} with the action \eqref{eq:closedaction}, we can immediately find the Hamiltonian of a closed string:
\begin{equation*}
\begin{aligned}
    H[X(\sigma),P(\sigma)] &=\oint\di\sigma\,  \frac{1}{2}\Big(g^{\mu\nu}(X)\big(P_\mu-B_{\mu\lambda}(X)X^{\prime\lambda}\big)\big(P_\nu-B_{\nu\lambda}(X)X^{\prime\lambda}\big) + g_{\mu\nu}(X)X^{\prime\mu}X^{\prime\nu}\Big)
\end{aligned}
\end{equation*}
We can formally pack together the momentum $P(\sigma)$ and the derivative $X'(\sigma)$ in the following doubled vector:
\begin{equation}
    \mathbb{P}^M(\sigma) := \begin{pmatrix}X^{\prime\mu}(\sigma)\\P_\mu(\sigma)\end{pmatrix}
\end{equation}
with $M=1,\dots,2n$. Notice that $\mathbb{P}^M(\sigma)$ is uniquely defined at any given loop $(X(\sigma),P(\sigma))$ in the phase space.
Thus, we can rewrite the Hamiltonian of the string as
\begin{equation}\label{eq:h}
    H[X(\sigma),P(\sigma)] = \oint\di\sigma\,\frac{1}{2} \mathbb{P}^M(\sigma)\,\mathcal{H}_{MN}(X(\sigma))\,\mathbb{P}^N(\sigma)
\end{equation}
where the matrix $\mathcal{H}_{MN}$ is defined by
\begin{equation}
    \mathcal{H}_{MN} \;:=\; \begin{pmatrix}g_{\mu\nu}- B_{\mu\lambda}g^{\lambda\rho}B_{\rho\nu} & B_{\mu\lambda}g^{\mu\nu} \\-g^{\mu\lambda}B_{\lambda\mu} & g^{\mu\nu} \end{pmatrix} . \label{genmet1}
\end{equation}
In conclusion, by putting everything together, we can see that a closed string is a classical system $(T^\ast\mathcal{L}M,\Omega,H)$, where $\Omega$ is the canonical symplectic form on $T^\ast\mathcal{L}M$ and the Hamiltonian $H$ is given by definition \eqref{eq:h}. We now see the appearance of the generalised metric, described by matrix \eqref{genmet1}. This metric is a representative of an $O(d,d)/O(d)\times O(d)$ coset, and defines the generalised metric of generalised geometry. As such the Hamiltonian \eqref{eq:h} has a manifest $O(d,d)$ symmetry. This is of course the T-duality symmetry of the string. As discussed in the introduction, the Hamiltonian will often exhibit the symmetries not present in the Lagrangian and T-duality is one of these symmetries. 

\subsection{Generalised coordinates and the Kalb-Ramond field}

\paragraph{Example: Generalised coordinates for a charged particle.}
In the geometric quantisation of an ordinary particle we have, in local Darboux coordinates, a local Liouville potential given by $\theta = p_\mu\di x^\mu$, where $p_\mu$ is the canonical momentum. In presence of an electromagnetic field with a minimally coupled $1$-form potential $A$, the canonical momentum $p_\mu$ which is defined from the Lagrangian perspective by $p_\mu= \frac{\partial \mathcal{L}}{\partial \dot{q^\mu}}$ is given by: $p_\mu=k_\mu + eA_\mu$. (We have used $k_\mu$ to denote the naive non-canonical momentum, also sometimes called the kinetic momentum). 

Then the Liouville potential can be rewritten as $\theta=k_\mu\di x^\mu + eA$, with $A$ is the pullback of the electromagnetic potential to the phase space. Consequently the symplectic form takes the form $\omega=\di k_\mu \wedge \di x^\mu + eF$. Let us call the $2$-form $\omega_{A=0} := \di k_\mu\wedge\di x^\mu$. Thus, the geometric prequantisation condition $[\omega]=[\omega_{A=0}]+e[F]\in H^2(T^\ast M,\mathbb{Z})$ of the symplectic form on the phase space implies the Dirac quantisation condition $e[F]\in H^2(M,\mathbb{Z})$ of the electromagnetic field on spacetime. 

Notice that, in canonical coordinates, we have a Hamiltonian $H = g^{\mu\nu}(p_\mu-eA_\mu)(p_\nu-eA_\nu)$ and the commutation relations
\begin{equation}
    [\hat{x}^\mu,\,\hat{x}^\nu] = 0, \quad [\hat{p}_\mu,\,\hat{x}^\nu] = i\hbar\delta_\mu^\nu, \quad [\hat{p}_\mu,\,\hat{p}_\nu] = 0.
\end{equation}
On the other hand, in terms of the kinetic non-canonical coordinates, we have the Hamiltonian $H = g^{\mu\nu}k_\mu k_\nu$ and commutation relations
\begin{equation}
    [\hat{x}^\mu,\,\hat{x}^\nu] = 0, \quad [\hat{k}_\mu,\,\hat{x}^\nu] = i\hbar\delta_\mu^\nu, \quad [\hat{k}_\mu,\,\hat{k}_\nu] = i\hbar eF_{\mu\nu}.
\end{equation}
It is worth observing that the space coordinates do not commute anymore and the non-commutativity term is proportional to the field strength of the electromagnetic field. A similar picture will hold for strings.

\paragraph{Kinetic coordinates for a charged string.}
Similarly to the charged particle, for a string we require our $\Omega$ defined by equation \eqref{eq:omega} to be quantised as $[\Omega]\in H^2(T^\ast\mathcal{L}M,\mathbb{Z})$. We will now see that this implies, similarly to the electromagnetic field, the quantisation of the Kalb-Ramond field flux $[H]\in H^3(M,\mathbb{Z})$.

Let us recall that an abelian gerbe with Dixmier-Douady class $[H]\in H^3(M,\mathbb{Z})$ on the base manifold $M$ is encoded by the following patching conditions:
\begin{equation}
\begin{aligned}
    H \,&=\, \di B_\alpha &\;\in\Omega^3_{\mathrm{cl}}(M)\\
    B_\alpha - B_\beta \,&=\, \di \Lambda_{\alpha\beta} &\;\in\Omega^2(U_\alpha\cap U_\beta)\\
    \Lambda_{\alpha\beta} + \Lambda_{\beta\gamma} + \Lambda_{\gamma\alpha} \,&=\, \di G_{\alpha\beta\gamma} &\;\in\Omega^1(U_\alpha\cap U_\beta \cap U_\gamma)
\end{aligned}
\end{equation}
(see \cite{Alvarez:1984es} and \cite{Brylinski:1993ab} for details). By using the properties of the transgression functor from $M$ to its loop space $\mathcal{L}M$, we immediately obtain the new patching conditions
\begin{equation}
\begin{aligned}
    \mathfrak{T}H \,&=\, \delta (\mathfrak{T}B_\alpha) &\;\in\Omega^2_{\mathrm{cl}}(\mathcal{L}M)\\
    \mathfrak{T}B_\alpha - \mathfrak{T}B_\beta \,&=\, \delta (\mathfrak{T}\Lambda_{\alpha\beta}) &\;\in\Omega^1(\mathcal{L}U_\alpha\cap \mathcal{L}U_\beta)
\end{aligned}
\end{equation}
on $\mathcal{L}M$.
Therefore, the transgression functor sends a gerbe on a manifold $M$ to a circle bundle on its loop space $\mathcal{L}M$, i.e. in other words 
\begin{equation}
    \mathfrak{T}:\, \mathrm{Gerbes}(M) \,\xrightarrow{\;\;\;\cong\;\;\;}\, U(1)\mathrm{Bundles}(\mathcal{L}M),
\end{equation}
where the first Chern class of the circle bundle is $\mathfrak{T}H\in H^2(\mathcal{L}M,\mathbb{Z})$.

Now we can decompose the canonical symplectic form of the phase space of the closed string $(T^\ast\mathcal{L}M,\Omega)$ by
\begin{equation}
    \Omega \;=\; \Omega_{B=0} + \mathfrak{T}H.
\end{equation}
We can write $\Omega_{B=0}:=  \oint\di\sigma\, \delta K_\mu(\sigma) \wedge\delta X^\mu(\sigma) $, so we find that the symplectic form can be expressed by
\begin{equation}
    \Omega \;=\; \oint\di\sigma\, \delta\Big( K_\nu(\sigma) + B_{\mu\nu}\big(X(\sigma)\big)X^{\prime \mu}(\sigma)\Big)\wedge\delta X^\nu(\sigma) 
\end{equation}
where $K_{\nu}(\sigma) := P_\nu(\sigma) - B_{\mu\nu}\big(X(\sigma)\big)X^{\prime \mu}(\sigma)$ is the \textit{non-canonical momentum} of the string and $P_\mu(\sigma)$ is its canonical momentum. 
The relevance of the transgression of gerbes to the loop space in dealing with T-duality and, more generally, with Double Field Theory was underlined by \cite{BelHulMin07}.

For a clarification on the relation between the phase space of a closed string, seen as a loop space, and the Courant algebroids of supergravity, see \cite{Ost19}.

\subsection{The algebra of operators of a closed string}

The machinery of geometric quantisation can now be applied on the phase space $(T^\ast\mathcal{L}M,\Omega)$ of the closed string. See also \cite{SaeSza12} for a different quantisation approach on a loop space.
Given our choice of gauge for the Liouville potential $\Theta=\oint\di\sigma P_\mu(\sigma)\delta X^\mu(\sigma)$, we can now determine the algebra $\mathfrak{heis}(T^\ast\mathcal{L}M,\Omega)$ of quantum observables defined by
\begin{equation}
    \hat{f} \;=\; i\hbar\nabla_{V_f} + f
\end{equation}
for any classical observable $f\in\Coo(T^\ast\mathcal{L}M)$. 
For the classical observables corresponding to the canonical coordinates $X^\mu(\sigma),P_\mu(\sigma)$ of the phase space of the closed string, we have the following quantum observables:
\begin{equation}
    \hat{P}_\mu(\sigma) \,=\, -i\hbar\frac{\delta}{\delta X^\mu(\sigma)}, \qquad \hat{X}^\mu(\sigma) \,=\, i\hbar\frac{\delta}{\delta P_\mu(\sigma)} + X^\mu(\sigma)
\end{equation}
If we choose the polarisation determined by the Lagrangian subbundle $L=T\mathcal{L}M$ with the corresponding basis $\big\{\Ket{X(\sigma)}\big\}_{X(\sigma)\in\mathcal{L}M}$, we have the following operators acting on wave-functional $\Psi[X(\sigma)]=\Braket{X(\sigma)|\Psi}$
\begin{equation}
    \begin{aligned}
    \Braket{X(\sigma)|\hat{P}_\mu(\sigma)|\Psi} \,&=\, -i\hbar\frac{\delta}{\delta X^\mu(\sigma)} \Psi[X^\mu(\sigma)] \\
     \Braket{X(\sigma)|\hat{X}^\mu(\sigma)|\Psi} \,&=\, X^\mu(\sigma)\Psi[X^\mu(\sigma)]
    \end{aligned}
\end{equation}
The commutation relations of these operators will then as follows:
\begin{equation}
    \begin{aligned}
    \big[\hat{P}_\mu(\sigma),\, \hat{X}^\nu(\sigma')\big] \,&=\, 2\pi i\hbar\delta_{\mu}^{\;\nu}\,\delta(\sigma-\sigma'), \\
    \big[\hat{X}^\mu(\sigma),\, \hat{X}^\nu(\sigma')\big] \,&=\, 0, \\
    \big[\hat{P}_\mu(\sigma),\, \hat{P}_\nu(\sigma')\big] \,&=\, 0.
    \end{aligned}
\end{equation}
These define the Heisenberg algebra $\mathfrak{heis}(T^\ast\mathcal{L}M,\Omega)$ of quantum observables on the phase space of the closed string.

On the other hand, if we use the non-canonical, kinetic coordinates $(K(\sigma), X(\sigma))$, we obtain commutation relations of the form
\begin{equation}
    \begin{aligned}
    \big[\hat{K}_\mu(\sigma),\, \hat{X}^\nu(\sigma')\big] \,&=\, 2\pi i\hbar\delta_{\mu}^{\;\nu}\,\delta(\sigma-\sigma'), \\
    \big[\hat{X}^\mu(\sigma),\, \hat{X}^\nu(\sigma')\big] \,&=\, 0, \\
    \big[\hat{K}_\mu(\sigma),\, \hat{K}_\nu(\sigma')\big] \,&=\, H_{\mu\nu\lambda}\big(X(\sigma)\big)X^{\prime\lambda}(\sigma)\,\delta(\sigma-\sigma').
    \end{aligned}
\end{equation}

\subsection{The phase space of the closed string on a torus}

\paragraph{Closed string on the torus.}
Let us consider a closed string propagating in the background $M=T^n$ with constant metric $g_{\mu\nu}$ and constant Kalb-Ramond field $B_{\mu\nu}$. As explained by \cite{KugZwi92}, the compactification condition $x^\mu = x^\mu + 2\pi$ of a torus target space is background-independent, i.e. it does not depend on the bosonic background fields $g_{\mu\nu}$ and $B_{\mu\nu}$. In fact, the coordinates $x^\mu$ are periodic with radius $2\pi$ and the physical radii of the compactification are given by $R_\mu := \int_{S^1_\mu}\!\sqrt{g_{\mu\mu}}\,\di x^\mu = \sqrt{g_{\mu\mu}}$.

Consider the phase space $T^\ast\mathcal{L}M\cong \mathcal{L}(T^n\times \mathbb{R}^n)$ of a closed string on a torus background. Let us call the matrix $E:=g+B$, so that its transpose is $E^\mathrm{T}= g-B$. Let us also define the generalised metric by the constant matrix
\begin{equation}
    \mathcal{H}_{MN} \;=\; \begin{pmatrix}g_{\mu\nu} - B_{\mu\lambda}g^{\lambda\rho}B_{\rho\nu} & -g^{\mu\lambda}B_{\lambda\nu}\\ B_{\mu\lambda}g^{\lambda\nu} & g^{\mu\nu}\end{pmatrix}.
\end{equation}
We can, now, explicitly expand position and momentum in $\sigma$ as it follows:
\begin{equation}
    \begin{aligned}
    X^\mu(\sigma) \; &=\; x^\mu + \alpha' w^\mu\sigma + \!\!\sum_{n\in\mathbb{N}\backslash\{0\}}\!\! \frac{1}{n}\Big( \alpha^\mu_n(\mathcal{H})e^{in\sigma} + \bar{\alpha}^\mu_n(\mathcal{H})e^{-in\sigma} \Big) \\[0.5em]
    P_\mu(\sigma) \; &=\; p_\mu + \!\!\sum_{n\in\mathbb{N}\backslash\{0\}}\!\! \Big( E_{\mu\nu}^{\mathrm{T}}\alpha^\nu_n(\mathcal{H})e^{in\sigma} + E_{\mu\nu}\bar{\alpha}^\nu_n(\mathcal{H})e^{-in\sigma} \Big)
    \end{aligned}
\end{equation}
where we used the notation $\alpha^\mu_n=\alpha^\mu_n(\mathcal{H})$  and $\bar{\alpha}^\mu_n=\bar{\alpha}^\mu_n(\mathcal{H})$ to indicate that the higher-modes depend on the background, encoded by the generalised metric. The operators $\hat{\alpha}^\mu_n$ and $\hat{\bar{\alpha}}^\mu_n$ are the creation and annihilation operators for the excited states of the string, which depend on the background. However, as pointed out by \cite{KugZwi92}, the operators $\hat{X}^\mu(\sigma)$ and $\hat{P}_\mu(\sigma)$ must be thought as background-independent objects. From the Fourier expansion, we also immediately obtain that the zero-modes coordinate $p_\mu$ and $w^\mu$ are integers because of the periodicity of $x^\mu$.

The T-dual coordinates $\widetilde{X}_i(\sigma)$ are defined by,
 \begin{equation}
\widetilde{X}^{\prime }_\mu(\sigma) \;=\; P_\mu(\sigma)\, ,
\end{equation} 
and so we must have $\widetilde{X}_\mu(\sigma)=\tilde{x}_\mu+\int_0^\sigma \di\sigma'P_\mu(\sigma')$. Therefore, we obtain the expressions
\begin{equation}
    \begin{aligned}
    \widetilde{X}_\mu(\sigma) \; &=\; \tilde{x}_\mu + \alpha' p_\mu\sigma + \!\!\sum_{n\in\mathbb{N}\backslash\{0\}}\!\! \frac{1}{n}\Big( -E_{\mu\nu}^{\mathrm{T}}\alpha^\nu_n(\mathcal{H})e^{in\sigma} + E_{\mu\nu}\bar{\alpha}^\nu_n(\mathcal{H})e^{-in\sigma} \Big) \\
    \widetilde{P}^\mu(\sigma) \; &=\; w^\mu + \!\!\sum_{n\in\mathbb{N}\backslash\{0\}}\!\! \frac{1}{n}\Big( \alpha^\mu_n(\mathcal{H})e^{in\sigma} + \bar{\alpha}^\mu_n(\mathcal{H})e^{-in\sigma} \Big)
    \end{aligned}
\end{equation}
Notice that the T-dual coordinate $\widetilde{X}_\mu(\sigma)$ is also periodic with period $2\pi$. Moreover the new coordinates $\widetilde{X}(\sigma)$ and $\widetilde{P}(\sigma)$ are independent from the background.
The commutation relations 
\begin{equation}
[\hat{X}^\mu(\sigma),\hat{P}_\nu(\sigma')]= i\delta^\mu_{\;\nu}\delta(\sigma-\sigma')
\end{equation} become, on zero and higher modes,
\begin{equation}
    \begin{aligned}
    [\hat{x}^\mu,\hat{p}_\nu] \,&=\, i\delta^\mu_{\;\nu} \\
    [\hat{\alpha}^\mu_n(\mathcal{H}),\hat{\alpha}^\nu_m(\mathcal{H})] \,&=\, n\delta_{n+m,0}\,g^{\mu\nu}
    \end{aligned}
\end{equation}

As we have seen in the first section, the action is given on the phase space by the equation $S:= \oint\di\sigma\,\dot{X}^{\mu}(\sigma,\tau)\,P_{\mu}(\sigma,\tau) - H[X(\sigma,\tau),P(\sigma,\tau)]$. We can, thus, rewrite the action of the closed string on a torus as
\begin{equation}
    S[X(\sigma,\tau),P(\sigma,\tau)] \;=\; \int\di\tau \oint\di\sigma\left( \dot{X}^\mu P_\mu -  \frac{1}{2}\mathbb{P}^M\mathcal{H}_{MN}\,\mathbb{P}^N \right)
\end{equation}

\subsection{T-duality and background independence}

\paragraph{T-duality as a symplectomorphism.}
Let $M$ be still a toroidal background with constant metric $g_{\mu\nu}$ and Kalb-Ramond field $B_{\mu\nu}$. We will now show that it is possible to interpret T-duality as a symplectomorphism of phase spaces of two closed strings of the form
\begin{equation}\begin{aligned} 
    f:\;\quad(T^\ast\mathcal{L}M,\,\Omega)\,&\longrightarrow\,(T^\ast\mathcal{L}\widetilde{M},\,\widetilde{\Omega}) \\
     \big(X^\mu(\sigma),P_\mu(\sigma)\big)\,&\longmapsto\, \big(\widetilde{X}_\mu(\sigma),\widetilde{P}^\mu(\sigma)\big).
\end{aligned}\end{equation}
In fact, in \cite{AAL94}, it was firstly argued that T-duality can be seen a canonical transformation. However, a canonical transformation with generating functional $F[X(\sigma),\widetilde{X}(\sigma)]$ is nothing but the symplectomorphism $f$ associated to the following Lagrangian correspondence of the form \eqref{diag:lagrangiancorr}
\begin{equation} \label{lagcor2}
    \begin{tikzcd}[row sep={13ex,between origins}, column sep={13ex,between origins}]
    &  \big(T^\ast\mathcal{L}(M\times\widetilde{M}),\, \pi^\ast\Omega-\widetilde{\pi}^{\ast}\widetilde{\Omega}\big) \arrow[rd, "\widetilde{\pi}", two heads]\arrow[ld, "\pi"', two heads] & \\
   (T^\ast\mathcal{L}M,\,\Omega) \arrow[rr, "f"] && (T^\ast\mathcal{L}\widetilde{M},\,\widetilde{\Omega})
    \end{tikzcd}
\end{equation}
which satisfies the following trivialisation condition for Liouville potential:
\begin{equation}\label{eq:trivcond}
    \begin{aligned}
        \pi^\ast\Theta \,-\, \widetilde{\pi}^\ast\widetilde{\Theta} \;=\; \delta F.
    \end{aligned}
\end{equation}
We can immediately check that, if we substitute the expression for the Liouville potentials, we get the expression
\begin{equation*}
    \begin{aligned}
        \oint\di\sigma\Big(P_\mu(\sigma)\delta X^\mu(\sigma)-\widetilde{P}^\mu(\sigma)\delta \widetilde{X}_\mu(\sigma) \Big) \;=\; \oint\di\sigma\left(\frac{\delta F}{\delta X^\mu(\sigma)}\delta X^\mu(\sigma) + \frac{\delta F}{\delta \widetilde{X}_\mu(\sigma)}\delta \widetilde{X}_\mu(\sigma)\right)
    \end{aligned}
\end{equation*}
and hence we recover exactly the equations of the canonical transformation
\begin{equation}
    \begin{aligned}
        P_\mu(\sigma) \;=\; \frac{\delta F}{\delta X^\mu(\sigma)}  , \qquad
        \widetilde{P}^\mu(\sigma) \;=\; -\frac{\delta F}{\delta \widetilde{X}_\mu(\sigma)}  .
        \end{aligned}
\end{equation}
By considering the generating functional
\begin{equation}\label{eq:genfunF}
    \begin{aligned}
    F[X(\sigma),\widetilde{X}(\sigma)] \;&:=\, \int_{D,\,\partial D=S^1}  \!\di \widetilde{X}_\mu \wedge \di X^\mu \\[0.1cm]
    \;&=\, \frac{1}{2} \oint\di\sigma\big( X^{\prime\mu}(\sigma)\widetilde{X}_{\mu}(\sigma) - X^{\mu}(\sigma)\widetilde{X}'_{\mu}(\sigma)\big)
    \end{aligned}
\end{equation}
which was originally proposed by \cite{AAL94, LAL94}, we obtain exactly T-duality on the phase space:
\begin{equation}
    \begin{aligned}
        P_\mu(\sigma) \;=\; \widetilde{X}_\mu'(\sigma)  , \qquad
        \widetilde{P}^\mu(\sigma) \;=\; X^{\prime\mu}(\sigma)  .
        \end{aligned}
\end{equation}
The Lagrangian correspondence space  in \eqref{lagcor2} is then the loop space of the doubled space of DFT. 
We can notice that, in this simple case, the doubled space can be identified with the correspondence space of a topological T-duality \cite{Bou03, Bou03x, Bou03xx, Bou04, Bou08} over a base point.
A similar observation was made in \cite{Papadopoulos:2014ifa}.

\paragraph{Relation with the symplectic form on the doubled space.}
Let us consider the symplectic $2$-form $\varpi:=\di x^\mu\wedge\di \widetilde{x}_\mu\in\Omega^2(M\times\widetilde{M})$ on the product space, where $\{x^\mu,\tilde{x}_\mu\}$ are local coordinates on $M\times\widetilde{M}$. Notice that, for such a symplectic form, we can choose a Liouville potential of the form $\frac{1}{2}(\tilde{x}_\mu\di x^\mu - x^\mu\di\tilde{x}_\mu)$. (This is like the choice in Weyl quantisation or when we construct a Fock space). Now, we can immediately recognise that the generating functional \eqref{eq:genfunF} is nothing but the transgression of this Liouville potential to the loop space $\mathcal{L}(M\times\widetilde{M})$ of the product space, i.e. we have
\begin{equation}
    F[X(\sigma),\widetilde{X}(\sigma)] \; = \; \frac{1}{2}\,\mathfrak{T}(\tilde{x}_\mu\di x^\mu - x^\mu\di\tilde{x}_\mu).
\end{equation}
By using the functorial property $\delta \mathfrak{T}=\mathfrak{T}\di$ of the transgression functor, we can rewrite the trivialisation condition \eqref{eq:trivcond} of T-duality by
\begin{equation}
    \pi^\ast\Theta \,-\, \widetilde{\pi}^\ast\widetilde{\Theta} \;=\; \mathfrak{T}(\varpi).
\end{equation}
Therefore, T-duality on the phase space is associated to the symplectic $2$-form $\varpi$ on the product space $M\times\widetilde{M}$, which may be identified with the \textit{doubled space}. Notice that this $2$-form is a particular and simple case of the fundamental $2$-form considered by \cite{Svo17, Svo18, Svo19}.

\paragraph{Background independence.}
Since the two loop phase spaces $T^\ast\mathcal{L}M$ and $T^\ast\mathcal{L}\widetilde{M}$ are symplectomorphic, they can be effectively considered the same symplectic $\infty$-dimensional Fr\'{e}chet manifold. In this "passive" symplectomorphism perspective, the T-duality from $(X(\sigma),P(\sigma))$ to $(\widetilde{X}(\sigma),\widetilde{P}(\sigma))$ can be interpreted as a change of coordinates on the phase space of the closed string. The Hamiltonian formulation of the closed string on the phase space is thus T-duality invariant.

\paragraph{T-duality as isomorphism of classical systems.}
T-duality, seen as a symplectomorphism $f:(T^\ast\mathcal{L}M,\,\Omega)\rightarrow(T^\ast\mathcal{L}\widetilde{M},\,\widetilde{\Omega})$ of the phase space of the closed string, does also preserve the Hamiltonian of the closed string, i.e. we have
\begin{equation}
    f^\ast H \,=\, H.
\end{equation}
In other words a T-duality is not just a symplectomorphism of our phase space $(T^\ast\mathcal{L}M,\Omega)$, but also an isomorphism of the classical system $(T^\ast\mathcal{L}M,\Omega,H)$ of the closed string.

In general we can T-dualise the Hamiltonian of the closed string by applying a transformation $\mathcal{O}\in O(n,n;\mathbb{Z})$ to the doubled metric $\mathcal{H}$. Notice that the Hamiltonian functional does not change under such transformations. We have, in fact,
\begin{equation*}
    {H}[\widetilde{X},\widetilde{P}] \,=\, \oint\di\sigma\,\frac{1}{2} (\mathcal{O}\mathbb{P})^{M}(\mathcal{O}^T\mathcal{H}\mathcal{O})_{MN}(\mathcal{O}\mathbb{P})^N \,=\, \oint\di\sigma\,\frac{1}{2} \mathbb{P}^M\mathcal{H}_{MN}\mathbb{P}^N \,=\, H[X,P].
\end{equation*}

\paragraph{T-duality as change of basis on the Hilbert space.}
The Lagrangian correspondence \eqref{lagcor2} induces a diagram of quantum Hilbert spaces
\begin{equation}
    \begin{tikzcd}[row sep={10ex,between origins}, column sep={10ex,between origins}]
    &  \mathbf{H}_{T\Gamma_f} \arrow[rd, "\pi^{\prime\ast}", hookleftarrow]\arrow[ld, "\pi^\ast"', hookleftarrow] & \\
    \mathbf{H}_L  && \mathbf{H}_{\widetilde{L}} \arrow[ll, "f^\ast"]
    \end{tikzcd}
\end{equation}
where $\mathbf{H}_L$ and $\mathbf{H}_{\widetilde{L}}$ are respectively polarised along the Lagrangian subbundles $L=T(\mathcal{L}M)$ and $\widetilde{L}=T(\mathcal{L}\widetilde{M})$. Now, as we have seen in \eqref{diag:hilbertcorr}, the map $(f^\ast)^{-1}$ $\mathbf{H}_L\cong\mathbf{H}_{\widetilde{L}}$ is an isomorphism $\mathbf{H}_L \cong \mathbf{H}_{\widetilde{L}}$ of Hilbert spaces. Therefore we can use just the notation $\mathbf{H}$ for the abstract quantum Hilbert space. 

Any quantum state $\Ket{\Psi}\in\mathbf{H}$ can be expressed in the two basses defined by the two different polarisations:
\begin{equation}
    \ket{\Psi} = \int\mathcal{D}X(\sigma)\,\Psi[X(\sigma)]\,\ket{X(\sigma)}, \qquad \ket{\Psi} = \int\mathcal{D}\widetilde{X}(\sigma)\,\widetilde{\Psi}[\widetilde{X}(\sigma)]\,\ket{\widetilde{X}(\sigma)}
\end{equation}
where we called the wave-functional
\begin{equation}
    \braket{X(\sigma)|\Psi} \;=:\, \Psi[X(\sigma)], \qquad \braket{\widetilde{X}(\sigma)|\Psi} \,=:\, \widetilde{\Psi}[\widetilde{X}(\sigma)].
\end{equation}
The expansions in different basses will be then related by the Fourier-like transformation $(f^\ast)^{-1}$ of string wave-functionals, given by
\begin{equation}\label{eq:tdualityasunitarytransformation}
    \widetilde{\Psi}[\widetilde{X}(\sigma)] \,=\, \int_{\mathcal{L}M} \mathcal{D}X(\sigma)\, e^{\frac{i}{\hbar}F[X(\sigma),\widetilde{X}(\sigma)]}\, \Psi[X(\sigma)],
\end{equation}
in accord with \cite{AAL94}. We can also explicitly write the matrix of the change of basis on $\mathbf{H}$ by 
\begin{equation}
    \braket{X(\sigma)|\widetilde{X}(\sigma)} \;=\;  e^{\frac{i}{\hbar}F[X(\sigma),\widetilde{X}(\sigma)]}.
\end{equation}
Interestingly, this isomorphism is naturally defined by lifting the polarised wave functionals $\Psi[X(\sigma)]\in\mathbf{H}_L$ and $\widetilde{\Psi}[\widetilde{X}(\sigma)]\in\mathbf{H}_{\widetilde{L}}$ to wave-functionals ${\Psi}[X(\sigma),\widetilde{X}(\sigma)]$ on the product space and by considering their Hermitian product in the Hilbert space of the doubled space. In double field theory solving the so called strong constraint provides the choice of polarisation. Here the quantisation procedure itself demands  a polarisation choice and the strong constraint is solved automatically. It is interesting to consider the weak constraint from this perspective but this is beyond the goals of this paper.

\paragraph{T-duality invariant dynamics.}
The dynamics of the quantised closed string is encoded by the background independent equation
\begin{equation}
    i\hbar\frac{\partial}{\partial \tau}\ket{\Psi} + \hat{H}\ket{\Psi} = 0.
\end{equation}
Let us consider, for simplicity, that we are starting from a Minkowski flat background with $g_{\mu\nu}=\eta_{\mu\nu}$ and $B_{\mu\nu}=0$. Then we will have a trivial doubled metric $\mathcal{H}_{MN}=\delta_{MN}$. Therefore, the equation of motion can be expressed in the basis $\big\{\ket{X(\sigma)}\big\}_{X(\sigma)\in\mathcal{L}M}$ by 
\begin{equation}
     i\hbar\frac{\partial}{\partial\tau}\Psi[X(\sigma)] + \oint\di\sigma\frac{1}{2}\left( -\hbar^2\frac{\delta^2\;\;}{\delta X(\sigma)^2} + X'(\sigma)^2 \right)\Psi[X(\sigma)] \,=\, 0,
\end{equation}
but immediately also in the T-dual basis $\big\{\ket{\widetilde{X}(\sigma)}\big\}_{X(\sigma)\in\mathcal{L}\widetilde{M}}$ by
\begin{equation}
     i\hbar\frac{\partial}{\partial\tau}\widetilde{\Psi}[\widetilde{X}(\sigma)] + \oint\di\sigma\frac{1}{2}\left( \widetilde{X}'(\sigma)^2 -\hbar^2\frac{\delta^2\;\;}{\delta \widetilde{X}(\sigma)^2} \right)\widetilde{\Psi}[\widetilde{X}(\sigma)] \,=\, 0.
\end{equation}

\paragraph{T-duality as a symplectomorphism for torus bundles.}
Let us conclude this section by considering a slightly more general class of examples, T-duality of torus bundles. 

Let $M\twoheadrightarrow N$ and $\widetilde{M}\twoheadrightarrow N$ be two principal $T^n$-bundles on a common base manifold $N$. T-duality can be still seen as a symplectomorphism between loop phase spaces $T^\ast\mathcal{L}M\rightarrow T^\ast\mathcal{L}\widetilde{M}$ and we can still employ the machinery of Lagrangian correspondence \eqref{diag:lagrangiancorr}. Now, the Lagrangian correspondence of the T-duality on the phase space of the closed string is
\begin{equation}\label{diag:lagcorstring}
    \begin{tikzcd}[row sep={13ex,between origins}, column sep={13ex,between origins}]
    &  \big(T^\ast \mathcal{L}(M\times_{N}\widetilde{M}), \, \pi^\ast\Omega-\widetilde{\pi}^{\ast}\widetilde{\Omega}\big) \arrow[rd, "\pi'", two heads]\arrow[ld, "\pi"', two heads] & \\
    (T^\ast \mathcal{L}M,\,\Omega) \arrow[rr, "f"] && (T^\ast \mathcal{L}\widetilde{M},\,\widetilde{\Omega})
    \end{tikzcd}
\end{equation}
where the fiber product $M\times_{N}\widetilde{M}$ can be naturally seen as the doubled torus bundle of the duality. 
For the torus fibration, we have the following equation for the Liouville potential:
\begin{equation}
    \pi^\ast\Theta \,-\, \widetilde{\pi}^\ast\widetilde{\Theta} \,=\, \mathfrak{T}(\varpi),
\end{equation}
where $\mathfrak{T}(\varpi)$ is the transgression to the loop space of the \textit{fundamental }$2$\textit{-form} $\varpi\in\Omega^2(M\times_{N}\widetilde{M})$, which lives on the doubled torus bundle $M\times_{N}\widetilde{M}$ and it is given by
\begin{equation}
    \varpi \;=\; (\di x^i + A^i) \wedge (\di \widetilde{x}_i + \widetilde{A}_i) \;\;\in\,\Omega^2(M\times_{N}\widetilde{M})
\end{equation}
If the bundle $M\times_{N}\widetilde{M}\twoheadrightarrow N$ is trivial, then we recover the symplectic form $\varpi = \di x^i\wedge\di \widetilde{x}_i$ from the previous paragraphs. Notice that, by moving to the generalised coordinates, we can easily recover the equation characterising topological T-duality by \cite{Bou04}, i.e. we have
\begin{equation}
    H - \widetilde{H} \;=\; \di (A^i\wedge \widetilde{A}_i)
\end{equation}
on the doubled torus bundle $M\times_N\widetilde{M}$.
In this class of cases, we notice that the doubled space may be identified with the correspondence space $M\times_N\widetilde{M}$ of a topological T-duality \cite{Bou03, Bou03x, Bou03xx, Bou04, Bou08} over a base manifold $M$.

\section{Quantum geometry of the doubled string}

\subsection{The generalised boundary conditions of the doubled string}

\paragraph{The phase space and the doubled space.}
Let us start from the phase space $T^\ast\mathcal{L}M$ of a closed string propagating in a general spacetime $M$, which we formalised in section \ref{sectionphasespace}. In this section we want to formalise the phase space of a general doubled string.

To describe doubled strings, we introduce new coordinates $\widetilde{X}_\mu(\sigma)$ which satisfy the equation $P_\mu(\sigma)=\widetilde{X}^{\prime}_\mu(\sigma)$. 
Let us define the following doubled loop-space vectors:
\begin{equation}
    \mathbb{X}^M(\sigma) \,:=\, \begin{pmatrix}X^\mu(\sigma)\\[0.5ex] \widetilde{X}_\mu(\sigma)\end{pmatrix}, \qquad \mathbb{P}^M(\sigma) \,:=\, \begin{pmatrix}X^{\prime\mu}(\sigma)\\[0.5ex] {P}_\mu(\sigma)\end{pmatrix} \,=\, \mathbb{X}^{\prime M}(\sigma).
\end{equation}
Notice that, since we started from a general spacetime, this time we will not generally have that $\mathbb{X}^M(\sigma)$ takes values in a product space.
Therefore, for a doubled string the doubled momentum $\mathbb{P}^M(\sigma)$ coincides with the derivative along the circle of the doubled position vector $\mathbb{X}^M(\sigma)$. 
Thus, instead of encoding the $\sigma$-model of the closed string by an embedding $\left(X^\mu(\sigma),\, P_\mu(\sigma)\right)$ into the phase space, we can encode it by an embedding $\mathbb{X}^M(\sigma)=(X^\mu(\sigma),\, \widetilde{X}_\mu(\sigma))$ into a doubled position space. Our objective is, then, be able to reformulate a string wave-functional $\Psi\!\left[X^\mu(\sigma),P_\mu(\sigma)\right]$ in terms of doubled fields as a wave-functional of the form $\Psi\!\left[\mathbb{X}^M(\sigma)\right]$.

However, notice that, since the new coordinates $\widetilde{X}_\mu(\sigma)$ are the integral of the momenta of the string, specifying $\widetilde{X}_\mu(\sigma)$ is a stronger statement than specifying $P_\mu(\sigma)=\widetilde{X}_\mu'(\sigma)$. This observation is crucial when considering the possible boundary conditions of the doubled string $\sigma$-model.

Let us define the following zero-modes of the doubled loop-space vectors:
\begin{equation}
    \mathbbvar{x}^M \,:=\, \frac{1}{2\pi}\oint\di\sigma\, \mathbb{X}^M(\sigma), \qquad \mathbbvar{p}^M \,:=\,\frac{1}{2\pi\alpha'} \oint\di\sigma\, \mathbb{X}^{\prime M}(\sigma),
\end{equation}
which, in components, read as it follows:
\begin{equation}
    \mathbbvar{x}^M \,=\, \begin{pmatrix}x^\mu\\[0.1ex] \tilde{x}_\mu\end{pmatrix}, \qquad \mathbbvar{p}^M \,=\, \begin{pmatrix}\tilde{p}^{\mu}\\[0.1ex] p_\mu\end{pmatrix} \,\equiv\, \begin{pmatrix}w^{\mu}\\[0.1ex] \tilde{w}_\mu\end{pmatrix}.
\end{equation}
By using the new coordinate $\widetilde{X}_\mu$, we can rewrite the action of a closed string by
\begin{equation}
    S_{\mathrm{string}}[X(\sigma,\tau),\widetilde{X}(\sigma,\tau)] \;=\; \frac{1}{2\pi\alpha'}\int\di\tau \oint\di\sigma\left( \dot{X}^\mu \widetilde{X}'_\mu -  \frac{1}{2}\mathbb{X}^{\prime M}\mathcal{H}_{MN}\,\mathbb{X}^{\prime N} \right)  \label{stringdoubledaction}
\end{equation}

Let us use the following notation for the derivatives
\begin{equation}
    \dot{\mathbb{X}}(\sigma,\tau) \,:=\, \frac{\partial \mathbb{X}(\sigma,\tau)}{\partial \tau} \qquad {\mathbb{X}}'(\sigma,\tau) \,:=\, \frac{\partial \mathbb{X}(\sigma,\tau)}{\partial \sigma}
\end{equation}

\paragraph{The generalised boundary conditions.}
Since in the action of the closed string the field $\mathbb{X}^{M}(\sigma)$ never appears, but only its derivatives $\mathbb{X}^{\prime M}(\sigma)$, we only need to require that the latter are periodic, i.e.
\begin{equation}
    \mathbb{X}^{\prime M}(\sigma+2\pi) \;=\; \mathbb{X}^{\prime M}(\sigma)
\end{equation}
This implies that the generalised boundary conditions are
\begin{equation}
    \mathbb{X}^M(\sigma+2\pi,\tau) \;=\; \mathbb{X}^M(\sigma,\tau) + 2\pi\alpha' \mathbbvar{p}^M(\tau),
\end{equation}
where the quasi-period $\mathbbvar{p}^M(\tau)$ can, in general, be dynamical and depend on proper time.

Let us define the \textit{quasi-loop space} $\mathcal{L}_{\mathrm{Q}}\mathcal{M}$ of a manifold $\mathcal{M}$ as it follows:
\begin{equation}
    \mathcal{L}_{\mathrm{Q}}\mathcal{M} \;:=\; \big\{\, \mathbb{X}:\mathbb{R}\longrightarrow \mathcal{M} \,\;\big|\;\, \di\mathbb{X}(\sigma + 2\pi)=\di\mathbb{X}(\sigma)\,\big\},
\end{equation}
where we used the simple identity $\mathbb{X}^{\prime M}\!(\sigma) \,\di\sigma = \di\mathbb{X}^M(\sigma)$.

The phase space of the doubled string will be a symplectic manifold $(\mathcal{L}_{\mathrm{Q}}\mathcal{M}, \mathbbvar{\Omega})$ where the symplectic form $\mathbbvar{\Omega}\in\Omega^2(\mathcal{L}_{\mathrm{Q}}\mathcal{M})$ will be determined in the following subsection.

\paragraph{A remark on the global geometry of the doubled space.}
Given local coordinates $\mathbbvar{x}^M$ on the doubled space we can express the vector $\mathbb{X}'(\sigma)$ by
\begin{equation}
    \mathbb{X}^{\prime M}\!(\sigma) \,\di\sigma \;=\; \di\mathbb{X}^M(\sigma)  \;=\; \mathbb{X}^\ast\!\left(\di \mathbbvar{x}^M\right) \, ,
\end{equation}
where $\mathbb{X}^\ast\!\left(\di \mathbbvar{x}^M\right)$ denotes the pullback of $\di \mathbbvar{x}^M$ to the quasi-loop space. Therefore, the requirement that $\mathbb{X}'(\sigma)$ is periodic can be immediately recasted as the requirement that the pullback of $\di\mathbbvar{x}^M$ is periodic. We remark that, in this more general setting, the doubled space $\mathcal{M}$ is not required to be a product space.
References \cite{Alf20} and \cite{Alf20b} explore the idea that a general doubled space $\mathcal{M}$ is globally not a smooth manifold, but a more generalised geometric object (see there for details). In particular, in the references, it is derived that the patching conditions for local coordinate patches $\mathcal{U}_{(\alpha)}$ and $\mathcal{U}_{(\beta)}$ of the doubled space $\mathcal{M}$ should be of the form $\mathbbvar{x}^M_{(\beta)}=\mathbbvar{x}^M_{(\alpha)} + \Lambda_{(\alpha\beta)}^M + \partial^M\!\phi_{(\alpha\beta)}$, where we have an additional gauge-like transformation $\phi_{(\alpha\beta)}$ on the overlap of patches. Notice that this implies the patching conditions $\di\mathbbvar{x}^M_{(\beta)}=\di\mathbbvar{x}^M_{(\alpha)} + \di\Lambda_{(\alpha\beta)}^M$ and therefore the gauge transformations $\phi_{(\alpha\beta)}$ do not appear for the differential. Since the \v{C}ech cocycle condition $\di\Lambda_{(\alpha\beta)}^M + \di\Lambda_{(\beta\gamma)}^M + \di\Lambda_{(\gamma\alpha)}^M = 0$ is satisfied, we do not encounter problems for $\mathbb{X}^\ast\!\left(\di \mathbbvar{x}^M\right)=\di\mathbb{X}(\sigma)$ being periodic.
In other words, a doubled string can naturally live on a doubled space $\mathcal{M}$ that is patched in a more general way than a manifold (like the proposal by \cite{Alf20} and \cite{Alf20b}) exactly because $\mathbb{X}(\sigma)$ does not appear in the action, but only $\mathbb{X}'(\sigma)$ does.


\subsection{The symplectic structure of the doubled string.}
Recall that the Lagrangian density is related to the Liouville potential $\mathbbvar{\Theta}$ by
\begin{equation}
\begin{aligned}
    \mathfrak{L}_H \;=\; ( \iota_{V_H}\mathbbvar{\Theta} - H)\di\tau.
\end{aligned}
\end{equation}
Therefore, we can find the symplectic structure $\mathbbvar{\Omega}=\delta\mathbbvar{\Theta}$ on the phase space of the doubled string from its full Lagrangian.
One should also note here that the different choices of Liouville potential will give different results corresponding to either a particular choice of duality frame or a duality symmetric frame.

\paragraph{The Tseytlin action.}
The doubled string $\sigma$-model  as first constructed by Tseytlin is by now well known \cite{Tse90, Tse90b} and there are many routes one might take to its construction. Here, since we already have the doubled perspective in place for the Hamiltonian, the immediate method is to take the action as given by \eqref{stringdoubledaction} and then allow the dual variables to also be dynamical by augmenting the term $\dot{X}^\mu P_\mu$ term in the action with its dual equivalent: $\dot{\widetilde{X}}_\mu \widetilde{P}^\mu $. This is like picking a duality symmetric choice for the Liouville potential. Once this term is included then one can simply substitute the expressions for the doubled vectors into the action. It is these $\dot{X} P$ terms that produce the Legendre transformation between the Hamiltonian and the Lagrangian. They are sometimes called the {\it{abbreviated action}} and from now on we will adopt this nomenclature. The duality-augmented abbreviated action is then:
\begin{equation}
\begin{aligned}
     S_{abb}&=\;  \frac{1}{4\pi\alpha'}\int\di\tau \oint\di\sigma\left(\dot{{X}}^\mu {P}_\mu + \dot{\widetilde{X}}_\mu \widetilde{P}^\mu \right)
    \end{aligned}
\end{equation}
which using the expressions for the doubled vectors becomes (up to total derivative  terms of which we will discuss more later) the $O(d,d)$ manifestly symmetric term:
\begin{equation}\label{eq:kin3}
\begin{aligned}
   S_{abb}=  \int\di\tau \oint\di\sigma\,\frac{1}{4\pi\alpha'}\!\left( \dot{\mathbb{X}}^{M}\eta_{MN}\mathbb{X}^{\prime N} \right) \; .
    \end{aligned}
\end{equation}
Combining this with the Hamiltonian to produce the total action $S=S_{abb}-H$ gives the Tseytlin action \cite{Tse90, Tse90b}: 
\begin{equation}
    S_{\mathrm{Tsey}}[\mathbb{X}(\sigma,\tau)] \;=\; \frac{1}{4\pi\alpha'}\int\di\tau \oint\di\sigma\left( \dot{\mathbb{X}}^{M}\eta_{MN}\mathbb{X}^{\prime N} -  \mathbb{X}^{\prime M}\mathcal{H}_{MN}\mathbb{X}^{\prime N}\right) \, .
\end{equation}
This action has been the subject of much study and we will return to the quantum equivalence to the usual string action later. Let us first examine this action taking care with the important property that the fields are quasi-periodic in $\sigma$ with quasi-period $\mathbbvar{p}^M$ which implies $\mathbb{X}^M(\sigma+2\pi,\tau)=\mathbb{X}^M(\sigma,\tau)+2\pi\alpha'\mathbbvar{p}^M(\tau)$.
This means that although the world sheet is periodic and has no boundary the total derivative terms of such quasi periodic fields can contribute to the action. These contributions to the Hamiltonian produce the important zero mode contributions to the Hamiltonian from the winding and momenta.
What follows is an analysis of the doubled abbreviated action with quasi-periodic fields.

\paragraph{The total derivative contributions to the Tseytlin action.}

When integrating with respect to $\sigma$ we may write $\oint\di\sigma$ as $\int_{\sigma_0}^{\sigma_0+2\pi} \di\sigma$. Then the integral of a total derivative is $ \int \di\sigma \frac{\di}{\di \sigma} f(\sigma)= f(2\pi+\sigma_0)- f(\sigma_0)$. For any periodic function $f(\sigma)$ this then vanishes as it should. However for a quasi-periodic function this integral will be non zero  e.g. $\mathbbvar{p}^M=\frac{1}{2\pi\alpha'}\oint\di\sigma\,\mathbb{X}^{\prime M}(\sigma)$. Note, that this is still independent of $\sigma_0$ as it should be since $\sigma_0$ is an entirely arbitrary choice of coordinate origin in the loop.

Let us write the doubled abbreviated action explicitly with the manifest dependence on $\sigma_0$ as follows:
\begin{equation}
    S_{\mathrm{Tsey}}(\sigma_0) \;=\; \frac{1}{4\pi\alpha'}\int\di\tau\int_{\sigma_0}^{\sigma_0+2\pi}\!\!\!\!\!\di\sigma \,\dot{\mathbb{X}}^M(\sigma,\tau)\eta_{MN}\mathbb{X}^{\prime N}(\sigma,\tau) \, .
\end{equation}
Then taking the derivative with respect to $\sigma_0$ produces:
\begin{equation}
\begin{aligned}
    \frac{\di S_{\mathrm{Tsey}}}{\di \sigma_0} \;&=\; \frac{1}{4\pi\alpha'}\int\di\tau \left(\dot{\mathbb{X}}^M(\sigma_0+2\pi,\tau)- \dot{\mathbb{X}}^M(\sigma_0,\tau)\right)\eta_{MN}\mathbb{X}^{\prime N}(\sigma_0,\tau) \\
    &=\; \frac{1}{2}\int\di\tau\, \dot{\mathbbvar{p}}^M(\tau)\eta_{MN}\mathbb{X}^{\prime N}(\sigma_0,\tau) \, .
\end{aligned}
\end{equation}
To get to the second line we have used the periodicity of $\mathbb{X}^{\prime N}(\sigma+2\pi,\tau) = \mathbb{X}^{\prime N}(\sigma,\tau)$ and the quasi-periodicity of $\mathbb{X}^{N}(\sigma)$.
We thus have an anomaly. The action now depends on on the arbitrary choice of $\sigma_0$ when we allow quasi-periodic fields to encode the zero modes in the doubled space. The proposal in \cite{Bla14} is then to add an explicit "boundary term" to the Tseytlin action to cancel this piece as follows:
\begin{equation}\label{TseyBoundaryTerm}
    S_{\partial\mathrm{Tsey}}[\mathbbvar{p}(\tau),\,\mathbb{X}(\sigma_0,\tau)] \;:=\; -\frac{1}{2}\int\di\tau\, \dot{\mathbbvar{p}}^M(\tau)\eta_{MN}\mathbb{X}^{N}(\sigma_0,\tau).
\end{equation}
The full action, therefore, does not depend on the choice of $\sigma_0$, i.e.
\begin{equation}
    \frac{\di}{\di \sigma_0} (S_{\mathrm{Tsey}} + S_{\partial\mathrm{Tsey}}) \;=\; 0
\end{equation}
and thus diffeomorphism-invariance of the doubled string is restored. 

Putting together all these terms we obtain the action
\begin{equation}
    \begin{aligned}  \label{tcorrect}
    {S}[\mathbb{X}(\sigma,\tau)] \,&=\, \frac{1}{4\pi\alpha'}\int\di\tau \!\oint\di\sigma \left( \dot{\mathbb{X}}^{M}(\sigma,\tau)\eta_{MN}\mathbb{X}^{\prime N}(\sigma,\tau) -  \mathbb{X}^{\prime M}(\sigma,\tau)\mathcal{H}_{MN}\mathbb{X}^{\prime N}(\sigma,\tau)\right) \\ 
    \;&-\, \frac{1}{2}\int\di\tau\,\mathbbvar{p}^{M}(\tau)\eta_{MN}\dot{\mathbb{X}}^N(\sigma_0,\tau) \, .
    \end{aligned}
\end{equation}

This doubled action is now world sheet diffeomorphism invariant for quasi-periodic fields but how does it relate to the original string action? Recall that the in writing down the Tseytlin action total derivative terms were neglected. We will now examine the relationship between the Tseytlin string and ordinary string with quasi-periodic fields.

\paragraph{The Relation between Tseytlin string and ordinary string.}
Let us begin with the usual abbreviated action for the ordinary string, and then integrate by parts keeping the $\sigma$ total derivatives, we may neglect the total derivatives: in $\tau$ since there is no quasi periodicity in this variable:
\begin{equation}
\begin{aligned}
    \frac{1}{2\pi\alpha'}\int\di\tau\oint\di\sigma\, \dot{X}^\mu P_\mu &= \frac{1}{2\pi\alpha'}\int\di\tau\oint\di\sigma\, \dot{X}^\mu \widetilde{X}'_\mu \\
    &= \frac{1}{4\pi\alpha'}\int\di\tau\oint\di\sigma\, ( \dot{X}^\mu \widetilde{X}'_\mu + \dot{X}^\mu\widetilde{X}'_\mu )\\
    &= \frac{1}{4\pi\alpha'}\int\di\tau\oint\di\sigma\, (\dot{X}^\mu \widetilde{X}'_\mu + X'^\mu \dot{\widetilde{X}}_\mu + \frac{\di}{\di \sigma} (\dot{X}^\mu \widetilde{X}_\mu) ) \\
    &= \frac{1}{4\pi\alpha'}\int\di\tau\oint\di\sigma\, \dot{\mathbb{X}}^M\eta_{MN}\mathbb{X}^{\prime N} \\
&+ \frac{1}{4\pi\alpha'}\int\di\tau\!\left( \dot{X}^\mu(2\pi+\sigma_0)\widetilde{X}_\mu(2\pi+\sigma_0)- \dot{X}^\mu(\sigma_0)\widetilde{X}_\mu(\sigma_0) \right)
    \end{aligned}  \label{tdaction}
\end{equation}
Here we remark that we have also here made a choice in boundary total derivative term for $\sigma$. We are free to exchange it with:
\begin{equation}
-\frac{\di}{\di \sigma} ({X}^\mu\dot{\widetilde{X}}_\mu) \, .
\end{equation}
This choice is simply related by a neglected total derivative in $\tau$. A natural possibility is also the duality symmetric combination:
\begin{equation}
\frac{1}{2}\frac{\di}{\di \sigma} \left(\dot{X}^\mu{\widetilde{X}}_\mu- {X}^\mu\dot{\widetilde{X}}_\mu \right) \, .
\end{equation}
Now, the quasi-periodicity of the fields is:
\begin{equation}
    \mathbb{X}^M(2\pi+\sigma_0) = \mathbb{X}^M(\sigma_0) + 2\pi\alpha'\mathbbvar{p}^M 
\end{equation}
so that we may evaluate the final term in \eqref{tdaction} as follows:
\begin{equation}
    \dot{X}(2\pi+\sigma_0)\widetilde{X}(2\pi+\sigma_0)- \dot{X}(\sigma_0)\widetilde{X}(\sigma_0) \;=\; 2\pi\alpha'\!\left(\dot{\tilde{p}}{\widetilde{X}}(\sigma_0) + {p}\dot{X}(\sigma_0) + 2\pi\alpha'\dot{\tilde{p}}p\right) \, .
\end{equation}
Substituting this into \eqref{tdaction} produces:
\begin{equation}\label{eq:totderivative}
\begin{aligned}
    \frac{1}{2\pi\alpha'}\int\di\tau\oint\di\sigma\, \dot{X}^\mu P_\mu 
    \;&=\; \frac{1}{4\pi\alpha'}\int\di\tau\oint\di\sigma\, \dot{\mathbb{X}}^M\eta_{MN}\mathbb{X}^{\prime N}\,+ \\ &+ \int\di\tau\!\left(\frac{1}{2}\dot{\tilde{p}}{\widetilde{X}}(\sigma_0) + \frac{1}{2}{p}\dot{X}(\sigma_0) + \pi\alpha'\dot{\tilde{p}}p\right)
    \end{aligned}
\end{equation}

Now, $\dot{X}(2\pi) = \dot{X}(2\pi) + 2\pi\alpha'\dot{\tilde{p}}$ is quasi-periodic, so the initial term $\int\di\tau\oint\di\sigma\, \dot{X}P $ suffers from the same anomaly as we described in the previous section. We need to also add a "boundary term" for $\int\di\tau\oint\di\sigma\, \dot{X}P $ so that there is no dependence on $\sigma_0$. This implies that the abbreviated action on the  LHS of \eqref{tdaction} must be changed to
\begin{equation}\label{Sabb}
    S_\mathrm{abb} \;:=\; \frac{1}{2\pi\alpha'}\int\di\tau\oint\di\sigma\, \dot{X}^\mu P_\mu - \int\di\tau\, \dot{\tilde{p}}^{\mu}\widetilde{X}_\mu(\sigma_0)
\end{equation}
so there is no overall $\sigma_0$ dependence.
Then including the same term to the RHS of \eqref{tdaction} will produce:
\begin{equation}\label{SabbEquation}
\begin{aligned}
    S_\mathrm{abb} \;&=\; \frac{1}{4\pi\alpha'}\int\di\tau\oint\di\sigma\, \dot{\mathbb{X}}^M\eta_{MN}\mathbb{X}^{\prime N} \,+\\ & \;+ \; \int\di\tau\!\left(-\frac{1}{2}\dot{\tilde{p}}^\mu{\widetilde{X}}_\mu(\sigma_0) \; + \;\frac{1}{2}{p}_\mu\dot{X}^\mu(\sigma_0) + \pi\alpha'\dot{\tilde{p}}^\mu p_\mu \right)
    \end{aligned}
\end{equation}
We may now write this, using integration by parts and deleting total derivatives in $\tau$, or by using the duality symmetric choice of total derivative terms in $\sigma$, to give a duality symmetric action
\begin{equation}
\begin{aligned}
    S_\mathrm{abb} \;&=\; \frac{1}{4\pi\alpha'}\int\di\tau\oint\di\sigma\, \dot{\mathbb{X}}^M\eta_{MN}\mathbb{X}^{\prime N} \,+\\ & \;+ \;  \int\di\tau\!\left(+\frac{1}{2}{\tilde{p}}\dot{\widetilde{X}}(\sigma_0) + \frac{1}{2}{p}\dot{X}(\sigma_0) + \frac{\pi\alpha'}{2}(\dot{\tilde{p}}p - \tilde{p}\dot{p})\right)
    \end{aligned}
\end{equation}
that we may write equivalently write in a doubled fashion by 
\begin{equation}
\begin{aligned}
    S_\mathrm{abb} \;&=\; \frac{1}{4\pi\alpha'}\int\di\tau\oint\di\sigma\, \dot{\mathbb{X}}^M\eta_{MN}\mathbb{X}^{\prime N} \,+\\ & \;+ \;  \int\di\tau\!\left(\frac{1}{2}\mathbbvar{p}^M\eta_{MN}\dot{\mathbb{X}}^N(\sigma_0) + \frac{\pi\alpha'}{2}\dot{\mathbbvar{p}}^M\omega_{MN}\mathbbvar{p}^N\right)\!,  \label{tcorrect1}
    \end{aligned}
\end{equation}
where $\omega_{MN}$ is again the $2n$-dimensional standard symplectic matrix.

We may now identify the $\int\di\tau\frac{1}{2}\mathbbvar{p}^M\eta_{MN}\dot{\mathbb{X}}^N(0)$ term as the "boundary term" \eqref{TseyBoundaryTerm} we introduced earlier to make the Tseytlin action independent of $\sigma_0$. Thus when the dust settles we see that the abbreviated action of the string, including the "boundary" piece, produces the Tseytlin string (including the doubled "boundary" term) and a correction term from the final term in \eqref{tcorrect1}.
Thus, the relation between $S_\mathrm{abb}$ of the usual string and the Tseytlin abbreviated action $S_{\mathrm{Tsey,abb}}$ (including "boundary" pieces) is
\begin{equation}
    S_\mathrm{abb} \;=\; S_{\mathrm{Tsey,abb}} +\int\di\tau\,\frac{\pi\alpha'}{2}\dot{\mathbbvar{p}}^M\omega_{MN}\mathbbvar{p}^N \, .
\end{equation}

\paragraph{The dual picture.}
 Let us define the abbreviated action of the T-dual string, including the "boundary" piece, as follows: 
\begin{equation}
    \widetilde{S}_\mathrm{abb} \;:=\; \frac{1}{2\pi\alpha'}\int\di\tau\oint\di\sigma\, \dot{\widetilde{X}}_\mu \widetilde{P}^\mu - \int\di\tau\, \dot{{p}}_\mu{X}^\mu(\sigma_0)  \, .
\end{equation}
Now, we can repeat all this procedure with this "dual" abbreviated action. Then the difference between the two duality related frames for the action will be the difference of these abbreviated actions (the Hamiltonian being invariant), thus: 
\begin{equation}
\begin{aligned}
    S_\mathrm{abb} - \widetilde{S}_\mathrm{abb}\;=\; \pi\alpha'\int\di\tau\,\dot{\mathbbvar{p}}^M\omega_{MN}\mathbbvar{p}^N  \, .
    \end{aligned}
\end{equation}
The implication is that ordinary string and its dual are related by a phase shift:
\begin{equation}
\begin{aligned}
    \exp\!\left({\frac{i}{\hbar}S_\mathrm{abb}}\right)\;=\; \exp\!\left({\frac{i\pi\alpha'}{\hbar}\int\di\tau\,\dot{\mathbbvar{p}}^M\omega_{MN}\mathbbvar{p}^N }\right) \exp\!\left({\frac{i}{\hbar}\widetilde{S}_\mathrm{abb}} \right) .
    \end{aligned}
\end{equation}

Let us make a sanity check. For an ordinary toroidal space $\mathbbvar{p}$ is constant in $\tau$ so it is zero for ordinary strings on a torus. The reader at this point may feel frustrated that after some considerable care with total derivatives and quasi-periodic fields we have generated a term that vanishes. However, crucially this term will not vanish for strings in backgrounds where the winding number is not conserved. Such a situation is exactly where double field theory is most useful captures the dynamic nature of winding. For example this occurs with a string in a Kaluza-Klein monopole background; a set up that was first considered in \cite{Gregory:1997te} and studied using double field theory in \cite{Jensen:2011jna}. This term will also make a contribution if there is no globally defined duality frame and so one needs to form a good cover over the space and choose a duality frame in each patch. Such spaces with no globally defined T-duality frame are called T-folds. The above phase shift will then be part of the transition function between different patches acting on the string wavefunction. From the geometric quantisation perspective this is reminiscent of the Maslov correction.


\paragraph{The Hull "topological" term and its role.}
In addition to the discussion above, Hull proposed the addition of a "topological" to Tseytlin action based on global requirements for a gauging procedure \cite{Hull:2006va}. The importance of this term for the partition function was emphasized in \cite{Berman:2007vi,Tan:2014mba}. This term is given by:
\begin{equation}\label{eq:topterm}
    S_{\mathrm{top}}[\mathbb{X}(\sigma,\tau)] \;=\; \frac{1}{4\pi\alpha'}\int\di\tau \oint\di\sigma\, \dot{\mathbb{X}}^M\omega_{MN} \mathbb{X}^{\prime N}.
\end{equation}
Now, we can use Stokes' theorem as it follows:
\begin{gather}\label{eq:topExpanded}
    \begin{aligned}
    S_{\mathrm{top}}[\mathbb{X}(\sigma,\tau)] \;&=\; \frac{1}{4\pi\alpha'}\int\di\tau \oint\di\sigma\, \dot{\mathbb{X}}^{M}\omega_{MN}\mathbb{X}^{\prime N} \\
    \;&=\; \frac{1}{2\pi\alpha'}\int\!\!\!\!\oint_\Sigma \di{X}^{\mu}\wedge\di\widetilde{X}_\mu   \\
    \;&=\; \frac{1}{2\pi\alpha'} \int\di\tau\left({X}^{\mu}(\sigma_0+2\pi,\tau)\dot{\widetilde{X}}_\mu(\sigma_0+2\pi,\tau) - {X}^{\mu}(\sigma_0,\tau)\dot{\widetilde{X}}_\mu(\sigma_0,\tau)  \right)  \\
    &= \frac{1}{2} \int\di\tau\left(\dot{\tilde{p}}{\widetilde{X}}(\sigma_0) + {p}\dot{X}(\sigma_0) + 2\pi\alpha'\dot{\tilde{p}}p\right),
    \end{aligned}\raisetag{0.7cm}
\end{gather}
where, without any loss of generality, we have chosen the gauge $X^{\mu}\di\widetilde{X}_\mu$ for the potential of the $2$-form $\di{X}^{\mu}\wedge\di\widetilde{X}_\mu$.

Just like for the Tseytlin action, we can define a boundary term for the topological term
\begin{equation}
     S_{\partial\mathrm{top}}[\mathbb{X}(0,\tau),\mathbbvar{p}(\tau)] \;=\; -\frac{1}{2}\int\di\tau \,\dot{\mathbbvar{p}}^M(\tau)\omega_{MN}{\mathbb{X}}^N(\sigma_0,\tau)
\end{equation}
to remove the dependence on the cut $\sigma_0$.

Recall the equation \eqref{eq:totderivative} relating the Tseytlin abbreviated action of a doubled string with the term $\int\di\tau\oint\di\sigma\,\dot{X}^\mu P_\mu$. By combining equation \eqref{eq:totderivative} with equation \eqref{eq:topExpanded}, we immediately obtain the relation
\begin{equation}
     \frac{1}{2\pi\alpha'}\int\di\tau\oint\di\sigma\,\dot{X}^\mu P_\mu \;=\; \frac{1}{4\pi\alpha'}\int\di\tau \oint\di\sigma\, \dot{\mathbb{X}}^M(\eta_{MN} + \omega_{MN}) \mathbb{X}^{\prime N} .
\end{equation}
At this point, we can provide both sides of the equation with the boundary term, so that we have
\begin{equation}\label{SabbTop}
     S_{\mathrm{abb}} \;=\; \frac{1}{4\pi\alpha'}\int\di\tau \oint\di\sigma\, \dot{\mathbb{X}}^M(\eta_{MN} + \omega_{MN}) \mathbb{X}^{\prime N} - \frac{1}{2}\int\di\tau\,\mathbbvar{p}^M(\eta_{MN}+\omega_{MN})\dot{\mathbb{X}}^N(\sigma_0),
\end{equation}
where we recall the definition \eqref{Sabb} for the abbreviated action (including the boundary term) of the ordinary string:
\begin{equation}
    S_\mathrm{abb} \;:=\; \frac{1}{2\pi\alpha'}\int\di\tau\oint\di\sigma\, \dot{X}^\mu P_\mu - \int\di\tau\, \dot{\tilde{p}}^{\mu}\widetilde{X}_\mu(\sigma_0)
\end{equation}
This equation can be interpreted as the fact that we need to add the boundary term $-\int\di\tau\, \dot{\tilde{p}}^{\mu}\widetilde{X}_\mu(\sigma_0)$ to the usual abbreviated action $\frac{1}{2\pi\alpha'}\int\di\tau\oint\di\sigma\, \dot{X}^\mu P_\mu$ of an ordinary string to obtain the manifestly $O(n,n)$-covariant abbreviated action \eqref{SabbTop}.

\paragraph{The total action.}
Finally, putting all this together, the total action of the doubled string will be given as follows
\begin{gather}
    \begin{aligned}
    \mathbb{S}[\mathbb{X}(\sigma,\tau)] &= \frac{1}{4\pi\alpha'}\!\int\!\di\tau \!\oint\!\di\sigma\!\left( \dot{\mathbb{X}}^{M}(\sigma,\tau)(\eta_{MN}+\omega_{MN} )\mathbb{X}^{\prime N}(\sigma,\tau) -  \mathbb{X}^{\prime M}(\sigma,\tau)\mathcal{H}_{MN}\mathbb{X}^{\prime N}(\sigma,\tau)\right) \\
    \,&-\, \frac{1}{2}\int\di\tau\,\mathbbvar{p}^M(\eta_{MN}+\omega_{MN})\dot{\mathbb{X}}^N(\sigma_0).
    \end{aligned}
    \raisetag{0.7cm}
\end{gather}

\paragraph{Fourier expansion of the kinetic term.}
Now, recall that the Hamiltonian of the doubled string is the functional
\begin{equation}
    H[\mathbb{X}(\sigma)] \;=\; \frac{1}{4\pi\alpha'}\oint\di\sigma\,\mathbb{X}^{\prime M}(\sigma)\mathcal{H}_{MN}\big(\mathbb{X}(\sigma)\big)\mathbb{X}^{\prime N}(\sigma).
\end{equation}
We can now expand $\mathbb{X}^M(\sigma)$ in $\sigma$ by
\begin{equation}
    \mathbb{X}^M(\sigma) \;=\; \mathbbvar{x}^M + \alpha'\sigma\mathbbvar{p}^M + \!\!\sum_{n\in\mathbb{Z}\backslash\{0\}}\!\frac{1}{n}\bbalpha^M_n e^{in\sigma},
\end{equation}
where $\bbalpha^M_n$ satisfies the identity $\bbalpha^M_{-n}= \bar{\bbalpha}^M_n$ and it can be decomposed as $\bbalpha^M_n = (\alpha_n^\mu,\,\widetilde{\alpha}_{n\mu})$ with
\begin{equation}
    \widetilde{\alpha}_{n\mu} \;=\; \begin{cases}-E_{\mu\nu}^{\mathrm{T}}\,\alpha_n^\nu, & n>0\\+E_{\mu\nu}\,\alpha_n^\nu, & n<0\end{cases}
\end{equation}
The coordinates of the phase space can be taken as $\{\mathbbvar{x}^M,\mathbbvar{p}^M,\alpha_k^\mu,\bar{\alpha}_k^\mu\}_{k\in\mathbb{N}\backslash\{0\}}$.
Now we can explicitly express the kinetic part of the action of the doubled string in these coordinates.
Firstly, we calculate the mode expansion of the abbreviated Tseytlin action together with the topological term:
\begin{equation}
\begin{aligned}
    &\frac{1}{4\pi\alpha'}\int\di\tau \oint\di\sigma\, \dot{\mathbb{X}}^{M}(\eta_{MN}+\omega_{MN})\mathbb{X}^{\prime N} \,= \, \\ 
    &=\, \frac{1}{2} \int\di\tau\!\left(p_\mu\dot{x}^\mu + \pi\alpha'\dot{\tilde{p}}^\mu p_\mu + \dot{\tilde{p}}^\mu\!\!\!\sum_{n\in\mathbb{Z}\backslash\{0\}}\!\frac{1}{n}\dot{\tilde{\alpha}}_{n\mu} +i \!\!\!\sum_{n\in\mathbb{Z}\backslash\{0\}}\!\frac{1}{n}\,\omega _{MN}\,\dot{\bbalpha}_{-n}^M\bbalpha_n^N \right)\!.
\end{aligned}
\end{equation}
Then, we expand the boundary terms:
\begin{equation}
    \begin{aligned}
    (S_{\partial\mathrm{Tsey}}+S_{\partial\mathrm{top}})[\mathbb{X}(0,\tau),\mathbbvar{p}(\tau)] \;&=\; -\int\di\tau \,\frac{1}{2}\mathbbvar{p}^M(\tau)(\eta_{MN}+\omega_{MN} )\dot{\mathbb{X}}^N(\sigma_0,\tau) \\
    \;&=\; -\int\di\tau\,\frac{1}{2}\left( \tilde{p}^\mu\dot{\tilde{x}}_\mu + \tilde{p}^\mu\!\!\!\sum_{n\in\mathbb{Z}\backslash\{0\}}\!\frac{1}{n}\dot{\tilde{\alpha}}_{n\mu} \right)\!.
    \end{aligned}
\end{equation}
By adding these terms together, we obtain the the mode expansion of the abbreviated action of the doubled string, which is the following:
\begin{equation*}
\begin{aligned}
    \mathbb{S}[\mathbb{X}(\sigma,\tau)]+  \int\!\di\tau H[\mathbb{X}(\sigma,\tau)] \,&=\, \int\!\di\tau \!\left( \mathbbvar{p}_M\dot{\mathbbvar{x}}^M - \frac{\pi\alpha'}{2}\omega^{MN}\mathbbvar{p}_M\dot{\mathbbvar{p}}_N +i \!\!\!\sum_{n\in\mathbb{Z}\backslash\{0\}}\!\frac{1}{n}\,\omega _{MN}\,\dot{\bbalpha}_{-n}^M\bbalpha_n^N \right)\!.
\end{aligned}
\end{equation*}

\paragraph{The symplectic structure of the doubled string.}
Now we can use the equation
\begin{equation}\label{eq:Liouvillepot}
\begin{aligned}
    \mathbb{S}[\mathbb{X}(\sigma,\tau)]+ \int\di\tau H[\mathbb{X}(\sigma,\tau)] \;&=\; \int\di\tau\,\iota_{V_H}\mathbbvar{\Theta}
\end{aligned}
\end{equation}
to determine the Liouville potential $\mathbbvar{\Theta}$ on the phase space of the doubled string and, hence, the its symplectic structure. 
To solve the equation, we choose again the Hamiltonian vector $V_H$ associated to the time flow, which, in the new coordinates $\{\mathbbvar{x}^M,\mathbbvar{p}_M,\alpha_k^\mu,\bar{\alpha}_k^\mu\}_{n\in\mathbb{N}\backslash\{0\}}$ of the phase space, takes the form
\begin{equation}
\begin{aligned}
    V_H \;&=\; \frac{\di}{\di\tau} \\
    &=\; \dot{\mathbbvar{x}}^M\frac{\partial}{\partial\mathbbvar{x}^M} + \dot{\mathbbvar{p}}^M\frac{\partial}{\partial\mathbbvar{p}^M} + \sum_{n>0} \left(\dot{\alpha}_n^\mu \frac{\partial}{\partial\alpha_n^\mu} + \dot{\bar{\alpha}}_n^\mu \frac{\partial}{\partial\bar{\alpha}_n^\mu} \right)
    \end{aligned}.
\end{equation}
Now, by solving the equation \eqref{eq:Liouvillepot} we obtain the following Liouville potential:
\begin{equation}
\begin{aligned}
    \mathbbvar{\Theta} \;&=\; \mathbbvar{p}_M \di\mathbbvar{x}^M -\frac{\pi\alpha'}{2} \omega^{MN}\mathbbvar{p}_M\di{\mathbbvar{p}}_N +i \!\!\!\sum_{n\in\mathbb{Z}\backslash\{0\}}\!\frac{1}{n}\,\omega _{MN}\,\di\bbalpha_{-n}^M\bbalpha_n^N.
\end{aligned}
\end{equation}
By calculating the differential $\mathbbvar{\Omega}=\delta\mathbbvar{\Theta}$, we finally obtain the symplectic form
\begin{equation}
    \mathbbvar{\Omega} \;=\; \di\mathbbvar{p}_M\wedge\di\mathbbvar{x}^M - \frac{\pi\alpha'}{2}\omega^{MN}  \di\mathbbvar{p}_M\wedge\di\mathbbvar{p}_N +i \!\!\!\sum_{n\in\mathbb{N}\backslash\{0\}}\!\frac{1}{n}\,\omega _{MN}\,\di\bbalpha_{-n}^M\wedge\di\bbalpha_n^N.
\end{equation}
This is, therefore, the symplectic form of the phase space $(\mathcal{L}_\mathrm{Q}\mathcal{M},\,\mathbbvar{\Omega})$ of the doubled string.

Notice that we can also rewrite this symplectic form as
\begin{equation}
    \mathbbvar{\Omega} \;=\; \oint\di\sigma\,\frac{1}{2} \omega_{MN}\,\delta\mathbb{X}^M(\sigma)\wedge\delta\mathbb{X}^{\prime N}(\sigma) + \di\mathbbvar{p}_M\wedge\di\mathbb{X}^M(\sigma_0).
\end{equation}
Notice that, if we ignore the second term encoding the boundary, the first term can be immediately given by a potential $\oint\di\sigma\,\frac{1}{2} \omega_{MN}\,\delta\mathbb{X}^M(\sigma)\mathbb{X}^{\prime N}(\sigma)$, which is nothing but the transgression of a symplectic form
$\varpi := \frac{1}{2}\omega_{MN}\,\di\mathbbvar{x}^M\wedge\di\mathbbvar{x}^{N} = \di x^\mu\wedge \di \tilde{x}_\mu$ defined on the doubled space. Notice that this is still a particular example of the fundamental $2$-form which appears in Born geometry in \cite{Svo17, Svo18, Svo19} for a background without fluxes.

\subsection{Algebra of observables}
We want to determine the algebra $\mathfrak{heis}(\mathcal{L}_\mathrm{Q}\mathcal{M},\mathbbvar{\Omega})$ of quantum observables of the phase space of the doubled string.
\begin{equation}
    \hat{f} \;=\; i\hbar\nabla_{V_f} + f.
\end{equation}
We, thus, obtain the following commutation relations:
\begin{equation}\label{eq:commsigma}
    \big[\hat{\mathbb{X}}^M(\sigma)\,,\hat{\mathbb{X}}^N(\sigma')\big] \;=\; i\pi\hbar\alpha'\omega^{MN} -i\hbar\eta^{MN}\varepsilon(\sigma-\sigma') 
\end{equation}
where the function $\varepsilon(\sigma)$ is the quasi-periodic function defined by
\begin{equation}
    \varepsilon(\sigma) \;:=\; \sigma - i\!\!\!\sum_{n\in\mathbb{Z}\backslash\{0\}}\!\!\!\frac{e^{in\sigma}}{n}
\end{equation}
and it satisfies the following properties: firstly, its derivative $\varepsilon'(\sigma)=\delta(\sigma)$ is the Dirac comb; secondly, it satisfies the boundary condition $\varepsilon(\sigma+2\pi n)=\varepsilon(\sigma)+2\pi n$ and, finally, it is an odd function, i.e. $\varepsilon(-\sigma)=-\varepsilon(\sigma)$.

The fact that the operators associated with $X^\mu(\sigma)$ and $\widetilde{X}_\mu(\sigma')$ do not commute by a term $\propto\epsilon(\sigma-\sigma')$ was already observed as far as in \cite{DGMP13}. However, the commutation relations \eqref{eq:commsigma} contain a new term: the skew-symmetric constant matrix $\propto\pi\hbar\alpha'\omega^{MN}$, originating from the topological term \eqref{eq:topterm} in the total action of the doubled string and totally analogous of the one recently observed by \cite{Fre17a, Fre17b}.

We can also easily derive the commutation relations for the higher modes
\begin{equation}
    \left[\hat{\alpha}_n^\mu, \hat{\alpha}_m^\nu\right] \;=\; ng^{\mu\nu}\delta_{m+n,0},
\end{equation}
and we have the identity $\bar{\alpha}_n^\mu = \alpha_{-n}^\mu$.

\paragraph{Limits of the algebra of observables} It is worth remarking the role of the dimensionful constants $\hbar$ and $\alpha'$ in providing the deformation to the classical algebra. We are used to seeing $\hbar$ as a quantum deformation parameter but here we also see $\hbar \alpha'$ as another quantum deformation parameter. This suggests interesting limits. The classical limit is the obvious limit given by $\hbar \rightarrow 0$. The particle limit is $\hbar$ fixed but $\alpha' \rightarrow 0$. This algebra suggests a new limit:
\begin{equation}
\hbar \rightarrow 0 \, ; \qquad \alpha' \rightarrow \infty\, ; \qquad \alpha' \hbar \,\,\, \text{fixed}.
\end{equation}
This would be a classical stringy limit where we keep the stringy deformation but remove the quantum deformation. It would be interesting to study the system further in this limit to identify the pure string deformation based effects.

\section{Geometric quantisation on the doubled space}

\subsection{The phase space of the zero-mode string}
Recall that we can expand the fields $\mathbb{X}^M(\sigma)$ of our doubled string $\sigma$-model by
\begin{equation}
    \mathbb{X}^M(\sigma) \;=\; \mathbbvar{x}^M + \alpha'\sigma\mathbbvar{p}^M + \!\!\sum_{n\in\mathbb{Z}\backslash\{0\}}\!\frac{1}{n}\bbalpha^M_n e^{in\sigma},
\end{equation}
where the coordinates of the phase space of a doubled string are $\{\mathbbvar{x}^M,\mathbbvar{p}^M,\alpha_k^\mu,\bar{\alpha}_k^\mu\}_{k\in\mathbb{N}\backslash\{0\}}$, with the former $\{\mathbbvar{x}^M,\mathbbvar{p}^M\}$ co-ordinatising the zero-modes of the string and the latter $\{\alpha_k^\mu,\bar{\alpha}_k^\mu\}_{k\in\mathbb{N}\backslash\{0\}}$ co-ordinatising its higher-modes.
A zero-mode truncated doubled string is a doubled string where we are neglecting the higher-modes and it can be seen as a simple embedding of the form
\begin{equation}
    \mathbb{X}^M(\sigma) \;=\; \mathbbvar{x}^M + \alpha'\sigma\mathbbvar{p}^M.
\end{equation}
The zero-modes of a doubled string $\mathbb{X}^M(\sigma,\tau)$ can be thought as a particle in a doubled phase space $(\mathbbvar{x}^M(\tau), \mathbbvar{p}_M(\tau))$. Similarly we expect that the wave-functional $\Psi[\mathbb{X}(\sigma)]$ at zero modes is just a wave-function $ \psi(\mathbbvar{x},\mathbbvar{p})$ on the doubled phase space of zero modes:
\begin{equation}
    \Psi[\mathbb{X}(\sigma)] \;\; \xrightarrow{\;{0\text{ modes}}\;} \;\; \psi(\mathbbvar{x},\mathbbvar{p}), \qquad \mathbbvar{\Omega} \;\; \xrightarrow{\;{0\text{ modes}}\;} \;\; \bbomega.
\end{equation}
The phase space of the zero-modes of a doubled string is, therefore, a $4n$-dimensional symplectic manifold $(\mathcal{P},\bbomega)$ with symplectic form
\begin{equation}
    \bbomega \;=\; \eta_{MN}\, \di\mathbbvar{p}^M\wedge\di\mathbbvar{x}^N - \frac{\pi\alpha'}{2}\omega_{MN} \, \di\mathbbvar{p}^M\wedge\di\mathbbvar{p}^N
\end{equation}
and underlying smooth manifold $\mathcal{P}=\mathbb{R}^{4n}$. Notice that this $2$-form, obtained by a Hamiltonian treatment of the total action of a doubled string $\sigma$-model, exactly agrees with the symplectic form found by \cite{Fre17a, Fre17b} by starting from vertex algebra arguments. 

Now, we can apply the machinery of geometric quantisation to this symplectic manifold $(\mathcal{P},\bbomega)$ to quantise the zero-modes of a doubled string.

\paragraph{Kinetic coordinates for the doubled phase space.}
Let us change the canonical momentum coordinates $\{\mathbbvar{p}^M\}$ with the untwisted non-canonical momentum coordinates $\{\mathbbvar{k}^M\}$ given by $\mathbbvar{k}^M=(e^{-B})^M_{\;\;\,N}\mathbbvar{p}^N$, where the matrix $(e^{-B})^M_{\;\;\,N}$ is given by the embedding of $B_{\mu\nu}$ seen as a nilpotent matrix, i.e.
\begin{equation}
    (e^{-B})^M_{\;\;\,N} \;=\; \begin{pmatrix}
\delta^\mu_{\;\nu} & 0\\
-B_{\mu\nu} & \delta_\mu^{\;\nu}
\end{pmatrix}.
\end{equation}
Given a the fields $\mathbb{X}(\sigma,\tau)$ of a doubled $\sigma$-model, these will be related with $\mathbbvar{k}^M$ by
\begin{equation}
    k_\mu(\sigma,\tau) = \oint \di\sigma\, g_{\mu\nu}\dot{X}^\nu(\sigma,\tau), \qquad \tilde{k}^\mu(\sigma,\tau) = \oint \di\sigma\, g^{\mu\nu}\dot{\widetilde{X}}_\nu(\sigma,\tau).
\end{equation}
We can rotate the doubled coordinates, accordingly $\mathbbvar{x}^M\mapsto(e^{-B})^M_{\;\;\,N}\mathbbvar{x}^N$ to the untwisted frame.
We can now rewrite the symplectic form in the kinetic coordinates $\{\mathbbvar{x}^M,\mathbbvar{k}^M\}$, so that we find
\begin{equation}
    \bbomega \;=\; \eta_{MN}\, \di\mathbbvar{k}^M\wedge\di\mathbbvar{x}^N - \frac{\pi\alpha'}{2}\omega_{MN}^{(B)}\, \di\mathbbvar{k}^M\wedge\di\mathbbvar{k}^N,
\end{equation}
where we called the matrix
\begin{equation}
    \omega^{(B)}_{MN} \;=\;  \begin{pmatrix}B_{\mu\nu} & \delta_\mu^{\;\,\nu} \\-\delta_{\;\,\nu}^{\mu} & 0 \end{pmatrix}.
\end{equation}
Then, we can choose the following gauge for the Liouville potential:
\begin{equation}
    \bbtheta \;=\; \eta_{MN}\, \mathbbvar{k}^M\di\mathbbvar{x}^N - \frac{\pi\alpha'}{2}\omega_{MN}^{(B)}\, \mathbbvar{k}^M\di\mathbbvar{k}^N.
\end{equation}

\paragraph{The action of the zero-mode string.}
As we remarked, in geometric quantisation the Lagrangian density $\mathfrak{L}\in\Omega^1(\gamma)$ of a particle is related to the Liouville potential $\bbtheta\in\Omega^1(\mathcal{P})$ by the equation 
\begin{equation}
    \mathfrak{L}_H \,=\, (\iota_{V_H}\bbtheta-H)\di\tau.
\end{equation}
We can then immediately use it, in the form
\begin{equation}
    S[\mathbbvar{x}(\tau),\mathbbvar{k}(\tau)] \,=\, \int_\gamma\di\tau\big(\iota_{V_H}\bbtheta-H\big) \qquad \text{with} \qquad V_H=\dot{\mathbbvar{x}}^M\frac{\partial}{\partial \mathbbvar{x}^M} + \dot{\mathbbvar{k}}^M\frac{\partial}{\partial \mathbbvar{k}^M},
\end{equation}
to find the action of the zero-mode doubled string: 
\begin{equation}
    S[\mathbbvar{x}(\tau),\mathbbvar{k}(\tau)] \,=\, \int_\gamma \di\tau \left(\eta_{MN}\mathbbvar{k}^M\dot{\mathbbvar{x}}^N- \frac{\pi\alpha'}{2}\omega_{MN}^{(B)}\, \mathbbvar{k}^M\dot{\mathbbvar{k}}^N -\mathcal{H}_{MN}^{(0)} \mathbbvar{k}^M\mathbbvar{k}^N\right)\!,
\end{equation}
where we called the matrix
\begin{equation}
    \mathcal{H}^{(0)}_{MN} \;=\;  \begin{pmatrix}g_{\mu\nu} & 0 \\0 & g^{\mu\nu} \end{pmatrix}.
\end{equation}

\subsection{Algebra of the observables}

Recall that in geometric quantisation a quantum observable $\hat{f}\in\mathrm{Aut}(\mathbf{H})$ is a linear automorphism of the Hilbert space, obtained from the corresponding classic observable $f\in\mathcal{C}^\infty(\mathcal{P})$ by the following identification:
\begin{equation}\label{eq:op}
    \hat{f} \;:=\; i\hbar \nabla_{V_f} + f,
\end{equation}
where the vector $V_f\in\mathfrak{X}(\mathcal{P})$ is the Hamiltonian vector with Hamiltonian function $f$, i.e. the vector which solves the Hamilton equation
\begin{equation}\label{eq:hamilton}
    \iota_{V_f}\bbomega\,=\,\di f.
\end{equation}
In this subsection we want to determine the Lie algebra of quantum observables  $\mathfrak{heis}(\mathcal{P},\bbomega)$ on the doubled phase space.

\paragraph{Hamiltonian vector fields.}
Let us first solve the Hamilton equation \eqref{eq:hamilton} for a generic Hamiltonian function $f\in\mathcal{C}^\infty(\mathcal{P})$. We expand the vector $V_f\in\mathfrak{X}(\mathcal{P})$ in the kinetic coordinates
\begin{equation}
    V_f \;=\; V_{f,\mathbbvar{x}}^M\,\frac{\partial}{\partial\mathbbvar{x}^M} \,+\, V_{f,\mathbbvar{k}}^M\,\frac{\partial}{\partial\mathbbvar{k}^M}.
\end{equation}
Hence, the Hamilton equation \eqref{eq:hamilton}, in coordinates, becomes
\begin{equation}\begin{aligned}
   \eta_{MN}\,( V_{f,\mathbbvar{k}}^M \,\di\mathbbvar{x}^N -V_{f,\mathbbvar{x}}^M \,\di\mathbbvar{k}^N ) - \pi\alpha'\omega_{MN}^{(B)}\, V_{f,\mathbbvar{k}}^M\,\di\mathbbvar{k}^N \;&=\; \frac{\partial f}{\partial\mathbbvar{x}^M}\di\mathbbvar{x}^M + \frac{\partial f}{\partial\mathbbvar{k}^M}\di\mathbbvar{k}^M.
\end{aligned}\end{equation}
Therefore, the Hamiltonian vector field $V_f$ with Hamiltonian $f$ is given by
\begin{equation}
    \begin{aligned}
    V_{f} \;=\; \Big( \eta^{MN}\frac{\partial f}{\partial\mathbbvar{x}^N} \Big) \frac{\partial}{\partial\mathbbvar{k}^M} + \Big( -\eta^{MN}\frac{\partial f}{\partial\mathbbvar{k}^N} - \pi\alpha'\omega^{MN}_{(B)}\, \frac{\partial f}{\partial\mathbbvar{x}^N} \Big) \frac{\partial}{\partial\mathbbvar{x}^M},
    \end{aligned}
\end{equation}
where we called $\omega^{MN}_{(B)}:=\eta^{ML}\, \omega_{LP}^{(B)}\, \eta^{PN}$. 
In particular the Hamiltonian vector fields corresponding to the classical observables of the kinetic coordinates $\mathbbvar{x}^M$ and $\mathbbvar{k}^M$ are
\begin{equation}
    \begin{aligned}
    V_{f=\mathbbvar{x}^N} \;&=\; \eta^{MN}\frac{\partial }{\partial\mathbbvar{k}^M} - \pi\alpha'\omega^{MN}_{(B)}\, \frac{\partial }{\partial\mathbbvar{x}^M},\\
    V_{f=\mathbbvar{k}^N}  \;&=\; -\eta^{MN}\frac{\partial }{\partial\mathbbvar{x}^M}.
    \end{aligned}
\end{equation}

\paragraph{Non-commutative Heisenberg algebra.}
By applying the definition \eqref{eq:op} of quantum observable, we find that the operators associated to the kinetic coordinates are the following:
\begin{equation}
\begin{aligned}
        \hat{\mathbbvar{x}}^M \,&=\,  i\hbar\eta^{NM}\frac{\partial}{\partial\mathbbvar{k}^N}  -i\hbar\frac{\pi\alpha'}{2}\omega_{(B)}^{NM}\frac{\partial}{\partial\mathbbvar{x}^N} + \mathbbvar{x}^M, \\
        \hat{\mathbbvar{k}}^M \,&=\,  -i\hbar\eta^{NM}\frac{\partial}{\partial\mathbbvar{x}^N}.
\end{aligned}
\end{equation}
Therefore the commutation relations between the coordinates operators are the following:
\begin{equation}
    [\hat{\mathbbvar{x}}^M,\hat{\mathbbvar{x}}^N] \,=\, \pi i\hbar\alpha' \omega^{MN}_{(B)}, \quad [\hat{\mathbbvar{x}}^M,\hat{\mathbbvar{k}}^N] \,=\, i\hbar\eta^{MN}, \quad [\hat{\mathbbvar{k}}^M,\hat{\mathbbvar{k}}^N] \,=\,0.
\end{equation}
Thus, the $4n$-dimensional Lie algebra $\mathfrak{heis}(\mathcal{P},\bbomega)$ can be regarded as a non-commutative version of the usual Heisenberg algebra, where the position operators do not generally commute.

Explicitly, in undoubled notation, we have the following commutation relations:
\begin{equation}\label{eq:commrelund}
    \begin{aligned}
    [\hat{x}^\mu,\hat{x}^\nu]&=0,&\quad [\hat{x}^\mu,\hat{\tilde{x}}_\nu]&=\pi i\hbar \alpha'\delta^\mu_{\;\nu},& \quad [\hat{\tilde{x}}_\mu,\hat{\tilde{x}}_\nu]&=-2\pi i\hbar\alpha'B_{\mu\nu}, \\[0.5ex]
    [\hat{k}_\mu,\hat{k}_\nu]&=0,&\quad [\hat{k}_\mu,\hat{\tilde{k}}^\nu]&=0,& \quad [\hat{\tilde{k}}^\mu,\hat{\tilde{k}}^\nu]&=0, \\[0.5ex]
    [\hat{x}^\mu,\hat{\tilde{k}}^\nu]=[\hat{\tilde{x}}_\mu,\hat{k}_\nu]&=0,&\quad [\hat{x}^\mu,\hat{k}_\nu] &= i\hbar\delta^\mu_{\;\nu},& \quad [\hat{\tilde{x}}_\mu,\hat{\tilde{k}}^\nu] &= i\hbar\delta_\mu^{\;\nu}.
    \end{aligned}
\end{equation}

Examining this algebra from the perspective of the limits we discussed earlier we see that $\hbar$ controls the noncommutativity of the position with the momentum and $\hbar \alpha'$ the noncommutativity of the coordinates and their duals. Finally, $\alpha' B$ the noncommutativity of the spacetime coordinates. Thus when the B-field is included we have three noncommutativity parameters.

\paragraph{Uncertainty principle on the doubled space.}
Following standard text book techniques applied to the commutation relations \eqref{eq:commrelund}, we can immediately show that any position coordinate $x^\mu$ and its dual $\tilde{x}_\mu$ satisfy the following uncertainty relation:
\begin{equation}
    \Delta x \, \Delta\tilde{x} \;\geq\; \frac{\pi\hbar}{2}\alpha'.
\end{equation}
This means that $x^\mu$ and $\tilde{x}_\mu$ cannot be measured with absolute precision at the same time, but there will be always a minimum uncertainty proportional to the area $\hbar\alpha'$. This provides support to the intuition of a minimal distance scale in string theory. The standard lore is that for small distances one goes to the T-dual frame and the distances will always be larger than the string scale.

In addition, both the couples $(x,p)$ and $(\tilde{x},\tilde{p})$ satisfy the usual uncertainty relation between position and momentum:
\begin{equation}
    \Delta x \, \Delta p \;\geq\; \frac{\hbar}{2}, \qquad \Delta \tilde{x} \, \Delta\tilde{p} \;\geq\; \frac{\hbar}{2}.
\end{equation}
However, it is interesting to notice that the momentum and its dual can be measured at the same time, i.e.
\begin{equation}
    \Delta p \, \Delta\tilde{p} \;\geq\; 0.
\end{equation}

\paragraph{Hamiltonian.}
Notice that the Hamiltonian operator of the zero-mode doubled string will be given by
\begin{equation}
    \hat{H} \,=\, \mathcal{H}_{MN}^{(0)} \hat{\mathbbvar{k}}^M\hat{\mathbbvar{k}}^N \,=\, -\hbar^2\, \mathcal{H}^{MN}_{(0)} \frac{\partial}{\partial\mathbbvar{x}^M}\frac{\partial}{\partial\mathbbvar{x}^N}
\end{equation}
where we called $\mathcal{H}^{MN}_{(0)}:=\eta^{ML}\eta^{NP}\mathcal{H}_{LP}^{(0)}$.

\paragraph{Non-commutative Heisenberg algebra in canonical coordinates.}
In the zero-mode string canonical coordinates $\{\mathbbvar{x}^M,\mathbbvar{p}_M\}$ we obtain the following operators:
\begin{equation}
\begin{aligned}
        \hat{\mathbbvar{x}}^M \,&=\,  i\hbar\frac{\partial}{\partial\mathbbvar{p}_M}  -i\hbar\frac{\pi\alpha'}{2}\omega^{NM}\frac{\partial}{\partial\mathbbvar{x}^N} + \mathbbvar{x}^M \\
        \hat{\mathbbvar{p}}_M \,&=\,  -i\hbar\frac{\partial}{\partial\mathbbvar{x}^M}
\end{aligned}
\end{equation}
Therefore the commutation relations between the canonical coordinates observables are 
\begin{equation}\label{eq:canonicalhesialgebra}
    [\hat{\mathbbvar{x}}^M,\hat{\mathbbvar{x}}^N] \,=\, \pi i\hbar\alpha' \omega^{MN}, \quad [\hat{\mathbbvar{x}}^M,\hat{\mathbbvar{p}}_N] \,=\, i\hbar\delta^{M}_{\;\;N}, \quad [\hat{\mathbbvar{p}}_M,\hat{\mathbbvar{p}}_N] \,=\,0.
\end{equation}

\paragraph{Relation with the symplectic structure of the doubled space.}
Let us now focus on the subalgebra generated by the operators $\hat{x}^\mu$ and $\hat{\tilde{x}}_\mu$. This will be given by the following commutation relations:
\begin{equation}
    \begin{aligned}
    [\hat{x}^\mu,\hat{x}^\nu]=0,\quad [\hat{x}^\mu,\hat{\tilde{x}}_\nu]=\pi i\hbar \alpha'\delta^\mu_{\;\nu}, \quad [\hat{\tilde{x}}_\mu,\hat{\tilde{x}}_\nu]=0.
    \end{aligned}
\end{equation}
Notice that this can be seen as an ordinary $2n$-dimensional Heisenberg algebra $\mathfrak{h}(2n)$. This means that such an algebra is immediately given by a symplectic manifold $(\mathcal{M},\varpi)$ with $\mathcal{M}\cong\mathbb{R}^{2n}$ and symplectic form $\varpi := \pi\hbar\alpha'\di x^\mu\wedge \di \tilde{x}_\mu$. This symplectic structure on the doubled space is exactly the one introduced by \cite{Vai12}.

\subsection{T-duality and the string deformed Fourier transform}

In ordinary quantum mechanics we choose to represent the wavefunctions in either the position or the momentum basis and it is the Fourier transform that maps the wavefunction in one basis to the other basis. From the persepective of geometric quantisation this is the transformation between elements of the Hilbert spaces constructed with different choices of Lagrangian submanifold ie. different polarisations. T-duality is a change in our choice of polarisation. We can then follow the pairing construction used in \cite{WeiGQ, Kostant, Souriau, Woodhouse:1980pa} to construct the transformation for the string wavefunction moving between different duality frames. This will produce a string deformed Fourier transform (that reduces to the usual Fourier transform in the $\alpha' \rightarrow 0$ limit). From the geometric quantisation perspective these transformations are known as Blatter-Kostant-Sternberg \cite{WeiGQ} kernel's. 

\paragraph{Polarisations.}
In the geometric quantisation of a symplectic space $(\mathcal{P},\bbomega)$, a polarisation corresponds to a choice of an integrable Lagrangian subspace $L\subset \mathcal{P}$. 
Since $\mathcal{P}$ is a vector space, the first Chern class of the prequantum $U(1)$-bundle whose curvature is the symplectic form $\bbomega\in\Omega^2(\mathcal{P})$, is necessarily trivial.
In geometric quantisation, this implies that the Hilbert space of the quantised system is defined by the space of the complex $\mathrm{L}^2$-functions on the Lagrangian submanifold $L\subset\mathcal{P}$, i.e. by
\begin{equation}
    \mathbf{H} \;:=\; \mathrm{L}^2(L,\mathbb{C}).
\end{equation}
Remarkably, this does not depend on the choice of polarisation $L\subset\mathcal{P}$ and it is possible to prove that, for any other Lagrangian subspace $L'\subset\mathcal{P}$, we would have an isomorphism of Hilbert spaces $\mathbf{H} \cong \mathrm{L}^2(L',\mathbb{C})$.

\paragraph{T-duality as a change of polarisation.}
Let us rewrite the symplectic form $\bbomega\in\Omega^2(\mathcal{P})$ in canonical coordinates $(\mathbbvar{x}^M,\mathbbvar{p}^M)$, i.e.
\begin{equation}
    \bbomega \;=\; \eta_{MN}\,\di\mathbbvar{p}^M\wedge\di\mathbbvar{x}^N - \frac{\pi\alpha'}{2}\omega_{MN}\, \di\mathbbvar{p}^M\wedge\di\mathbbvar{p}^N,
\end{equation}
and let us recall that the momenta doubled vector can be interpreted as the doubled vector of winding numbers $\mathbbvar{p}^M=(w^\mu,\,\tilde{w}_\mu)$. 
It is now immediate that the vector spaces
\begin{equation}
    L \;:=\; \mathrm{Span}(x^\mu,w^\mu), \qquad \widetilde{L} \;:=\; \mathrm{Span}(\tilde{x}_\mu,\tilde{w}_\mu)
\end{equation}
are Lagrangian subspaces of the symplectic space $(\mathcal{P},\bbomega)$. This means that we will have two polarisations corresponding to the two T-duality frames $(x^\mu,w^\mu)$ and $(\tilde{x}_\mu,\tilde{w}_\mu)$.
We can thus define two basis $\{\Ket{{x},{w}}\}_{({x},{w})\in{L}}$ and $\{\Ket{\tilde{x},\tilde{w}}\}_{(\tilde{x},\tilde{w})\in\widetilde{L}}$ for our Hilbert space $\mathbf{H}$.
If we consider a generic state $\Ket{\psi}\in\mathbf{H}$ of our Hilbert space, we can now express it in the basis associated to both the T-duality frames by
\begin{equation}
    \psi_w(x) \;:=\; \Braket{x,w|\psi}, \qquad \widetilde{\psi}_{\tilde{w}}(\tilde{x}) \;:=\; \Braket{\tilde{x},\tilde{w}|\psi}.
\end{equation}
Now, we want to explicitly find the isomorphism $\mathrm{L}^{\!2}(L;\mathbb{C})\cong\mathrm{L}^2(\widetilde{L};\mathbb{C})$ between wave-functions in the two T-duality frames. Let us expand our zero-mode truncated string by $X^\mu(\sigma)=x^\mu + \alpha'\sigma \tilde{p}^\mu$ and $\widetilde{X}_\mu(\sigma)=\tilde{x}_\mu +\alpha'\sigma p_\mu$ and call $\mathbb{X}^M(\sigma)=\big(X^\mu(\sigma),\,\widetilde{X}_\mu(\sigma)\big)$. A zero-mode truncated string $\mathbb{X}^M(\sigma)=\mathbbvar{x}^M+\alpha'\sigma\mathbbvar{p}^M$ is represented by a point $(\mathbbvar{x}^M,\mathbbvar{p}_M)\in\mathcal{P}$ of the phase space of the zero-mode doubled string. 

\noindent Now, we notice notice that the symplectic form $\bbomega$ immediately satisfies the following equation of the form \eqref{eq:genfunandpol}:
\begin{equation}
\begin{aligned}
    \bbomega \;=\;  \di_L\di_{\widetilde{L}}\Big(\tilde{p}^\mu\tilde{x}_\mu - p_\mu x^\mu + \pi\alpha'p_\mu\tilde{p}^\mu\Big),
    \end{aligned}
\end{equation}
where $\di_L$ and $\di_{\widetilde{L}}$ are respectively the differentials on the Lagrangian subspaces $L = \mathrm{Span}(x^\mu,\tilde{p}^\mu)$ and $\widetilde{L} = \mathrm{Span}(\tilde{x}_\mu,p_\mu)$.
Thus, we can express the symplectomorphism $f:\mathcal{P}\rightarrow\mathcal{P}$ encoding T-duality on the phase space of the zero-mode doubled string by the generating function
\begin{equation}\label{eq:genfundou}
\begin{aligned}
    F(\mathbbvar{x},\mathbbvar{p})\;&=\; \tilde{p}^\mu\tilde{x}_\mu - p_\mu x^\mu + \pi\alpha'p_\mu\tilde{p}^\mu \\
    \;&=\; \omega_{MN} \mathbbvar{p}^M\mathbbvar{x}^N + \frac{\pi\alpha'}{2}\eta_{MN}\mathbbvar{p}^M\mathbbvar{p}^N,
    \end{aligned}
\end{equation}
which is nothing but the zero-mode truncation of the lift to the phase space of the doubled string of the action functional \eqref{eq:genfunF}.
Such a symplectomorphism is simply the $O(n,n)$ transformation of the doubled coordinates and momenta by $(\mathbbvar{x}^M,\mathbbvar{p}^M) \mapsto (\eta_{MN}\mathbbvar{x}^N,\,\eta_{MN}\mathbbvar{p}^N)$.
Now, by applying the machinery of geometric quantisation, the matrix of the change of basis on the Hilbert space $\mathbf{H}$ will be given by the generating function \eqref{eq:genfundou} as it follows:
\begin{equation}
\begin{aligned}
    \Braket{x,w|\tilde{x},\tilde{w}} \;&=\; \exp\frac{i}{\hbar}\!\left(p_\mu x^\mu - \tilde{p}^\mu\tilde{x}_\mu + \pi\alpha'p_\mu\tilde{p}^\mu\right) \\[0.5ex]
    \;&=\; \exp\frac{i}{\hbar}\!\left(\omega_{MN} \mathbbvar{p}^M\mathbbvar{x}^N + \frac{\pi\alpha'}{2}\eta_{MN}\mathbbvar{p}^M\mathbbvar{p}^N\right),
    \end{aligned}
\end{equation}
where $w^\mu \equiv \tilde{p}^\mu$ and $\tilde{w}_\mu \equiv p_\mu$. 
Therefore we can equivalently rewrite the transformation
\begin{equation}
    \Braket{\tilde{x},\tilde{w}|\psi} \;=\; \int_L\di^n x\, \di^n w \Braket{\tilde{x},\tilde{w}|x,w}\Braket{x,w|\psi}
\end{equation}
as the following, {\bf{{{stringy}} Fourier transformation}}:
\begin{equation}
    \widetilde{\psi}_{\tilde{w}}(\tilde{x}) \;=\; \int_L \di^n x\, \di^n w \exp\frac{i}{\hbar}\!\left(\omega_{MN} \mathbbvar{p}^M\mathbbvar{x}^N  + \frac{\pi\alpha'}{2}\eta_{MN}\mathbbvar{p}^M\mathbbvar{p}^N\right)\psi_w(x)  \, .
\end{equation}
This is the transformation between the wavefunctions in different duality frames. Mathematically it is the isomorphism $\mathrm{L}^{\!2}(L,\mathbb{C})\cong\mathrm{L}^2(\widetilde{L},\mathbb{C})$. In undoubled coordinates we can explicitly rewrite such a stringy Fourier transformation as it follows: 
\begin{equation}
    \widetilde{\psi}_{\tilde{w}}(\tilde{x}) \;=\; \int_L \di^n x\, \di^n w \exp\frac{i}{\hbar}\!\left(\tilde{w}_\mu x^\mu - w^\mu\tilde{x}_\mu + \pi\alpha'\tilde{w}_\mu w^\mu\right)\psi_w(x),
\end{equation}
where we used the identities $w^\mu\equiv \tilde{p}^\mu$ and $\tilde{w}_\mu\equiv p_\mu$. Notice that, even if the form of the symplectomorphism $f$ is particularly simple, the transformation for wave-function is more complicated than just a Fourier transform. The difference from the usual Fourier transformation is given by the additional $\frac{\pi\alpha'}{2}\eta_{MN}\mathbbvar{p}^M\mathbbvar{p}^N$ term. This then will reduces to a standard Fourier transform in the limit $\alpha'\rightarrow 0$.

In terms of basis, this transformation can be also be expressed by
\begin{equation}
    \Ket{\tilde{x},\tilde{w}} \;=\; \int_L \di^n x\, \di^n w \exp\frac{i}{\hbar}\!\left(\omega_{MN} \mathbbvar{p}^M\mathbbvar{x}^N  + \frac{\pi\alpha'}{2}\eta_{MN}\mathbbvar{p}^M\mathbbvar{p}^N\right) \Ket{x,w}.
\end{equation}

\paragraph{A phase term in the change of polarisation.} 
Finally, notice that, if we restrict our generalised winding to ordinary integer winding $w,\tilde{w}\in\mathbb{Z}^{n}$, we will obtain a change of polarisation of the form
\begin{equation}
    \widetilde{\psi}_{\tilde{w}}(\tilde{x}) \,=\, \sum_{w\in\mathbb{Z}^n}e^{\frac{i}{\hbar}\pi\alpha'\tilde{w}_\mu w^\mu} \! \int_M \di^n x\, e^{\frac{i}{\hbar}\!\left(\tilde{w}_\mu x^\mu - w^\mu\tilde{x}_\mu\right)}\psi_w(x).
\end{equation}
In this context, as firstly noticed with different arguments by \cite{Fre17b}, T-duality does not simply act as a "double" Fourier transformation of the wave-function of a string, because there will be an extra phase contribution given by $\exp\!\big(i\pi{\frac{\alpha'}{\hbar}\tilde{w}_\mu w^\mu}\big)$ for any term with $w,\tilde{w}\neq 0$.
Since we are restricting now to the case where $w,\tilde{w}$ are integers and $\sqrt{\hbar/\alpha'}$ is just the unit of momentum, we immediately conclude that the only possible phase contributions are $\exp\!\big(i\pi{\frac{\alpha'}{\hbar}\tilde{w}_\mu w^\mu}\big)\in \{+1,-1\}$, depending on the product $\tilde{w}_\mu w^\mu\equiv p_\mu w^\mu$ being even or odd.
Notice that the presence of the topological term in the action induces a very similar phase term in the partition function of a string with an analogous role, as seen by \cite{Ber07}.

\paragraph{Darboux coordinates for the zero-mode string.}
Let us find the Darboux coordinates on the manifold $\mathcal{P}$ for the symplectic form $\bbomega\in\Omega^2(\mathcal{P})$. If we define the new coordinates
\begin{equation}
    \begin{aligned}
        q^\mu \;&:=\; x^\mu \\
        \tilde{q}_\mu \;&:=\; \tilde{x}_\mu - \pi\alpha'p_\mu
    \end{aligned}
\end{equation}
and we pack them together as $\mathbbvar{q}^M:=(q^\mu,\tilde{q}_\mu)$, we can rewrite the symplectic form simply as
\begin{equation}
    \bbomega \;=\; \di\mathbbvar{p}_M\wedge\di\mathbbvar{q}^M
\end{equation}
with $\mathbbvar{p}_M=\eta_{MN}\mathbbvar{p}^N=(p_\mu,\tilde{p}^\mu)$. Therefore the conjugate variable on the phase to the canonical momenta $\mathbbvar{p}_M$ of the zero-mode string is the new coordinate $\mathbbvar{q}^M$. Notice that this variable is not the proper position $\mathbbvar{x}^M$ on the doubled space, but a mix of position and momentum. This change of coordinates is intimately related to what is known as Bopp's shift in non-commutative quantum mechanics.

\subsection{Relation with non-commutativity induced by fluxes}

The non-commutativity we are exploring follows that in \cite{Bla14}, but is different from (though close to) the one introduced by \cite{Lus10} and further explored by \cite{ALLP13}, where the non-commutativity of the doubled space is induced by the presence of fluxes. For a more recent account see \cite{Sza18} and \cite{Ost19}. The notion of non-commutativity we are considering is completely independent by the presence of fluxes and characterises even flat and topologically trivial doubled spaces. As we will see in the next section, the non-commutativity between a physical coordinate and its T-dual is intrinsic and linked to the existence of a minimal length $\ell_s =\sqrt{\hbar\alpha'}$ on the doubled space.

The link between the two notions of non-commutativity is provided by \cite{Bla14}.
The presence of flux implies monodromies for the generalised metric of the form
\begin{equation}
    \mathcal{H}(x + 2\pi) \;=\; \mathcal{O}\mathcal{H}(x)\mathcal{O}^\mathrm{T}
\end{equation}
with monodromy matrix $\mathcal{O}\in O(n,n)$. Thus, as explained by \cite{Bla14}, we need to consider the further generalised boundary conditions
\begin{equation}
    \mathbb{X}^M(\sigma+2\pi) \;=\; \mathcal{O}^M_{\;\;\;N}  \mathbb{X}^N(\sigma) + 2\pi\mathbbvar{p}^M
\end{equation}
for our doubled string $\sigma$-model. When we write the Tseytlin action, we then have to generalise its boundary term accordingly. As seen by \cite{Bla14}, the new action produces the non-commutativity given by the fluxes on the doubled phase space.

\section{Non-commutative QM of the zero-mode string}

Non-commutative quantum mechanics was introduced as far as in 1947 by \cite{Sny47}. The fundamental idea, at the time, was to quantize flat spacetime by introducing a minimal length and generalising the uncertainty principle to make it fuzzy.

As we saw in the previous section, the zero-mode truncation of the pre-quantised wave-functional $\Psi[\mathbb{X}(\sigma)]$ of a doubled string can be seen as a conventional pre-quantised wave-function $\psi(\mathbbvar{x},\mathbbvar{p})$ of a particle in a doubled space. In other words, the zero-modes of strings behave like particles in a double space.
However, such a doubled space is intrinsically non-commutative. As we derived, indeed, the non-commutative Heisenberg algebra \label{eq:canonicalhesialgebra} includes commutation relations of the position operators of the form
\begin{equation}
    [\hat{\mathbbvar{x}}^M,\hat{\mathbbvar{x}}^N] \,=\, i\vartheta^{MN}  \quad\text{ with }\quad \vartheta^{MN} := \pi\hbar\alpha'\omega^{MN} .
\end{equation}
This means that the Quantum Mechanics of the string zero-modes will be non-commutative (NCQM). Let us choose units where we only require $c=1$. This way the two physical dimensions of length and energy are explicitly parametrised by the two universal constants as it follows:
\begin{equation}
    \left[\hbar\alpha'\right] \,=\, \mathrm{length}^2, \qquad \left[\frac{\hbar}{\alpha'}\right] \,=\, \mathrm{energy}^2
\end{equation}
and thus the string scale must be expressed as $\ell_s=\sqrt{\hbar\alpha'}$. We notice that any couple of T-dual coordinates fails in commuting by an area which is proportional to the string scale, i.e. we have
\begin{equation}
    \big[\hat{x}^\mu,\,\hat{\tilde{x}}_\mu\big] \,=\, i\pi\ell_s^2
\end{equation}
for any fixed $\mu=1,\dots,n$.
In this context $\pi\ell^2_s$ can be interpreted as a minimal area of the doubled space.


\subsection{Non-commutative coherent states of zero-mode strings}

Let us start from the non-commutative Heisenberg algebra $\mathfrak{heis}(\bbomega,\mathcal{P})$ of the phase space $(\bbomega,\mathcal{P})$ of the zero-modes truncated doubled string:
\begin{equation}\label{eq:heisenbergalgebra}
    [\hat{\mathbbvar{x}}^M,\hat{\mathbbvar{x}}^N] \,=\, \pi i\hbar\alpha' \omega^{MN} , \quad [\hat{\mathbbvar{x}}^M,\hat{\mathbbvar{p}}_N] \,=\, i\hbar\delta^{M}_{\;\, N}, \quad [\hat{\mathbbvar{p}}_M,\hat{\mathbbvar{p}}_N] \,=\,0.
\end{equation}
Notice that the subspace of $\mathcal{P}=\mathbb{R}^{4n}$ spanned by $(x^\mu,\tilde{x}_\mu)$ is not a Lagrangian subspace, therefore there is no well-defined notion of wave-function of the form $\psi(x^\mu,\tilde{x}_\mu)$. In quantum mechanical terms, since T-dual coordinates $[\hat{x}^\mu,\hat{\tilde{x}}_\nu]\neq 0$ do not generally commute, there exists no basis $\big\{\Ket{x^\mu,\tilde{x}_\mu}\big\}\not\subset \mathbf{H}$ of eigenstates of the doubled position operators.
However, we can overcome this obstacle by defining a basis of coherent states on the doubled space.

\paragraph{Coherent states.}
We can define the following annihilation and creation operators
\begin{equation}
    \begin{aligned}
    \hat{\mathbbvar{z}}^\mu \;&=\; \frac{1}{\sqrt{2\pi\hbar}}\left(\hat{x}^\mu + i \hat{\tilde{x}}_\mu\right), \\
    \hat{\mathbbvar{z}}^{\dagger\mu} \;&=\; \frac{1}{\sqrt{2\pi\hbar}}\left(\hat{x}^\mu - i \hat{\tilde{x}}_\mu\right).
    \end{aligned}
\end{equation}
By using the commutation relations \eqref{eq:heisenbergalgebra}, we immediately find that the commutator
\begin{equation}
    \big[\hat{\mathbbvar{z}}^\mu,\,\hat{\mathbbvar{z}}^{\dagger\nu}\big] \;=\; \alpha'\delta^{\mu\nu}
\end{equation}
satisfies the commutation relations of the Fock algebra.
Thus the non-commutative quantum configuration space is a bosonic Fock space
\begin{equation}
    \mathbf{F}_{\mathrm{cs}} \;:=\; \bigodot_{k\in\mathbb{N}} \mathbb{C}^n \;=\;  \mathbb{C} \oplus \mathbb{C}^n \oplus (\mathbb{C}^n\odot\mathbb{C}^n) \oplus (\mathbb{C}^n\odot\mathbb{C}^n\odot\mathbb{C}^n) \oplus \dots
\end{equation}
generated by vectors of the form
\begin{equation}
    \Ket{0}, \quad\hat{\mathbbvar{z}}^{\dagger\mu}\Ket{0}, \quad\frac{1}{\sqrt{2}} \hat{\mathbbvar{z}}^{\dagger\mu_1}\hat{\mathbbvar{z}}^{\dagger\mu_2}\Ket{0}, \quad \frac{1}{\sqrt{3!}} \hat{\mathbbvar{z}}^{\dagger\mu_1}\hat{\mathbbvar{z}}^{\dagger\mu_2}\hat{\mathbbvar{z}}^{\dagger\mu_3}\Ket{0}, \quad \dots
\end{equation}
where the vacuum state $\Ket{0}$ is defined by the equation $\hat{\mathbbvar{z}}^\mu\Ket{0}=0$ for all $\mu=1,\dots,n$.

The important aspect of working with the creation and annihilation operators is that there exist eigenstates $\Ket{\mathbbvar{z}^1,\cdots,\mathbbvar{z}^n}$ for all the operators $\hat{\mathbbvar{z}}^\mu$ with $\mu=1,\dots,n$. These satisfy the following defining properties:
\begin{equation}
    \begin{aligned}
    \hat{\mathbbvar{z}}^\mu\Ket{\mathbbvar{z}^1,\cdots,\mathbbvar{z}^n} \;&=\; \mathbbvar{z}^\mu\Ket{\mathbbvar{z}^1,\cdots,\mathbbvar{z}^n} \\[0.7ex]
   \Bra{\mathbbvar{z}^1,\cdots,\mathbbvar{z}^d} \hat{\mathbbvar{z}}^{\dagger\mu} \;&=\; \Bra{\mathbbvar{z}^1,\cdots,\mathbbvar{z}^d} \bar{\mathbbvar{z}}^{\mu}
    \end{aligned}
\end{equation}
for eigenvalues $(\mathbbvar{z}^1, \cdots, \mathbbvar{z}^n)\in\mathbb{C}^n$. These states are called \textit{coherent states}.

Let us now use the compact notation $\Ket{\mathbbvar{z}} := \Ket{\mathbbvar{z}^1,\cdots,\mathbbvar{z}^n}$ for coherent states. A normalised coherent state can be expressed by
\begin{equation}
    \Ket{\mathbbvar{z}} \;:=\; \frac{1}{(2\pi\alpha')^{\frac{n}{2}}}\exp\!\left(-\frac{\delta_{\mu\nu}}{2\alpha'}\,\mathbbvar{z}^\mu\,\bar{\mathbbvar{z}}^\nu\right) \exp\!\left(\frac{\delta_{\mu\nu}}{\alpha'}\,\mathbbvar{z}^\mu\,\hat{\mathbbvar{z}}^{\dagger\nu}\right) \Ket{0},
\end{equation}
where the vacuum state $\Ket{0}\in\mathbf{F}_{\mathrm{cs}}$ is defined as previously.
These states constitute a complete basis on the Fock space space $\mathbf{F}_{\mathrm{cs}}$, since they satisfy the property
\begin{equation}
   \int_{\mathbb{C}^n}\di^n\mathbbvar{z}\,\di^n\bar{\mathbbvar{z}} \,\Ket{\mathbbvar{z}}\Bra{\mathbbvar{z}} \;=\; 1
\end{equation}
There is an isomorphism between the non-commutative quantum configuration space $\mathbf{F}_{\mathrm{cs}}$ and the quantum Hilbert space $\mathbf{H}$, i.e.
\begin{equation}
    \mathbf{F}_{\mathrm{cs}} \,\cong\, \mathbf{H}.
\end{equation}
Such an isomorphism will be explicitly presented in equation \eqref{eq:NCFourier}.

\paragraph{Mean position of a coherent state.}
The expectation value of the non-commutative position operators on a coherent state $\Ket{\mathbbvar{z}}$ can be found by
\begin{equation}
    \braket{\mathbbvar{z}|\hat{x}^\mu|\mathbbvar{z}} \,=\, \sqrt{2\pi\hbar}\;\mathfrak{Re}(\mathbbvar{z}^\mu) \,=:\, x^\mu, \qquad \braket{\mathbbvar{z}|\hat{\tilde{x}}_\mu|\mathbbvar{z}} \,=\, \sqrt{2\pi\hbar}\;\mathfrak{Im}(\mathbbvar{z}^\mu) \,=:\, \tilde{x}_\mu
\end{equation}
The doubled vector $\mathbbvar{x}^M=(x^\mu,\,\tilde{x}_\mu)$ is the \textit{mean position} of the coherent state $\ket{\mathbbvar{z}}$ on the doubled space $\mathbb{R}^{2d}$, also known as quasi-coordinate vector. It is important to remark that $\mathbbvar{x}^M$ are not coordinates, i.e. they are not eigenvalues of the operators $\hat{\mathbbvar{x}}^M$. Thus, by working with coherent states $\ket{\mathbbvar{z}}$ with mean position $\mathbbvar{x}^M$, we can bypass the problem of not being able to work with eigenstates of the position operators. In general, any operator $\hat{f}(\hat{\mathbbvar{x}}^M)$ can be expressed as a function of the mean positions of a coherent state by $F({\mathbbvar{x}}^M) := \braket{\mathbbvar{z}|\hat{f}(\hat{\mathbbvar{x}}^M)|\mathbbvar{z}}$.

\paragraph{Minimal uncertainty.} Coherent states minimize the uncertainty between a coordinate operator of the doubled space and its T-dual, i.e.
\begin{equation}
    \Delta x^\mu\, \Delta\tilde{x}_\nu=\frac{\pi\ell^2_s}{2}\delta^\mu_{\;\nu}
\end{equation}
The coherent states of the quantum configuration space can then be interpreted as states which are approximately localised at a point $\mathbbvar{x}^M$ of the doubled space, the mean position. 

\subsection{Free particles on the doubled space}

\paragraph{Plane waves.}
The mean value of a plane wave operator on a coherent state is given by
\begin{equation}
    \Braket{\mathbbvar{z}|\exp\!\left(\frac{i}{\hbar}\mathbbvar{p}_M\hat{\mathbbvar{x}}^M\right)|\mathbbvar{z}} \;=\; \exp\!\left(\frac{i}{\hbar}\mathbbvar{p}_M\mathbbvar{x}^M-\frac{\pi\alpha'}{4\hbar}\delta^{MN}\mathbbvar{p}_M\mathbbvar{p}_N \right)
\end{equation}
This can be immediately proved by defining the complex momentum operators
\begin{equation}
    \begin{aligned}
    \hat{\mathbbvar{p}}_\mu^+ \;&=\; \sqrt{\frac{\pi}{{2\hbar}}}\left(\hat{p}_\mu + i \hat{\tilde{p}}^\mu\right) \\
    \hat{\mathbbvar{p}}^{-}_\mu \;&=\; \sqrt{\frac{\pi}{{2\hbar}}}\left(\hat{p}_\mu - i \hat{\tilde{p}}^\mu\right) 
    \end{aligned}
\end{equation}
and by applying the Baker–Campbell–Hausdorff formula as it follows:
\begin{equation}
    \begin{aligned}
    \exp\!\left(\frac{i}{\hbar}\mathbbvar{p}_M\hat{\mathbbvar{x}}^M\right) \;&=\; \exp\!\left(i\mathbbvar{p}_\mu^+\hat{\mathbbvar{z}}^{\dagger\mu} + i\mathbbvar{p}_\mu^-\hat{\mathbbvar{z}}^{\mu} \right) \\
    \;&=\; \exp\!\left(i\mathbbvar{p}_\mu^+\hat{\mathbbvar{z}}^{\dagger\mu} \right) \exp\!\left(i\mathbbvar{p}_\mu^-\hat{\mathbbvar{z}}^{\mu} \right) \exp\!\left(-\frac{\pi\alpha'}{4\hbar}\delta^{MN}\mathbbvar{p}_M\mathbbvar{p}_N \right).
    \end{aligned}
\end{equation}

\paragraph{The Hilbert space of a free particle.}
The subspace $L_\mathbbvar{p}=\mathrm{Span}(\mathbbvar{p}^M)\subset\mathcal{P}$ is Lagrangian. This can be immediately understood by writing the symplectic form in Darboux coordinates as $\bbomega = \di\mathbbvar{p}_M\wedge \di \mathbbvar{q}^M$. Therefore we can express our Hilbert space $\mathbf{H}$ as the space of complex $\mathrm{L}^2$-functions on $L_\mathbbvar{p}$, i.e. as
\begin{equation}
    \mathbf{H} \;\cong\; \mathrm{L}^{\!2}(L_{\mathbbvar{p}};\mathbb{C})
\end{equation}
Since the doubled momenta commute, i.e. $[\hat{\mathbbvar{p}}_M,\hat{\mathbbvar{p}}_N] \,=\,0$, we can define a basis of eigenstates $\{\Ket{\mathbbvar{p}}\}_{\mathbbvar{p}\in L_\mathbbvar{p}}\subset \mathbf{H}$ of the doubled momentum operators $\hat{\mathbbvar{p}}_M$ without the problems encountered with the doubled position operator. These satisfy $\hat{\mathbbvar{p}}_M\Ket{\mathbbvar{p}} = \mathbbvar{p}_M\Ket{\mathbbvar{p}}$ for any $M=1,\dots,2d$. If we choose the basis of coherent states $\left\{\Ket{\mathbbvar{z}}\right\}_{\mathbbvar{z}\in\mathbb{C}^n}$, a doubled momentum eigenstate can then be expressed by
\begin{equation}\label{eq:NCfreeparticle}
    \Braket{\mathbbvar{z}|\mathbbvar{p}} \;=\; \frac{1}{(2\pi\hbar)^{n}} \exp\!\left(\frac{i}{\hbar}\mathbbvar{p}_M\mathbbvar{x}^M-\frac{\pi\alpha'}{4\hbar}\delta^{MN}\mathbbvar{p}_M\mathbbvar{p}_N \right)
\end{equation}
where $\mathbbvar{x}^M$ is the mean position of the coherent state $\Ket{\mathbbvar{z}}$. This can be interpreted as the expression of a free particle state on the doubled space, where we are using the mean position as a variable.

\paragraph{Strings are waves.}
Interestingly, if we choose the basis $\Ket{x,w}$ with $w^\mu=\tilde{p}^\mu$, which diagonalizes the commuting operators of the physical position $\hat{x}^\mu$ and winding $\hat{\tilde{p}}^\mu$, a doubled momentum eigenstate can be expressed just as a free particle in the wave-function on the physical space
\begin{equation}\label{eq:planewave}
    \Braket{x,w|\mathbbvar{p}} \;=\; \frac{1}{(2\pi\hbar)^{\frac{n}{2}}}\exp\!\left(\frac{i}{\hbar} p_\mu x^\mu \right)
\end{equation}
where now $x^\mu$ are proper eigenvalues of the position operator.
Analogously, in the T-dual frame $\Ket{\tilde{x},\tilde{w}}$ with $\tilde{w}_\mu=p_\mu$, we recover a free particle on the T-dual space
\begin{equation}
    \Braket{\tilde{x},\tilde{w}|\mathbbvar{p}} \;=\; \frac{1}{(2\pi\hbar)^{\frac{n}{2}}}\exp\!\left(\frac{i}{\hbar} \tilde{p}^\mu \tilde{x}_\mu \right)
\end{equation}
The interpretation of this fact is that a free particle, i.e. a plane wave, on the doubled space with fixed doubled momentum $\mathbbvar{p}_M = (p_\mu,\,\tilde{p}^\mu)$ can be quantum-mechanically interpreted as 
\begin{itemize}
    \item a free string on the physical space with fixed momentum $p_\mu$ and winding $w^\mu = \tilde{p}^\mu$,
    \item a free string on the T-dual space with fixed momentum $\tilde{p}^\mu$ and winding $\tilde{w}_\mu = p_\mu$.
\end{itemize}
As classically derived in \cite{BBR14}, this implies that a plane wave with doubled momentum $\mathbbvar{p}_M = (p_\mu,\,0)$ on the doubled space reduces to a plane wave with momentum $p_\mu$ on the physical space and one with $\mathbbvar{p}_M = (0,\,\tilde{p}^\mu)$ reduces to a standing string with winding $w^\mu=\tilde{p}^\mu$.

\paragraph{Probability distribution.}
Let us calculate the probability distribution of the wave-function \eqref{eq:NCfreeparticle} of a free particle on the doubled space:
\begin{equation}
    \left|\Braket{\mathbbvar{z}|\mathbbvar{p}}\right|^2 \;=\; \frac{1}{(2\pi\hbar)^{2n}}\exp\!\left(-\frac{\pi\alpha'}{2\hbar}\delta^{MN}\mathbbvar{p}_M\mathbbvar{p}_N\right).
\end{equation}
Hence, the probability of measuring the doubled momentum $\mathbbvar{p}_M$ (or equivalently a string with momentum $p_\mu$ and winding $\tilde{p}^\mu$) is not uniform, but it exponentially decays far from zero.

\paragraph{Coherent state as superposition of strings.}
Now, the eigenstate $\Ket{\mathbbvar{p}}$ can be interpreted as a free string for which we know with certainty the momentum $p_\mu$ and the winding number $w^\mu=\tilde{p}^\mu$. The equation \eqref{eq:NCfreeparticle} can be immediately interpreted as the expansion of a coherent state $\Ket{\mathbbvar{z}}$ in the basis $\Ket{\mathbbvar{p}}$, i.e.
\begin{equation}\label{eq:NCFourier}
    \Ket{\mathbbvar{z}} \;=\; \int\frac{\di^{2n}\mathbbvar{p}}{(2\pi\hbar)^{n}}\exp\!\left(\frac{i}{\hbar}\mathbbvar{p}_M\mathbbvar{x}^M-\frac{\pi\alpha'}{4\hbar}\delta^{MN}\mathbbvar{p}_M\mathbbvar{p}_N \right) \Ket{\mathbbvar{p}}
\end{equation}
where $\mathbbvar{x}^M$ is the mean doubled position of the coherent state $\Ket{\mathbbvar{z}}$.

\paragraph{Hamiltonian as number operator.}
Observe that the Hamiltonian operator is given by
\begin{equation}
    \hat{H} \;=\; \mathcal{H}^{MN}\hat{\mathbbvar{p}}_M\hat{\mathbbvar{p}}_N
\end{equation}
Let us consider the simple case where the generalised metric is trivial, i.e. $\mathcal{H}^{MN}=\delta^{MN}$.
We can then define a number operator $\hat{N}_\mu \,:=\, \hat{\mathbbvar{p}}_\mu^-\hat{\mathbbvar{p}}_\mu^+$ for any fixed $\mu=1,\dots,n$ and the total number operator as a sum $\hat{N}= \sum_{\mu=1}^{n}\hat{N}_\mu$. What we obtain is that the Hamiltonian is proportional to the number operator by $\hat{H}=\frac{2}{\pi}\hbar\hat{N}$.

\subsection{Minimal scale of the doubled space}

\paragraph{Non-commutative Fourier transform.}
Let us consider a general string state $\Ket{\psi}\in\mathbf{H}$. We can express this state as a wave function $\psi(\mathbbvar{p})=\Braket{\mathbbvar{p}|\psi}$ on the momentum space. Thus, if we want to express it in the coherent states basis $\Ket{\mathbbvar{z}}$, we need to use equation \eqref{eq:NCFourier} as it follows:
\begin{equation}
    \Braket{\mathbbvar{z}|\psi} \;=\; \int\frac{\di^{2n}\mathbbvar{p}}{(2\pi\hbar)^{n}}\exp\!\left(\frac{i}{\hbar}\mathbbvar{p}_M\mathbbvar{x}^M-\frac{\pi\alpha'}{4\hbar}\delta^{MN}\mathbbvar{p}_M\mathbbvar{p}_N \right) \Braket{\mathbbvar{p}|\psi}.
\end{equation}
Now we can transform wave-functions $\psi(\mathbbvar{p}):=\Braket{\mathbbvar{p}|\psi}$ on the doubled momentum space to wavefunctions $\psi(\mathbbvar{x}):=\Braket{\mathbbvar{z}|\psi}$ expressed in the  basis of the coherent states. (Here $\mathbbvar{x}^M$ denotes the mean position of $\Ket{\mathbbvar{z}}$ and is not a coordinate.) This is effectively a non-commutative version of the Fourier transform. 

Let us now mention an example of this non-commutative Fourier transform which is useful to develop some intuition about the non-commutative nature of the doubled space. Let us choose a wave-function $\psi(\mathbbvar{p})=1/(2\pi\hbar)^n$ on the doubled momentum space, which, in some sense, means that the doubled momentum is maximally spread. The transformation \eqref{eq:NCFourier}, applied to $\psi(\mathbbvar{p})=1/(2\pi\hbar)^n$, gives
\begin{equation}
\begin{aligned}
    \psi(\mathbbvar{x}) \;&=\; \int\frac{\di^{2n}\mathbbvar{p}}{(2\pi\hbar)^{2n}}\exp\!\left(\frac{i}{\hbar}\mathbbvar{p}_M\mathbbvar{x}^M-\frac{\pi\alpha'}{4\hbar}\delta^{MN}\mathbbvar{p}_M\mathbbvar{p}_N \right) \\[0.1cm]
    \;&=\; \frac{1}{(\pi^2\hbar\alpha')^n}\exp\Bigg(-\frac{\big|\mathbbvar{x}^M\big|^2}{\pi\hbar\alpha'}\Bigg),
    \end{aligned}
\end{equation}
which is a Gaussian distribution on the doubled space and not a delta function. This, on an intuitive level, means that, even if the doubled momentum is maximally spread, the uncertainty on the doubled coordinates cannot be zero. This is because each couple of T-dual coordinates can shrink only to a minimal area proportional to $\ell^2_s=\hbar\alpha'$. This is an interesting manifestation of the fuzziness of doubled space between physical and T-dual coordinates, which is parametrised by $\alpha'$.

\paragraph{Amplitude between coherent states.}
Let us consider two coherent states $\Ket{\mathbbvar{z}_1}$ and $\Ket{\mathbbvar{z}_2}$, respectively with mean position $\mathbbvar{x}_1^M$ and $\mathbbvar{x}_2^M$. We want now to calculate the scattering amplitude $\Braket{\mathbbvar{z}_2|\mathbbvar{z}_1}$ between such states\footnote{We thank Kevin T. Grosvenor for extremely helpful discussion, which led to the improvement of this section. In particular, the calculation of the amplitude between coherent states will follow the proposal \cite{Kev21}.}.
First, notice that the following identity holds:
\begin{equation}
    \exp\Bigg(-\frac{\delta_{\mu\nu}}{\alpha'}\mathbbvar{z}^\mu\hat{\mathbbvar{z}}^{\dagger\nu}\Bigg) \hat{\mathbbvar{z}}^\lambda \exp\Bigg(+\frac{\delta_{\mu\nu}}{\alpha'}\mathbbvar{z}^\mu\hat{\mathbbvar{z}}^{\dagger\nu}\Bigg) \;=\; \hat{\mathbbvar{z}}^\lambda + \mathbbvar{z}^\lambda.
\end{equation}
By combining this identity with the definition of coherent state, we get the equation
\begin{equation*}
    \Braket{\mathbbvar{z}_2|\mathbbvar{z}_1} \;=\; \frac{1}{(2\pi\alpha')^{n}} \exp\Bigg(-\frac{\big|\mathbbvar{z}_2|^2+|\mathbbvar{z}_1\big|^2}{2\alpha'}\Bigg) \Braket{0| \exp\Bigg(\frac{\delta_{\mu\nu}}{\alpha'}\bar{\mathbbvar{z}}^\mu_2\hat{\mathbbvar{z}}^{\nu}_2\Bigg) \exp\Bigg(\frac{\delta_{\mu\nu}}{\alpha'}\mathbbvar{z}^\mu_1\hat{\mathbbvar{z}}^{\dagger\nu}_1\Bigg) |0}.
\end{equation*}
By using the Baker–Campbell–Hausdorff formula, we have
\begin{equation*}
    \exp\Bigg(\frac{\delta_{\mu\nu}}{\alpha'}\bar{\mathbbvar{z}}^\mu_2\hat{\mathbbvar{z}}^{\nu}_2\Bigg) \exp\Bigg(\frac{\delta_{\mu\nu}}{\alpha'}\mathbbvar{z}^\mu_1\hat{\mathbbvar{z}}^{\dagger\nu}_1\Bigg) \;=\; \exp\Bigg(\frac{\delta_{\mu\nu}}{\alpha'}\mathbbvar{z}^\mu_1\hat{\mathbbvar{z}}^{\dagger\nu}_1\Bigg) \exp\Bigg(\frac{\delta_{\mu\nu}}{\alpha'}\bar{\mathbbvar{z}}^\mu_2\hat{\mathbbvar{z}}^{\nu}_2\Bigg) \exp\Bigg(\frac{\delta_{\mu\nu}}{\alpha'}\bar{\mathbbvar{z}}^\mu_2\mathbbvar{z}^{\nu}_1\Bigg).
\end{equation*}
From this, we finally find the amplitude:
\begin{equation}
    \Braket{\mathbbvar{z}_2|\mathbbvar{z}_1} \;=\; \frac{1}{(2\pi\alpha')^{n}} \exp\Bigg(-\frac{\big|\mathbbvar{z}_2-\mathbbvar{z}_1\big|^2 + \delta_{\mu\nu}(\mathbbvar{z}_2^\mu\bar{\mathbbvar{z}}_1^\nu - \bar{\mathbbvar{z}}_2^\mu\mathbbvar{z}_1^\nu)}{2\alpha'}\Bigg).
\end{equation}
Thus, we immediately get the result that the amplitude between two coherent states with different mean positions $\mathbbvar{x}_1^M$ and $\mathbbvar{x}_2^M$ is given by
\begin{equation}\label{eq:amplitudez}
    \Braket{\mathbbvar{z}_2|\mathbbvar{z}_1} \;=\; \frac{1}{(2\pi\alpha')^{n}} \exp\Bigg(-\frac{\big|\mathbbvar{x}_2-\mathbbvar{x}_1\big|^2 - i\omega_{MN}\mathbbvar{x}_2^M\mathbbvar{x}_1^N}{\pi\hbar\alpha'}\Bigg).
\end{equation}
Thus, the ordinary Dirac delta function $\braket{x_2|x_1}=\delta^{(n)}(x_2-x_1)$ between eigenstates of the coordinates operator of commutative Quantum Mechanics is replaced by the function \eqref{eq:amplitudez}, whose width is proportional to string length scale $\ell_s=\sqrt{\hbar\alpha'}$.
The squared amplitude between two coherent states is, then, the Gaussian distribution
\begin{equation}
    |\Braket{\mathbbvar{z}_2|\mathbbvar{z}_1}|^2 \;=\; \frac{1}{(2\pi\alpha')^{2n}} \exp\bigg(-\frac{2}{\pi\hbar\alpha'}\big|\mathbbvar{x}_2-\mathbbvar{x}_1\big|^2\bigg).
\end{equation}
Physically, this means that the probability is high if the distance between the mean positions $\mathbbvar{x}_1$ and $\mathbbvar{x}_2$ of the respective coherent states $\Ket{\mathbbvar{z}_1}$ and $\Ket{\mathbbvar{z}_2}$ is smaller than $\sqrt{\pi}\ell_s$.

\paragraph{The $\alpha'\rightarrow 0$ limit.}
If we take the limit $\alpha'\rightarrow 0$ the fuzziness of the doubled space disappears. 
The non-commutative Heisenberg algebra of quantum observables reduces to an ordinary commutative $4n$-dimensional Heisenberg algebra, whose commutation relations are given by 
\begin{equation}
    \lim_{\alpha'\rightarrow\,0}\,[\hat{\mathbbvar{x}}^M,\hat{\mathbbvar{x}}^N] \,=\, 0 , \;\quad \lim_{\alpha'\rightarrow\,0}\,[\hat{\mathbbvar{x}}^M,\hat{\mathbbvar{p}}_N] \,=\, i\hbar\delta^{M}_{\;\, N}, \;\quad \lim_{\alpha'\rightarrow\,0}\,[\hat{\mathbbvar{p}}_M,\hat{\mathbbvar{p}}_N] \,=\,0.
\end{equation}
Consequently the minimal uncertainty in measuring a coordinate and its dual vanishes.
The basis of coherent states $\Ket{\mathbbvar{z}}$ reduces to a basis of eigenstates $\Ket{x,\tilde{x}}$ of the position operator $\hat{\mathbbvar{x}}$, which are now well-defined.
Moreover, the scattering amplitudes shrink to
\begin{equation}
    \lim_{\alpha'\rightarrow\,0}\Braket{\mathbbvar{z}_2|\mathbbvar{z}_1} \;\propto\; \delta^{(2n)}(\mathbbvar{x}_2-\mathbbvar{x}_1).
\end{equation}
In the limit $\alpha'\rightarrow 0$, the quantum mechanics on the doubled space becomes ordinary commutative quantum mechanics on a $2n$-dimensional spacetime.

\subsection{Polarisation of coherent states in a T-duality frame}

Let us consider on our Hilbert space the basis $\{\Ket{x,w}\}_{(x,w)\in L}\subset\mathbf{H}$, which corresponds to the T-duality frame given by the Lagrangian subspace $L\subset\mathcal{P}$ with coordinates $(x^\mu,w^\mu)$. Recall that, given a string state $\Ket{\psi}\in\mathbf{H}$, we can express it as a wave-function on the Lagrangian subspace $L\subset\mathcal{P}$ by $\psi_w(x) \,:=\, \Braket{x,w|\psi}$, where $w^\mu:=\tilde{p}^\mu$ is the generalised winding number. Thus $|\psi_w(x)|^2$ can be interpreted as the probability of measuring a string at the point $x^\mu$ on physical spacetime with winding number $w^\mu$. Now, we want express a coherent state $\Ket{\mathbbvar{z}}$ in this basis. In other words we want to calculate
\begin{equation}
    \psi^{\mathrm{coh}}_w(x) \;:=\; \Braket{x,w|\mathbbvar{z}}.
\end{equation}
To do that, we can use the fact that $\Ket{\mathbbvar{p}}$ are a complete basis for the Hilbert space $\mathbf{H}$ and write
\begin{equation}\label{eq:cohchangeofbasis}
    \Braket{x,w|\mathbbvar{z}} \;=\; \int \di^{2n}\mathbbvar{p}' \Braket{x,w|\mathbbvar{p}'} \Braket{\mathbbvar{p}'|\mathbbvar{z}}.
\end{equation}
Let us now use the notation $\braket{\mathbbvar{x}^M}:=\braket{\mathbbvar{z}|\hat{\mathbbvar{x}}^M|\mathbbvar{z}}$ for the mean position of a coherent sate $\Ket{\mathbbvar{z}}$ and again $(x^\mu,w^\mu)$ for the coordinates of the Lagrangian subspace $L\subset\mathcal{P}$ where the polarised wave-function lives. Let us use the expressions \eqref{eq:NCfreeparticle} and \eqref{eq:planewave} of a free particle in the doubled space to calculate the intermediate terms
\begin{equation}
    \begin{aligned}
    \Braket{x,w|\mathbbvar{p}'} \;&=\; \frac{1}{(2\pi\hbar)^{\frac{n}{2}}}\exp\!\left(\frac{i}{\hbar}p_\mu^{\prime}x^\mu\right)\delta^{(n)}(w - \tilde{p}^{\prime}), \\
    \Braket{\mathbbvar{p}'|\mathbbvar{z}} \;&=\; \frac{1}{(2\pi\hbar)^{n}} \exp\!\left(-\frac{i}{\hbar}\mathbbvar{p}'_M\!\Braket{\mathbbvar{x}^M}-\frac{\pi\alpha'}{4\hbar}\delta^{MN}\mathbbvar{p}'_M\mathbbvar{p}'_N \right).
    \end{aligned}
\end{equation}
Hence the integral \eqref{eq:cohchangeofbasis} becomes a Fourier transform in the physical momentum $p_\mu'$ only
\begin{equation*}
    \Braket{x,w|\mathbbvar{z}} = \left(\int\!\!\frac{\di^{n}p'}{(2\pi\hbar)^{\frac{3}{2}n}} \exp\!\left(\!-\frac{i}{\hbar}p'_\mu\!(x^\mu - \Braket{x^\mu}) - \frac{\pi\alpha'}{4\hbar}\left|p_\mu'\right|^2 \!\right)\!\right)\! \exp\!\left(\frac{i}{\hbar}w^\mu\!\Braket{\tilde{x}_\mu} - \frac{\pi\alpha'}{4\hbar}\left|w^\mu\right|^2 \!\right)\!.
\end{equation*}
Thus we obtain the following wave-function:
\begin{equation}\label{eq:cohwavefunction}
    \psi^{\mathrm{coh}}_w(x) \;=\; \frac{1}{(\pi^2\hbar\alpha')^{\frac{n}{2}}}\, \exp\!\left(-\frac{\left|x^\mu-\braket{x^\mu}\right|^2}{\pi\hbar\alpha'} \right)\frac{1}{(2\pi\hbar)^{\frac{n}{2}}}\, \exp\!\left(\frac{i}{\hbar}w^\mu\!\Braket{\tilde{x}_\mu} - \frac{\pi\alpha'}{4\hbar}\left|w^\mu\right|^2 \right)\!.
\end{equation}
We notice that the first term of this wave-function is a Gaussian on the physical position space and that the second term contains an exponential cut-off for large winding numbers.

\paragraph{Probability distribution.}
The probability of measuring a string with position $x^\mu$ and winding number $w^\mu$, for a given coherent state $\Ket{\mathbbvar{z}}$ with mean doubled position $\Braket{\mathbbvar{x}^M}$, will be immediately given by
\begin{equation}
    \left|\psi^{\mathrm{coh}}_w(x)\right|^2 \;=\; \frac{1}{(\pi^2\hbar\alpha')^{n}}\frac{1}{(2\pi\hbar)^{n}}\,\exp\!\left(-\frac{2}{\pi\hbar\alpha'}\left|x^\mu-\braket{x^\mu}\right|^2  - \frac{\pi\alpha'}{2\hbar}\left|w^\mu\right|^2 \right)
\end{equation}
This probability distribution exponentially decays by going away from the mean position $\Braket{x^\mu}$ on the physical position space and from zero on the winding number space. 

\paragraph{Limit $\alpha'\rightarrow 0$.}
It is immediate to notice that, in the limit $\alpha'\rightarrow 0$, the probability distribution spreads in the winding space and localizes in the physical position space. In other words, our probability distribution shrinks to a Dirac delta $\left|\psi^{\mathrm{coh}}_w(x)\right|^2\propto\delta^{(n)}(x-\Braket{x})$.

\paragraph{Change of T-duality frame.}
Now we want to find the T-dual wave-function of \eqref{eq:cohwavefunction}. To do so, we only need to express the same coherent state $\Ket{\mathbbvar{z}}\in\mathbf{H}$ in another basis of our Hilbert space, the basis $\Ket{\tilde{x},\tilde{w}}$ corresponding to the complementary Lagrangian subspace $\widetilde{L}\subset\mathcal{P}$. In other words we must calculate $\widetilde{\psi}^{\mathrm{coh}}_{\tilde{w}}(\tilde{x}) \;:=\; \Braket{\tilde{x},\tilde{w}|\mathbbvar{z}}$. 
We immediately obtain the following wave-function on the dual Lagrangian subspace $\widetilde{L}\subset\mathcal{P}$
\begin{equation}
    \widetilde{\psi}^{\mathrm{coh}}_{\tilde{w}}(\tilde{x}) \;=\; \frac{1}{(\pi^2\hbar\alpha')^{\frac{n}{2}}}\, \exp\!\left(-\frac{\left|\tilde{x}_\mu-\!\braket{\tilde{x}_\mu}\right|^2}{\pi\hbar\alpha'} \right) \frac{1}{(2\pi\hbar)^{\frac{n}{2}}}\, \exp\!\left(\frac{i}{\hbar}\tilde{w}_\mu\Braket{x^\mu} - \frac{\pi\alpha'}{4\hbar}\left|\tilde{w}_\mu\right|^2 \right).
\end{equation}
where the role of the physical coordinates and the T-dual coordinates is exchanged.

\subsection{Coherent states for non-canonical K\"{a}hler structures}

Notice that, from the point of view of holomorphic quantisation, in this section we always assumed that we have (at least locally) on our doubled space $\mathcal{M}$ a K\"{a}hler structure given by the symplectic matrix and the metric
\begin{equation}
    \omega_{MN}\;=\; \begin{pmatrix}0&\delta_\mu^{\;\nu}\\-\delta^\mu_{\;\nu}&0\end{pmatrix}, \qquad \delta_{MN}\;=\;\begin{pmatrix}\delta_{\mu\nu}&0\\0&\delta^{\mu\nu}\end{pmatrix},
\end{equation}
with the complex structure $I^M_{\;\;N}:=\omega^{ML}\delta_{LN}$, i.e.
\begin{equation}
    I^{M}_{\;\;N}\;=\; \begin{pmatrix}0&\delta_{\mu\nu}\\-\delta_{\mu\nu}&0\end{pmatrix}.
\end{equation}
This choice is legitimate since, in the perspective of the doubled phase space $\mathcal{P}$, we are working with canonical coordinates $\{\mathbbvar{x}^M,\mathbbvar{p}_M\}$.
If we want to perform a consistency check, we can change frame on the doubled space via a geometric $O(n,n)$-transformation, so that the symplectic form $\omega_{MN}$ and metric $\delta_{MN}$ are transformed to
\begin{equation}
    \omega_{MN}^{(B)}\;=\; \begin{pmatrix}B_{\mu\nu}&\delta_\mu^{\;\nu}\\-\delta^\mu_{\;\nu}&0\end{pmatrix}, \qquad \mathcal{H}_{MN}\;=\;\begin{pmatrix}g_{\mu\nu}- B_{\mu\lambda}g^{\lambda\rho}B_{\rho\nu} & B_{\mu\lambda}g^{\mu\nu} \\-g^{\mu\lambda}B_{\lambda\mu} & g^{\mu\nu} \end{pmatrix},
\end{equation}
with a new complex structure given by $I^{\prime M}_{\;\;\;N}:=\omega^{ML}\mathcal{H}_{LN}$. Since we started from an integrable K\"{a}hler structure we end up with another integrable K\"{a}hler structure. In the perspective of the doubled phase space $\mathcal{P}$, this corresponds to the choice of non-canonical coordinates $\{\mathbbvar{x}^M,\mathbbvar{k}^M\}$. Notice that the K\"{a}hler structure $(\omega^{(B)},I',\mathcal{H})$ is exactly the one derived by \cite{Svo19} in the context of para-Hermitian geometry.

\section{Metaplectic structure}


Let us focus on the doubled space $\mathcal{M}$. We observed that it comes, at least locally, equipped with a canonical symplectic form $\varpi$.
Let now assume that $(\mathcal{M},\varpi)$ is simply a $2n$-dimensional symplectic manifold and $L\subset T\mathcal{M}$ be a Lagrangian subbundle.

\paragraph{The metaplectic structure.}
The \textit{metaplectic group} $Mp(2n,\mathbb{R})$ is the universal double cover of the symplectic group $Sp(2n,\mathbb{R})$. It is, thus, given by a group extension of the form
\begin{equation}
    \begin{tikzcd}[row sep=7ex, column sep=5ex]
    0\arrow[r, hook] & \mathbb{Z}_2 \arrow[r, hook] & Mp(2n,\mathbb{R})\arrow[r, two heads] & Sp(2n,\mathbb{R}). \arrow[r, two heads] & 0
    \end{tikzcd}
\end{equation}
A \textit{metaplectic structure} on a symplectic manifold $(\mathcal{M},\varpi)$ is defined as the lift of the structure group $Sp(2n,\mathbb{R})$ of the bundle $T\mathcal{M}$ along the group extension $Mp(2n,\mathbb{R})\twoheadrightarrow Sp(2n,\mathbb{R})$.

There is a lemma (see \cite{WeiGQ}) which states that $T\mathcal{M}$ admits a metaplectic structure if and only if $L$ admits a metalinear structure.
Another result states \cite{WeiGQ} that the existence of a metalinear structure on a bundle $E$ is equivalent to the existence of the square root bundle $\sqrt{\mathrm{det}(E)}$.
By putting these two lemmas together we obtain that $T\mathcal{M}$ admits a metaplectic structure if and only if $\sqrt{\mathrm{det}({L})}$ exists.

The existence of a metaplectic structure is intimately linked to the definition of the canonical $\mathrm{Spin}(d,d)$ spinor bundle
\begin{equation}
    S_\mathcal{M} \;=\; \wedge^\bullet \widetilde{L} \otimes \sqrt{\mathrm{det}({L})}
\end{equation}
If the Lagrangian subbundle $L$ is integrable, there exists a submanifold $M\subset\mathcal{M}$ such that $L=TM$, i.e. the physical spacetime. In this case, the  canonical $\mathrm{Spin}(d,d)$ spinor bundle is isomorphic to the spinor bundle of generalised geometry on $M$, which is defined in \cite{Gualtieri:2007ng}. Thus, the isomorphism $L\oplus L^\ast$ by a B-shift $L\oplus\widetilde{L} \xrightarrow{ e^{-B} } L\oplus L^\ast$ can be immediately extended to an isomorphism
\begin{equation}
    S_\mathcal{M} \;\cong\; \wedge^\bullet T^\ast M \otimes \sqrt{\mathrm{det}({TM})}
\end{equation}
given by the untwist $\Phi \mapsto e^{-B}\wedge\Phi$ on polyforms $\Phi\in \wedge^\bullet \widetilde{L} \otimes \sqrt{\mathrm{det}({L})}$. Notice that this recovers a construction which is analogous to \cite{GMPW09}.

\paragraph{The quantum Hilbert space.}
The physical necessity for the existence of  $\sqrt{\mathrm{det}({L})}$ is that it is this measure that is used to construct the quantum  Hilbert space. In half-form quantisation, one thinks of a state as the combination of the wavefunction with the half-form used to construct its norm.

Thus, the quantum Hilbert space of this symplectic manifold, which will be:
\begin{equation}
    \mathbf{H} \;=\; \bigg\{\psi\in\Gamma\big(\mathcal{M},\,\mathcal{E}\otimes\sqrt{\mathrm{det}({L})}\big) \;\bigg|\; \nabla_V\psi=0 \;\; \forall V\in L \bigg\}.
\end{equation}
Let us now consider sections of the form $e^{-\phi}\sqrt{\mathrm{vol}_M} \in \Gamma\big(\mathcal{M},\sqrt{\mathrm{det}({L})}\big)$, where the top form is the Riemannian volume form $\mathrm{vol}_M := \sqrt{\mathrm{det}(g)}\, \di x^1 \wedge \dots \wedge \di x^n$ and $\phi\in\Coo(M)$ is just a function. Any section $\Ket{\psi}\in\mathbf{H}$ can be uniquely written as: 
\begin{equation}
\Ket{\psi} = \psi e^{-\phi} \otimes \sqrt{\mathrm{vol}_M} \,.
\end{equation}
With $\psi$ and $\phi$ obeying the polarisation condition.

For $A\in GL(n;\mathbb{R})$ acting on the bundle $L$, we have that sections of the square root bundle transform accordingly by
\begin{equation}
    e^{-\phi}\sqrt{\mathrm{vol}_M} \;\longmapsto\; \sqrt{\mathrm{det}(A)}\,(e^{-\phi}\sqrt{\mathrm{vol}_M})
\end{equation}
Consider a state $\ket{\psi}\in\mathbf{H}$. Let us call simply $\psi$ the corresponding wave-function. We, thus, have a Hilbert product given by
\begin{equation}
    \Braket{\psi|\psi} \;=\; \int_M \psi^\dagger\psi \,\sqrt{\mathrm{det}(g)}\, e^{-2\phi} \di x^1 \wedge \dots \wedge \di x^n
\end{equation}
where $\sqrt{\mathrm{det}(g)}\, e^{-2\phi}$ is nothing but the string frame measure and it is T-duality invariant.
By following the literature we can, define a T-duality invariant dilaton by
\begin{equation}
    d \;:=\; \phi - \frac{1}{2}\ln{\mathrm{det}(g)},
\end{equation}
so that we can rewrite the measure as $\sqrt{\mathrm{det}(g)}\, e^{-2\phi} = e^{-2d}$. 
Now, notice that a Hilbert product $\Braket{\psi|\psi}$ does not depend on the choice of polarisation and, therefore, it must be invariant under change of T-duality frame. Under the symplectomorphism encoding T-duality we have the volume half form transforming by
\begin{equation}
    e^{-d}\sqrt{\di {x}^1 \wedge \dots \wedge \di {x}^n} \;\mapsto\; e^{-d}\sqrt{\di \tilde{x}^1 \wedge \dots \wedge \di \tilde{x}^n}
\end{equation}
Thus, we can express the same state $\ket{\psi}\in\mathbf{H}$ as an $\widetilde{L}$-polarised section $\tilde{\psi}e^{-\tilde{\phi}}\otimes \sqrt{\mathrm{vol}_{\widetilde{M}}}$, where the dual measure is $\mathrm{vol}_{\widetilde{M}} := \sqrt{\mathrm{det}(\tilde{g})}\,  \di \tilde{x}^1 \wedge \dots \wedge \di \tilde{x}^n$. In this T-duality frame the Hilbert product will immediately have the following form:
\begin{equation}
    \Braket{\psi|\psi} \;=\; \int_{\widetilde{M}} \tilde{\psi}^\dagger\tilde{\psi} \,\sqrt{\mathrm{det}(\tilde{g})}\, e^{-2\tilde{\phi}} \di \tilde{x}^1 \wedge \dots \wedge \di \tilde{x}^n
\end{equation}
Thus the dilaton transformation arises from the transformation of the measure in the half-form quantisation of the string.

\paragraph{The Metaplectic correction to observables}
There is one further effect associated to the Metaplectic structure of quantisation. When we move to the representation of observables the operators now act on states in $\mathbf{H}$ i.e. $ \psi e^{-\phi} \otimes \sqrt{\mathrm{vol}_M}$ not just on the wavefunctions $\psi$. Practically that means there may be in additional contribution to an operator given by the Lie derivative generated by the vector field associated to the observable acting on the half form. Contributions of this type occur with holomorphic polarizations in which case the Hamiltonian operator is shifted by $1/2$. For the simple harmonic oscillator in quantum mechanics this is just the usual "zero-point" energy shift. In this context, the Hamiltonian constructed in section 5.24 would receive a zero-point shift. This would be relevant for T-fold type configurations where the space time moves between $x$ and $\tilde{x}$ spaces. Of course, we have only dealt with the bosonic string, it is a open question as to whether Fermionic contributions might cancel this shift for the full superstring.

\paragraph{The Maslov correction}

A related effect is the Maslov quantisation condition \cite{Arn67} (also known as Einstein–Brillouin–Keller quantisation) is
\begin{equation}
    \frac{1}{2\pi}\oint_{\gamma}\bbtheta \;=\; \hbar\left( n + \frac{\mu(\gamma)}{4} \right),
\end{equation}
where $n\in\mathbb{Z}$ and $\mu(\gamma)$ is the Maslov index of the loop $\gamma$.
Notice that the prequantisation condition $[\bbomega]\in H^2(\mathcal{P},\mathbb{Z})$ alone implies only that $\frac{1}{2\pi}\oint_{\gamma}\bbtheta = \hbar n$ for some integer $n\in\mathbb{Z}$. The Maslov quantisation condition adds an explicit correction to the quantisation procedure depending on the Maslov index of the loop.

These metaplectic/Maslov type corrections really only appear when the polarisation is non-trivial by which we mean in the double field theory context a spacetime that moves between $x$ and $\tilde{x}$ spaces. One expects such a description is needed for a T-fold where no global T-duality frame exists. These subtle "quantum" effects will then change the string spectrum in the T-fold background. We leave the detailed study of the metaplectic/Maslov corrections for T-folds for future work.

\section{Discussion}
This paper follows the approach of geometric quantisation for strings and links to the doubled space in double field theory. A key result is the identification of the stringy effects linked to the noncommutativity of the doubled space controlled by the string length. The choice of polarisation in quantisation then becomes the choice of duality frame. Transformations between frames is then given geometrically by changing polarisations and constructing the non-local transforms acting on wavefunctions. 
The construction of a double coherent state gives a minimal distance state which we can examine from the point of view of traditional polarisations. Finally, the subtle metaplectic effects may have important consequences for quantising strings on T-folds.

All of this leads to some further questions far outside the scope of this paper. Exceptional field theory is the extension of double field theory to M-theory where the U-duality group becomes a manifest symmetry. See \cite{Berman:2020tqn, Berman:2019biz1} for a recent reviews. Usually the properties of double field theory are shared with exceptional field theory. Here though seems a mystery. If double field theory is just phase space and its subsequent quantisation then what is exceptional field theory. Is there some sense in which it can be thought of as a more general "quantisation" with the generalised "phase space" being related to the extended space. Spacetime would no longer be a Lagrangian submanifold. Perhaps some clue is available in the construction of the basic states of theory as given in \cite{Berman:2014hna} where the branes were again momentum states in the extended space but now also combined with a type of  generalised monopole to give a self-dual configuration. Other mysterious properties of M-theory phase space have been noticed in \cite{Lust:2017bwq}. Other exotica that would be curious to explain from the phase space perspective would be the recently discovered non-Riemannian phase to double and exceptional field theory as discussed in \cite{Morand:2017fnv,Cho:2018alk,Berman:2019izh,Park:2020ixf,Blair:2020gng,Gallegos:2020egk}; this is also somewhat of a mystery from the quantisation perspective. Any insight into such backgrounds from the quantisation approach developed here would be very interesting and we leave for future work.

\acknowledgments

The authors would like to thank Chris Blair, Laurent Friedel, Emanuel Malek, Malcolm Perry, Franco Pezzella, Paul Townsend and Alan Weinstein for fruitful discussions. In particular, the authors thank Chris Blair, for sharing unpublished notes on the noncommutativity of the doubled space through world sheet quantisation, and Kevin T. Grosvenor, for sharing unpublished notes on the string-deformed Fourier transform. DSB is supported by the UK Science and Technology Facilities Council (STFC) with consolidated grant ST/L000415/1, String Theory, Gauge Theory and Duality.

\medskip
\addcontentsline{toc}{section}{References}
\bibliographystyle{JHEP}
\bibliography{sample}
\end{document}